\documentclass[letterpaper,twocolumn,10pt]{article}
\usepackage{usenix}
\usepackage[available]{usenixbadges}

\usepackage{array}
\usepackage{hyperref}
\usepackage{pifont}

\usepackage{booktabs}
\usepackage{multirow}
\usepackage{multicol}
\usepackage{threeparttablex}
\usepackage{longtable}

\usepackage{tikz}

\usepackage{graphicx}

\usepackage{arydshln} 
\usepackage{overpic}
\usepackage{enumitem}
\usepackage{appendix}
\usepackage{float}
\usepackage{tcolorbox}

\usepackage{authblk}
\usepackage{makecell}

\newcommand{\ignore}[1]{}

\newtheorem{observation}{{\bf Insight}}

\newcommand{\cop}[4]{
\begin{table*}[!htbp]
    \vspace{#2}
    \scriptsize
    \begin{threeparttable}
        \scalebox{1}{
        \hspace{#1}
        \begin{tabular}{l}
            #4 \\
        \end{tabular}
        }
    \end{threeparttable}
    \vspace{#3}
\end{table*}}

\usepackage{listings}
\lstdefinelanguage{diff}{
    basicstyle=\ttfamily\scriptsize,
    columns=flexible,
    frame=single,
    framesep=1em,
}

\makeatletter
\let\old@lstKV@SwitchCases\lstKV@SwitchCases
\def\lstKV@SwitchCases#1#2#3{}
\makeatother
\usepackage{lstlinebgrd}
\makeatletter
\let\lstKV@SwitchCases\old@lstKV@SwitchCases

\lst@Key{numbers}{none}{%
    \def\lst@PlaceNumber{\lst@linebgrd}%
    \lstKV@SwitchCases{#1}%
    {none:\\%
     left:\def\lst@PlaceNumber{\llap{\normalfont
                \lst@numberstyle{\thelstnumber}\kern\lst@numbersep}\lst@linebgrd}\\%
     right:\def\lst@PlaceNumber{\rlap{\normalfont
                \kern\linewidth \kern\lst@numbersep
                \lst@numberstyle{\thelstnumber}}\lst@linebgrd}%
    }{\PackageError{Listings}{Numbers #1 unknown}\@ehc}}
\makeatother

\begin{document}
%-------------------------------------------------------------------------------

%don't want date printed
% \date{}


\title{\Large \bf SoK: Automated Vulnerability Repair: Methods, Tools, and Assessments}

\author[1, 2]{Yiwei Hu}
\author[1, 2, 5 \thanks{Zhen Li is the corresponding author.}]{Zhen Li}
\author[2]{Kedie Shu}
\author[1, 2]{Shenghua Guan}
\author[1, 2, 5]{Deqing Zou}
\author[4]{Shouhuai Xu}
\author[1, 2, 5]{Bin Yuan}
\author[1, 3]{Hai Jin}

\renewcommand\Authands{, }

\affil[1]{\it National Engineering Research Center for Big Data Technology and System, Services Computing Technology and System Lab, Hubei Key Laboratory of Distributed System Security, Hubei Engineering Research Center on Big Data Security, Cluster and Grid Computing Lab}
\affil[2]{\it School of Cyber Science and Engineering, Huazhong University of Science and Technology, China}
\affil[3]{\it School of Computer Science and Technology, Huazhong University of Science and Technology, China}
\affil[4]{\it Department of Computer Science, University of Colorado Colorado Springs, USA}
\affil[5]{\it JinYinHu Laboratory, China}

\maketitle

%-------------------------------------------------------------------------------
%\tableofcontents

\begin{abstract}
The increasing complexity of software has led to the steady growth of vulnerabilities.
{\em Vulnerability repair} investigates how to fix software vulnerabilities. Manual vulnerability repair is labor-intensive and time-consuming because it relies on human experts, highlighting the importance of {\em Automated Vulnerability Repair} (AVR). 
In this SoK, we present the systematization of AVR methods through the three steps of AVR workflow: vulnerability analysis, patch generation, and patch validation.  
We assess AVR tools for C/C++ and Java programs as they have been widely studied by the community. 
Since existing AVR tools for C/C++ programs are evaluated with different datasets, which often consist of a few vulnerabilities, we construct the first C/C++ vulnerability repair benchmark dataset, dubbed \textsc{Vul4C}, which contains 144 vulnerabilities as well as their exploits and patches.
We use \textsc{Vul4C} to evaluate seven AVR tools for C/C++ programs and use the third-party \textsc{Vul4J} dataset to evaluate two AVR tools for Java programs. We also discuss future research directions. 
\end{abstract}
%-------------------------------------------------------------------------------
% Your abstract text goes here. Just a few facts. Whet our appetites.
% Not more than 200 words, if possible, and preferably closer to 150.

%-------------------------------------------------------------------------------
%-------------------------------------------------------------------------------
\section{Introduction}
%-------------------------------------------------------------------------------

The number of software vulnerabilities has been growing rapidly. For example, in 2023 alone,
30,927 vulnerabilities are added to the {\em National Vulnerability Database} (NVD),
and a majority of them are critical  \cite{skybox2024}. The large number of vulnerabilities on an annual basis 
highlights the importance of {\em Automated Vulnerability Repair} (AVR), which aims to fix vulnerabilities while minimizing, if not completely eliminating, the reliance on domain experts.

In principle, AVR is a special case of
{\em Automated Program Repair} (APR) because the latter aims to fix software defects that include {\em security defects} (i.e., vulnerabilities) and {\em functional defects} (i.e., non-vulnerabilities).
It is known that an APR technique may not be an effective AVR technique
\cite{pinconschi2021comparative, wu2023effective}
or even applicable to AVR (e.g., the statistical fault localization technique \cite{agarwal2014fault}). 

The status quo of AVR can be summarized as follows. 
First, the most closely related prior study is an independent and concurrent SoK 
\cite{li2025sok}, which focuses on the {\em patch generation} step of the AVR workflow, but not the entire AVR workflow. Moreover, their
empirical analysis \cite{li2025sok} focuses on four patch generation methods \cite{huang2019senx, gao2021extractfix, zhang2022vulnfix, kulsum24vrpilot} for C/C++ programs via two vulnerability datasets  \cite{gao2021extractfix, shen2021vulnloc}, leaving other AVR tools unaddressed.
Second, there are 
surveys on APR that also briefly discuss AVR \cite{gazzola2019automatic,huang2024evolving,monperrus2018automatic} and on learning-based AVR \cite{zhou2024large}. However, these surveys do not deal with the entire AVR workflow or other approaches to AVR.
Third, there are many empirical studies, such as:
\cite{pinconschi2021comparative, wu2023effective, bui24apr4vul} analyze the effectiveness of APR tools;
\cite{zhang2024pre} analyzes the effectiveness of pre-trained models for vulnerability repair; and  \cite{pearce2023examining, liu2024exploring} analyze the effectiveness of {\em Large Language Model} (LLM) for vulnerability repair. 
However, these studies do not consider AVR tools.
Fourth, there is neither a benchmark nor a unified assessment process for evaluating AVR tools geared towards C/C++ programs because 
existing studies either use their own datasets \cite{huang2019senx, gao2021extractfix, zhang2022vulnfix, shariffdeen2024crashrepair, chen2023vrepair, fu2022vulrepair, zhou2024vulmaster}
or use their own evaluation methods, making it infeasible to compare their results (e.g., some evaluation methods require testing patch validity \cite{huang2019senx, gao2021extractfix, zhang2022vulnfix, shariffdeen2024crashrepair} while others do not \cite{chen2023vrepair, fu2022vulrepair, zhou2024vulmaster}). The unsatisfying status quo motivates the present SoK.

\noindent{\bf Our contributions}.
This paper makes four contributions.
First, we systematize the problem of, and existing solutions to, AVR. 
We systematize AVR through its three steps of workflow: {\em vulnerability analysis}, {\em patch generation}, and {\em vulnerability validation}. The systematization leads to a number of insights, such as: 
(i) existing {\em vulnerability analysis} methods cannot accurately localize vulnerabilities;
(ii) template-based {\em patch generation} methods perform well on specific types of vulnerabilities but lack general applicability; (iii) static analysis-based {\em patch validation} methods incur high false-negatives due to their limited rules and the path explosion problem of symbolic executions.

Second, to enable fair comparison between AVR tools, 
we create the first C/C++ vulnerability repair benchmark dataset, dubbed \textsc{Vul4C},
which contains 144 vulnerabilities associated with 23 software products
and 19 {\em Common Weakness Enumeration} (CWE) types \cite{CWE}.
When compared with previous datasets, \textsc{Vul4C} has two features:
(i) it provides exploits, vulnerable programs, 
patches, and methods for triggering vulnerabilities; (ii) it covers a broader range of vulnerabilities, software, and vulnerability types. \textsc{Vul4C} represents a new baseline for AVR research 
and is available at \url{https://doi.org/10.5281/zenodo.15609776}.

Third, 
we apply \textsc{Vul4C} to evaluate seven AVR tools for C/C++ programs and apply the existing \textsc{Vul4J} dataset to evaluate two AVR tools for Java programs.
Our findings include: 
(i) semantics-based AVR tools outperform learning-based ones in generating high-quality patches;
(ii) learning-based AVR tools lack rigorous evaluation methodologies;
(iii) the plausible patches generated by semantics-based {\em patch generation} methods cannot fix the corresponding vulnerabilities but can assist developers in addressing vulnerabilities.

Fourth, the preceding findings inspire us to propose a research roadmap towards tackling the AVR problem. The roadmap highlights the importance of developing effective vulnerability analysis techniques to serve as a foundation of patch generation and developing automated vulnerability validation techniques. 

% Conference Version
% As a side product (in the full version of the paper \cite{fullversion}, owing to space limit), we apply \textsc{Vul4C} to evaluate two APR tools for C/C++ programs and the third-party {\sc Vul4J} dataset to evaluate two APR tools for Java programs. This allows us to draw insights into the difference between AVR and APR, such as: (i) Models with a stronger code comprehension capability perform better, and leveraging multiple models together can perform even better. (ii) Employing detailed thought processes in the CoT framework can enhance LLMs' patch generation capabilities, while lacking vulnerability information leads to repair failures.

% Full Version
As a side product, we apply \textsc{Vul4C} to evaluate two APR tools for C/C++ programs and the third-party {\sc Vul4J} to evaluate two APR tools for Java programs. We find:
(i) APR models with a stronger code comprehension capability perform better in patch generation, and leveraging multiple models together can perform even better; (ii) APR models employing detailed thought processes in CoT achieve a higher patch generation capability, while lacking vulnerability information causes repair failures.

\noindent{\bf Paper Outline}. Section \ref{section: framework} describes our systematization methodology. Section \ref{section: vulnerability_analysis} systematizes automated vulnerability analysis. Section \ref{section: patch-generation} systematizes automated patch generation. Section \ref{section: patch-validation} systematizes automated patch validation. 
Section \ref{sec:benchmark dataset} presents \textsc{Vul4C}.
Section \ref{section: evaluation} describes our empirical study.
Section \ref{section: discussion} discusses future research directions.
Section \ref{section: conclusion} concludes the paper. 
% Conference Version
% We defer many details to the full version \cite{fullversion}.

\section{Systematization Methodology}
\label{section: framework}

\subsection{Terminology and Scope}

\noindent{\bf Terminology}.
We use the following standard terms \cite{gazzola2019automatic, shen2021vulnloc}. A {\em test case}, or {\em test suite}, is often prepared by a software developer to verify whether a program meets its specification. A test case that meets the specification is called a {\em positive} case; otherwise, it is a {\em negative} case.
An {\em exploit} is an input that leads to the compromise a security property.
A {\em vulnerability location} is the lines of code (or statements) where a vulnerabilty resides;
a vulnerability {\em fix location} refers to the lines of code that need to be modified in order to fix a vulnerability;
these two locations may or may not be the same (see examples in Appendix \ref{Appendix: fix location v.s vulnerability location}).
A {\em candidate patch} is a patch that has yet to be validated, meaning it may or may not be valid; 
a {\em plausible patch} is a candidate patch that has been validated from a security perspective, but may not truly fix
a vulnerability because patch validation may not be thorough enough (e.g., only considering a few test cases)  \cite{wang2020automated}.

\noindent{\bf Scope}. We focus on AVR, which often takes the output of a vulnerability detection tool as input.
Although AVR is a special case of APR, the literature does not always distinguish them \cite{monperrus2018living, monperrus2018automatic, gazzola2019automatic, huang2024evolving}, perhaps because some APR techniques can be leveraged for AVR purposes, as shown in the present paper.
Nevertheless, the following three aspects highlight the difference between APR and AVR.

\begin{itemize}
[leftmargin=.32cm,noitemsep,topsep=0pt]
\item{\bf Purpose}.
AVR deals with vulnerabilities that can be exploited to compromise security properties. 
APR deals with software defects that make programs behave unexpectedly but may not be exploited to compromise security properties.

\item{\bf Input}.
AVR often takes one exploit as input for triggering and validating a vulnerability, where the input is often obtained via fuzzing and often difficult to translate into test cases that are required by APR \cite{shen2021vulnloc}.
Whereas, APR often takes as input a set of positive and negative test cases, 
which are crafted by human developers to localize defects and validate correctness of repaired programs.

\item{\bf Analysis}.
AVR often starts with an analysis 
to localize vulnerabilities (e.g., via the violated security constraints \cite{gao2021extractfix, huang2019senx} or exploitation behaviors \cite{shen2021vulnloc}) and then generates patches (e.g., by resolving the violated security constraints \cite{shariffdeen2024crashrepair} or by leveraging exploitation behaviors
\cite{zhou2024vulmaster}). The analysis is often specific to vulnerability types, as different types exhibit different characteristics. On the other hand, APR is geared towards repairing the functionality of a program, which is orthogonal to security properties.
\end{itemize}

The preceding discussion highlights that APR solutions are not sufficient for AVR purposes, and thus the community should treat AVR as a separate research problem.

\subsection{AVR Workflow}
\label{subsection: proposed-workflow-and-framework}

\begin{figure}[!htbp]
    \centering
\includegraphics[width=0.9\linewidth]{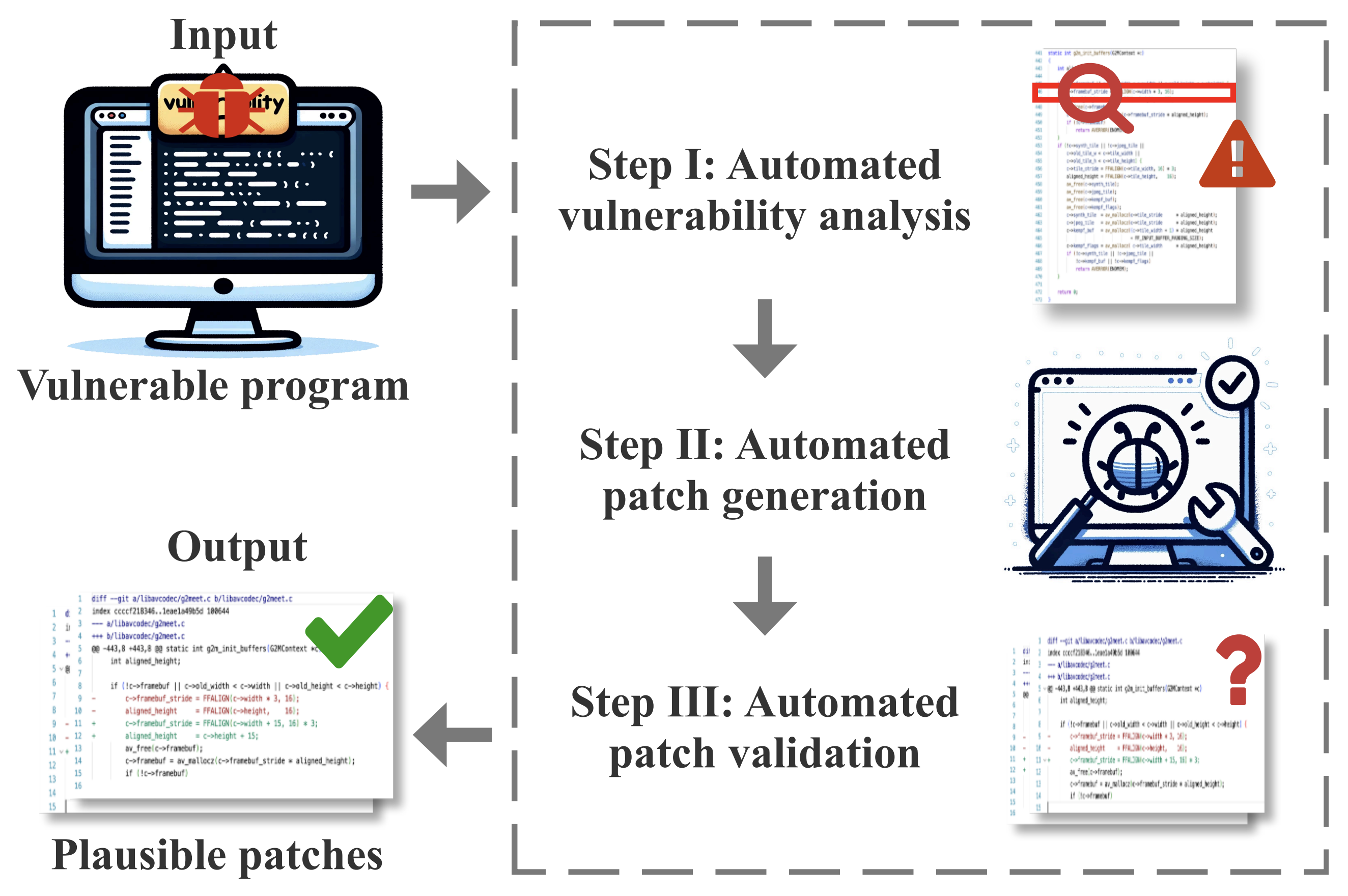}
    \caption{AVR workflow (three steps)}
    %: automated vulnerability localization, automated patch generation, and automated patch validation.}
    \label{fig: AVR Workflow}
\end{figure}

Figure \ref{fig: AVR Workflow} depicts the AVR workflow in three steps:
automated vulnerability analysis, automated patch generation, and automated patch validation, which are highlighted below and elaborated in subsequent sections.

\noindent{\bf Automated vulnerability analysis}. This step takes a vulnerable program as input (e.g., the output of a vulnerability detector), analyzes the vulnerable program, and outputs useful vulnerability characteristics (e.g., vulnerability type, violated security property, root cause, vulnerability location, fix location).
There are four approaches to automated vulnerability analysis: 
(i) {\em value-flow analysis}, which tracks the propagation of data and control in a program;
(ii) {\em formal analysis}, which leverages mathematical modeling and logical reasoning;
(iii) {\em symbolic execution}, which replaces program variables with symbolic expressions and systematically explores execution paths;
(iv) {\em statistical analysis}, which leverages 
statistical features to identify correlations.

\noindent{\bf Automated patch generation}.
This step generates one or multiple {\em candidate patches} for a given vulnerability.
There are four approaches: 
(i) {\em search-based}, which searches candidate patches in a pre-defined patch space; 
(ii) {\em template-based}, which leverages abstract patch templates;   
(iii) {\em semantics-based}, which leverages program semantics;
and (iv) {\em learning-based}, which leverages deep learning.

\noindent{\bf Automated patch validation}. 
This step determines the validity of a candidate patch generated in the preceding step, rejecting or validating it as a {\em plausible patch}.
There are two approaches to patch validation:  (i) {\em static analysis-based}, which verifies a candidate patch without executing the patched program;
(ii) {\em dynamic analysis-based}, which executes a patched program with the exploit and test cases in question (if available) and observes the program's runtime behavior.

%-------------------------------------------------------------------------------
%-------- Refer to SoK: Prudent Evaluation Practices for Fuzzing ---------------
\subsection{Identifying AVR Literature}
\label{subsection: select-literature}
%-------------------------------------------------------------------------------

We use Google Scholar to search for papers published in the CORE2023 A/A* conferences \cite{CORERanking} or top-tier journals (TDSC, TIFS, TSE, TOSEM, ESE, TC)
between 2000 and 2024
using the following eight keywords: {\em vulnerability repair}, {\em vulnerability fix}, {\em vulnerability patch}, {\em vulnerability patch generation}, {\em vulnerability fix generation}, {\em automated vulnerability repair}, {\em automated vulnerability fix}, and {\em automated vulnerability patch}. For each keyword, we consider the first 100 entries ranked by relevance given by Google Scholar. This leads to 800 papers for our manual examination based on their technical relevance, which results in 32 papers, including: (i) 27 papers that present 27 AVR tools, respectively; (ii) five empirical studies on vulnerability repair.

\begin{figure*}[!htbp]
    \centering
    \includegraphics[width=1\linewidth]{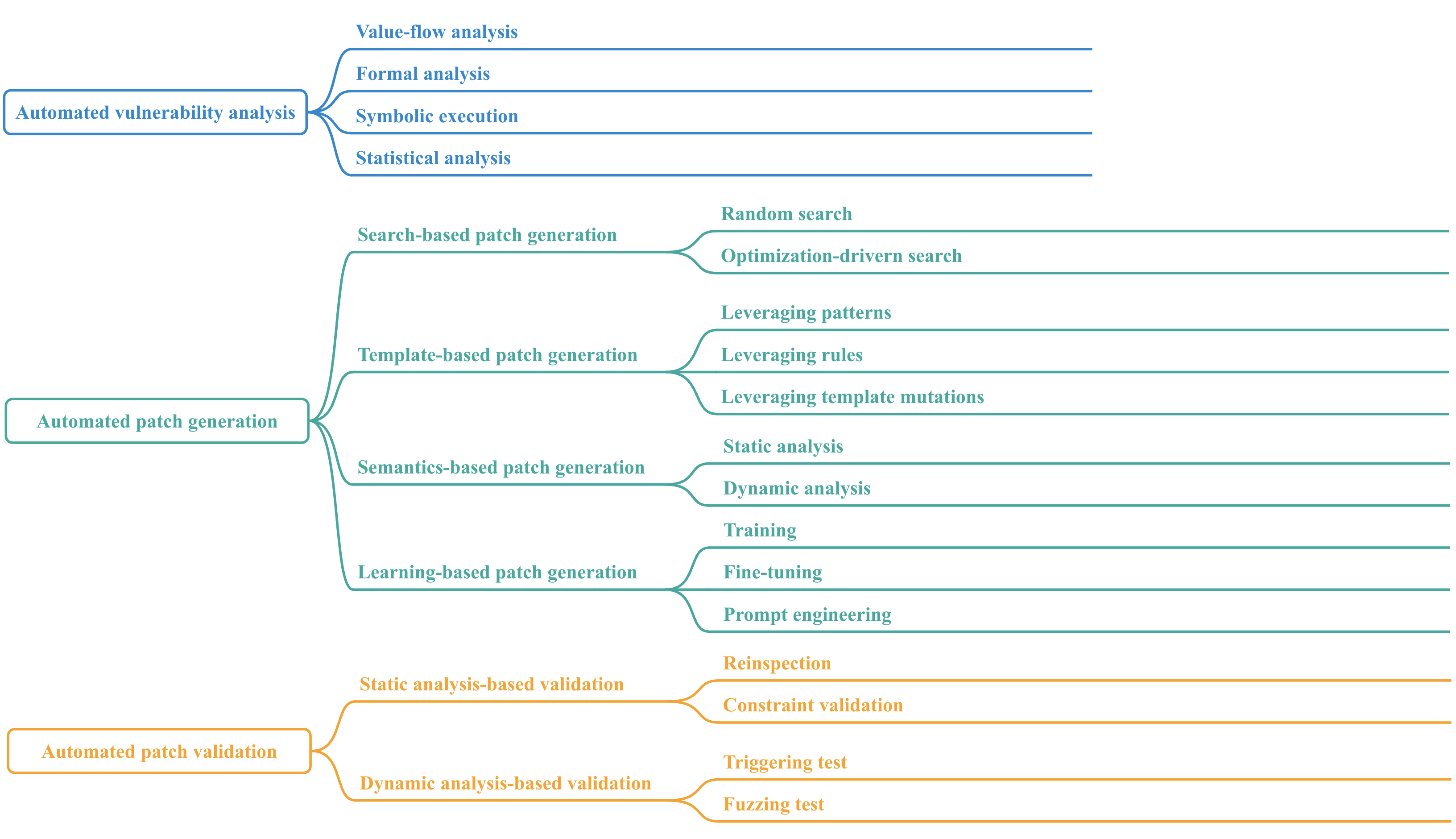}
    % 这个最后再补 analysis
    
    \caption{Characterizing the 64 AVR tool through AVR workflow: vulnerability analysis, patch generation, and patch validation}
    % \vspace{-0.6cm}
    \label{fig: AVR overview}
\end{figure*}

\cop{6.55 cm}{-11.24 cm}{10 cm}{\cite{gao2015leakfix, lee2018memfix, cheng2017intpti, zhang2010intpatch, wang2024contracttinker, zhang2020smartshield, hua2021rupair, son2013fix, shi2022skyport, shaw2014automatically, cheng2019cintfix, ferreira2022elysium, hong2020saver, Utture2023RLFixer, lin2007autopag}}
\cop{6.55 cm}{-10.40 cm}{10 cm}{\cite{li2022regexscalpel, fang2016web, chida2022remedy, yu2024tapfixer, xu2020vulmet, so2023smartfix, van2018footpatch, xing2024conch}}
\cop{6.55 cm}{-10.38 cm}{10 cm}{\cite{mechtaev2016angelix, tian2017errdoc, liu2023symlogrepair, chen2017karma, huang2019senx, duan2019osspatcher, tolmach2022definery, shariffdeen2020patchweave, nguyen2021sguard, gao2016bovinspector, gao2021extractfix, shariffdeen2024crashrepair}}
\cop{6.55 cm}{-10.40 cm}{10 cm}{\cite{novark2007exterminator, mesecan2021hypergi, weimer2009automatically, kim2013par, xu2019vfix}}

% Search based
\cop{11.8 cm}{-10.21 cm}{5.76 cm}{\cite{le2012genprog}}
\cop{11.8 cm}{-6.17 cm}{6 cm}{\cite{gao2019fix2fit, mesecan2021hypergi, tolmach2022definery, yu2024tapfixer}}
% Template based
\cop{11.8 cm}{-6.48 cm}{5.2 cm}{\cite{lin2007autopag, Coker2013cint, gao2016bovinspector, ma2017vurle, muntean2019intrepair, zhang2022Seader, lee2022NPEX, shahoor2023leakpair, Utture2023RLFixer, xing2024conch,so2023smartfix, zhang2020smartshield, nguyen2021sguard}\\ \cite{rodler2021evmpatch, ferreira2022elysium, yang2023tsbport, shariffdeen2020patchweave, shariffdeen2021fixmorph, shi2022skyport, son2013fix, li2022regexscalpel, shaw2014automatically, novark2007exterminator, chen2013safestack, duck2020e9patch, hua2021rupair}}
\cop{11.8 cm}{-5.46 cm}{5.7 cm}{\cite{xu2019vfix, Coker2013cint, ma2016cdrep, durieux2017npefix}}
\cop{11.8 cm}{-6.09 cm}{4.5 cm}{\cite{so2023smartfix, kim2013par}}
% Semantics based
\cop{11.8 cm}{-4.81 cm}{4.0 cm}{\cite{hong2020saver, van2018footpatch, cheng2017intpti, lee2018memfix, gao2015leakfix, zhang2010intpatch, fang2016web, chida2022remedy}}
\cop{11.8 cm}{-4.39 cm}{3.5 cm}{\cite{huang2019senx, gao2021extractfix, zhang2022vulnfix, shariffdeen2024crashrepair, shariffdeen2021cpr, chen2017karma, xu2020vulmet, duan2019osspatcher, liu2023symlogrepair, mechtaev2016angelix}}
%Learning based
\cop{11.8 cm}{-3.88 cm}{3.0 cm}{\cite{harer2018learning, zhou2022spvf, fu2024vqm}}
\cop{11.8 cm}{-3.39 cm}{2.5 cm}{\cite{chen2023vrepair, fu2022vulrepair, zhou2024vulmaster, chi2023seqtrans}}
\cop{11.8 cm}{-2.9 cm}{2.0 cm}{\cite{pan2024ppathf, wang2024contracttinker}}
% Static analysis
\cop{11.8 cm}{-2.31 cm}{1.5 cm}{\cite{gao2016bovinspector, van2018footpatch, Utture2023RLFixer, so2023smartfix, wang2024contracttinker, zhang2020smartshield, ferreira2022elysium}}
\cop{11.8 cm}{-1.9 cm}{1.5cm}{\cite{muntean2019intrepair, gao2021extractfix, lee2022NPEX, tolmach2022definery, son2013fix}}
%dynamic analysis
\cop{11.8 cm}{-1.97 cm}{1.5cm}{\cite{huang2019senx, lin2007autopag, shahoor2023leakpair, xu2019vfix, shariffdeen2021cpr, xu2020vulmet, chen2013safestack, duan2019osspatcher}\\ \cite{li2022regexscalpel, le2012genprog, gao2019fix2fit, cheng2019cintfix, durieux2017npefix, weimer2009automatically, kim2013par, rodler2021evmpatch, cheng2017intpti}}
\cop{11.8 cm}{-1.75 cm}{1.5cm}{\cite{shariffdeen2024crashrepair, shariffdeen2020patchweave, hua2021rupair}}

For each of the 32 papers, we use the snowballing approach \cite{wohlin2014snowballing} (as in \cite{lopez2024sok}) to identify other relevant papers published in CORE2023 A/A* conferences \cite{CORERanking} or top-tier journals, leading to 38 additional papers that present 37 AVR tools and one empirical study.
In total, we obtain $32+38=70$ papers for systematization, including 64 AVR tools and six empirical studies. 
Figure \ref{fig: AVR overview} highlights the 64 AVR tools in terms of their workflow.
Table \ref{tab: AVR-Comparison} compares the 64 AVR tools via attributes that will be described in Sections \ref{section: vulnerability_analysis}-\ref{section: patch-validation}.

%-------------------------------------------------------------------------------
\section{Automated Vulnerability Analysis}
\label{section: vulnerability_analysis}
%-------------------------------------------------------------------------------

%-------- Refer to SoK: Anti-Facial Recognition Technology ---------------

\begin{table*}[!htb]
    \centering
    \caption{Comparison of the 64 AVR tools}
   % \vspace{0.2cm}
    \begin{threeparttable}
    \resizebox{\textwidth}{!}{
    \begin{tabular}{|l|l|c|c|c|c c c c|c c c c|c c|c:c|c:c:c:c|c:c|} 
        \hline
        \multicolumn{1}{|c}{\textbf{Method}} 
        & \multicolumn{1}{|c}{\textbf{Venue}}
        & \multicolumn{1}{|c}{\textbf{Fix level}}
        & \multicolumn{1}{|c|}{\textbf{Language}} 
        & ~ 
        
        & \multicolumn{4}{c}{\textbf{Analysis}}             
        & \multicolumn{4}{|c}{\textbf{Generation}} 
        & \multicolumn{2}{|c}{\textbf{Validation}}

        & \multicolumn{2}{|c}{\textbf{Analysis attributes}}
        & \multicolumn{3}{|c}{\textbf{Generation attributes}}
        & \multicolumn{2}{|c|}{\textbf{Validation attributes}}
        \\
        
        ~  &  ~  & ~ & ~ & \rotatebox{90}{Accessibility\tnote{1}} 
        
        & \rotatebox{90}{Value-Flow Analysis}  
        & \rotatebox{90}{Formal Analysis}
        & \rotatebox{90}{Symbolic Execution}
        & \rotatebox{90}{Statistical Analysis}

        & \rotatebox{90}{Search-based}
        & \rotatebox{90}{Template-based} 
        & \rotatebox{90}{Semantics-based}
        & \rotatebox{90}{Learning-based}
        
        & \rotatebox{90}{Static Analysis-based}
        & \rotatebox{90}{Dynamic Analysis-based}
        
        & \rotatebox{90}{Applicability\tnote{2}} 
        & \rotatebox{90}{Input features\tnote{3}}
        
        & \rotatebox{90}{Applicability\tnote{2}}
        & \rotatebox{90}{Input features\tnote{3}}
        & \rotatebox{90}{Granularity\tnote{4}}
        & \rotatebox{90}{Edit type\tnote{5}}

        & \rotatebox{90}{Input features\tnote{3}}  
        & \rotatebox{90}{Rank algorithm\tnote{6}}
        
        \\ \hline

      Coker et al.\cite{Coker2013cint} & ICSE'13  & Source & C & \ding{56}                    
      &           &             &       &
      &           & \ding{52}   &       &            
      &           &
      & --
      & --    
      & IE        & P
      & Stat.     & --
      & --        
      & --
      \\ \hline
      
      BovInspector\cite{gao2016bovinspector} & ASE'16 & Source & C & \ding{52}   
      &           &             & \ding{52} &
      &           & \ding{52}   &           &               
      &\ding{52}  &
      & BO      
      & P        
      & BO        & P 
      & Stat.     & MH 
      & PP  
      & --
      \\ \hline
      
      VuRLE\cite{ma2017vurle} & ESORICS'17  & Source & Java    & \ding{56}                  
      &           &             &       &
      &           & \ding{52}   &       &  
      &           &  
      & --        
      & --     
      & GP        & P
      & Prog.     & MH  
      & --        
      & -- 
      \\ \hline

      Seader\cite{zhang2022Seader} & ICPC'22 &  Source & Java  & \ding{52}            
      &           &             &       &
      &           & \ding{52}   &       &
      &           &   
      & --         
      & --          
      & CM        & P 
      & Prog.     & MH 
      & --        
      & -- 
      \\ \hline

      CONCH\cite{xing2024conch} & USENIX SEC'24 & Source & C/Java & \ding{56}    
      &           & \ding{52}   &       & 
      &           & \ding{52}   &       &
      &           & 
      & NPD          
      & P              
      & NPD       & P
      & Prog.     &  MH   
      & --           
      & -- \\  \hline
    
      IntPTI\cite{cheng2017intpti}      & ASE'17  & Source & C    & \ding{52}        
      & \ding{52} &           &       &
      &           &           & \ding{52}   &     
      &           & \ding{52}
      & IE         
      & P             
      & IE        & P
      & Prog.     & SH
      & PP \& T     
      & --
      \\ \hline

      IntPatch\cite{zhang2010intpatch} &  ESORICS'10 & Source &  C/C++ &  \ding{56} 
      & \ding{52} &           &             &    
      &           &           & \ding{52}   &
      &           & 
      & IO, BO        
      & P             
      & IO, BO    & P
      & Stat.     & MH  
      & -- 
      & -- 
      \\ \hline

      FootPatch\cite{van2018footpatch} & ICSE'18 & Source &  C/Java  &  \ding{52}   
      &           & \ding{52} &             &  
      &           &           & \ding{52}   & 
      & \ding{52} &
      & Heap-Related          
      & P                  
      & NPD, ML, RL & P
      & Stat.     & MH
      & PP \& T         
      & --   
      \\ \hline

      SAVER\cite{hong2020saver} &  ICSE'20 & Source &  C  & \ding{52}  
      & \ding{52} &           &             &  
      &           &           & \ding{52}   &
      &           & 
      & Memory Errors             
      & P    
      & ML, DF, UAF & P
      & Prog.     & MH
      & --         
      & --
      \\ \hline

      MemFix\cite{lee2018memfix} & ESEC/FSE'18 & Source &  C  & \ding{52}  
      & \ding{52} &           &             & 
      &           &           & \ding{52}   &
      &           & 
      & ML, DF, UAF         
      & P               
      & ML, DF, UAF & P
      & Prog.     & MH
      & --          
      & \ding{52}
      \\ \hline

      LeakFix\cite{gao2015leakfix}  & ICSE'15 & Source & C &\ding{56}            
      & \ding{52} &           &             & 
      &           &           & \ding{52}   &         
      &           & 
      & ML
      & P    
      & ML        & P
      & Prog.     & MH
      & --        & -- \\ \hline

      LeakPair\cite{shahoor2023leakpair} & ASE'23 & Source & JavaScript & \ding{52}                
      &           &           &             &    
      &           & \ding{52} &             &
      &           & \ding{52}
      & --          
      & --          
      & ML        & P
      & Prog.     &  MH    
      & PP \& T      
      & --
      \\ \hline  

      RLFixer\cite{Utture2023RLFixer} & ESEC/FSE'23  & Source  & Java & \ding{56}              
      & \ding{52} &           &             & 
      &           & \ding{52} &             &
      &\ding{52}  & 
      & RL        
      & P         
      & RL        & P 
      & Prog.     & -- 
      & PP
      & -- 
      \\ \hline

      Remedy\cite{chida2022remedy} & S\&P'22 & Source & -- & \ding{52}              
      &           & \ding{52} &             & 
      &           &           & \ding{52}    &
      &           &    
      & DoS
      & P            
      & DoS       & P 
      & Stat.     & -- 
      & --           
      & --
      \\ \hline  

      AutoPaG \cite{lin2007autopag} & AsiaCCS'07 & Source & C/C++ & \ding{56}                 
      & \ding{52} &           &             &       
      &           & \ding{52}          &             &
      &           & \ding{52}     
      & OOB          
      & P \& E    
      & OOB       & P \& E
      & Stat.     & MH   
      & PP \& E          
      & --
      \\ \hline

       NPEX\cite{lee2022NPEX} & ICSE'22 & Source & Java  & \ding{52}                
      &           &           &             & 
      &           & \ding{52} &             &
      & \ding{52} & 
      & --        
      & --            
      & NPE       & P \& E
      & Stat.     & --   
      & PP             
      & --  
      \\ \hline 
      
      VFix\cite{xu2019vfix} & ICSE'19 & Source & C/C++/Java &  \ding{56}   
      &           &           &             & \ding{52}      
      &           & \ding{52} &             &
      &           & \ding{52}  
      & NPE    
      & P \& E
      & NPE       & P \& E
      & Prog.     & MH 
      & PP \& T 
      & --
      \\ \hline

      Yu et al.\cite{fang2016web}   & ISSTA'16 & Source  & PHP    &  \ding{56}         
      &           & \ding{52} &             & 
      &           &           & \ding{52}   &  
      &           & 
      & String Vulnerability          
      & P         
      & String Vulnerability    & P
      & Prog.     & MH 
      & --        
      & -- 
      \\ \hline

      Senx\cite{huang2019senx} &  S\&P'19 & Source &  C/C++ &  \ding{56}  
      &          &            & \ding{52}   & 
      &          &            & \ding{52}   &
      &          & \ding{52}      
      & BO, BC, IO          
      & P \& E  
      & BO, BC, IO  & P \& E
      & Prog.    & MH  
      & PP \& E         
      & --
      \\ \hline  

       ExtractFix\cite{gao2021extractfix} & TOSEM'20 & Source & C/C++ & \ding{52}                 
      &           &           & \ding{52}   & 
      &           &           & \ding{52}   &
      & \ding{52} & 
      & MRV        
      & P \& E       
      & MRV       & P \& E
      & Prog.     & MH   
      & --          
      & -- 
      \\ \hline

      CPR\cite{shariffdeen2021cpr} & PLDI'21 & Source & C/C++  & \ding{52}                 
      &           &           &             & 
      &           &           & \ding{52}   &
      &           & \ding{52} 
      & --    
      & --             
      & GP        & P \& T
      & Prog.     &  SH    
      & PP \& E           
      & \ding{52} 
      \\ \hline

      VulnFix\cite{zhang2022vulnfix} & ISSTA'22 & Source & C/C++    & \ding{52}        
      &           &           &             &        
      &           &           & \ding{52}   &
      &           &    
      & --
      & --          
      & GP        & P
      & Prog.     & MH 
      & --        
      & -- 
      \\ \hline

      CrashRepair\cite{shariffdeen2024crashrepair} & TOSEM'24 & Source & C/C++ & \ding{52}         
      &           &           & \ding{52}   & 
      &           &           & \ding{52}   &
      &           & \ding{52} 
      & MRV       
      & P \& E
      & MRV       & P \& E
      & Prog.     & SH    
      & PP \& E          
      & \ding{52}
      \\ \hline  
      
      Harer et al.\cite{harer2018learning}  & NeurIPS'18 & Source & C/C++ & \ding{56}             
      &           &           &             & 
      &           &           &             & \ding{52}
      &           & 
      & --        & --   
      & GP        & P
      & Func.     & --
      & --        
      & -- 
      \\ \hline

      SPVF\cite{zhou2022spvf} & ESE'22 & Source & C/C++/Python    & \ding{56}                
      &           &           &             & 
      &           &           &             & \ding{52}
      &           &            
      & --
      & --  
      & GP        & P
      & Func.  
      & --
      & --
      & --
      \\ \hline

      VulRepair\cite{fu2022vulrepair}  & ESEC/FSE'22 & Source & C/C++   & \ding{52}         
      &           &           &             & 
      &           &           &             & \ding{52}
      &           &    
      & --        & --   
      & GP        & P
      & Func.     & MH 
      & --        & -- \\ \hline

      VRepair\cite{chen2023vrepair}  & TSE'23 & Source & C/C++  & \ding{52}        
      &           &           &             & 
      &           &           &             & \ding{52}   
      &           & 
      & --        & --  
      & GP        & P
      & Func.     & MH
      & --        & -- \\ \hline

      SeqTrans\cite{chi2023seqtrans}  & TSE'23 & Source & Java  & \ding{52}            
      &           &          &             & 
      &           &          &              & \ding{52}
      &           & 
      & --        & -- 
      & GP        & P
      & Func.     & SH 
      & --        & -- \\ \hline

      VQM\cite{fu2024vqm}  & TOSEM'24 & Source & C/C++ & \ding{52}
      &           &          &             & 
      &           &          &              & \ding{52}
      &           & 
      & --        & -- 
      & GP        & P
      & Func.     & MH
      & --        & -- \\ \hline

      VulMaster\cite{zhou2024vulmaster}  & ICSE'24 & Source & C/C++/Java & \ding{52}
      &           &          &             & 
      &           &          &              & \ding{52}
      &           & 
      & --        & -- 
      & GP        & P
      & Func.     & MH
      & --        & -- \\ \hline

% ---------- Patch Transplantation -------------

      PatchWeave\cite{shariffdeen2020patchweave} & TOSEM'20 & Source & C/C++ & \ding{52}
      &           &           & \ding{52}   & 
      &           & \ding{52} &             & 
      &           & \ding{52}
      & GP        & P \& HP
      & GP        & P \& HP
      & Prog.     & MH
      & PP \& T   & --
      \\ \hline

      FixMorph\cite{shariffdeen2021fixmorph} & ISSTA'21 & Source & C & \ding{52}
      &           &           &             & 
      &           & \ding{52} &             & 
      &           &
      & --        & --
      & GP        & P \& HP
      & File.     & MH
      & --   & --
      \\ \hline

      SKYPORT\cite{shi2022skyport} & USENIX SEC'22 & Source & PHP & \ding{56}
      & \ding{52} &           &             & 
      &           & \ding{52} &             & 
      &           &
      & Injection Vulnerabilities  & P \& HP
      & Injection Vulnerabilities  & P \& HP
      & Prog.     & MH
      & --        & --
      \\ \hline

      TSBPORT\cite{yang2023tsbport} & CCS'23 & Source & C & \ding{52}
      &           &           &             & 
      &           & \ding{52} &             & 
      &           &
      & --        & --
      & GP        & P \& HP
      & File.     & MH
      & --        & \ding{52}
      \\ \hline

      PPatHF\cite{pan2024ppathf} &  ISSTA'24 & Source & C/C++ & \ding{52}
      &           &          &             & 
      &           &          &              & \ding{52}
      &           &
      & --        & --
      & GP        & P \& HP
      & Func.     & MH
      & --        & --
      \\ \hline

% ---------- HotPatch -------------
      KARMA\cite{chen2017karma} & USENIX SEC'17 & Binary & C/C++ & \ding{56}
      &           &          & \ding{52}    & 
      &           &          & \ding{52}    &
      &           &
      & GP        & P
      & GP        & P
      & Func.     & --
      & --          & --
      \\ \hline

      VULMET\cite{xu2020vulmet} & USENIX SEC'20 & Binary & C/C++ & \ding{56}
      &           & \ding{52}&             & 
      &           &          & \ding{52}     &
      &           & \ding{52}
      & GP        & P \& HP
      & GP        & P \& HP
      & Func.     & --
      & PP \& E   & --
      \\ \hline

% ---------- Binary -------------
     
     SafeStack\cite{chen2013safestack} & TDSC'13 & Binary & C/C++ & \ding{56}
      &           &           &             & 
      &           & \ding{52} &             &
      &           & \ding{52}
      & --        & --
      & BO        & P \& E
      & Prog.     & --
      & PP \& T   & --
      \\ \hline
     OSSPatcher\cite{duan2019osspatcher} & NDSS'19 & Binary & C/C++ & \ding{56}
      &           &          & \ding{52}    & 
      &           &          & \ding{52}    &
      &           & \ding{52}
      & GP        & P \& HP
      & GP        & P \& HP
      & Func.     & --
      & PP \& E   & --
      \\ \hline

     E9Patch\cite{duck2020e9patch} & PLDI'20 & Binary & C/C++/Fortran & \ding{52}
      &           &           &             & 
      &           & \ding{52} &             &
      &           &
      & --          & --
      & GP        & P
      & Prog.     & --
      &           & --
      \\ \hline
      
% ---------- WebApplication -------------
     FixMeUp\cite{son2013fix} & NDSS'13 & Source & PHP & \ding{56}
      & \ding{52} &           &             & 
      &           & \ding{52} &            &
      & \ding{52} &
      & AC        & P
      & AC        & P
      & Stat.     & SH
      & PP \& E   & --
      \\ \hline

      RegexScalpel\cite{li2022regexscalpel} & USENIX SEC'22 & Source & - & \ding{56}
      &           & \ding{52} &             & 
      &           & \ding{52} &             &
      &           & \ding{52}
      & DoS       & P
      & DoS       & P
      & Stat.     & --
      & PP \& T   & --
      \\ \hline
% ---------- Rust -------------
      Rupair\cite{hua2021rupair} & ACSAC'21 & Source & Rust & \ding{56}
      & \ding{52} &           &             & 
      &           & \ding{52} &             &
      &           & \ding{52}
      & BO        & P
      & BO        & P
      & Prog.     & --
      & PP \& T   & --
      \\ \hline
% --------- ??? --------
      TAPFixer\cite{yu2024tapfixer} & USENIX SEC'24 & -- & --  & \ding{52}
      &           & \ding{52}&             & 
      & \ding{52} &          &              &
      &           &
      & Trigger Action Programming          & --
      & Trigger Action Programming          & --
      & --          & --
      & --          & --
      \\ \hline

      GenProg\cite{le2012genprog} & TSE'12 & Source & C/C++  & \ding{52}
      &           &          &             & 
      & \ding{52} &          &              & 
      &           & \ding{52}
      & --        & --
      & GP        & P \& T
      & Stat.     & MH
      & PP \& T   & --
      \\ \hline

      Errdoc\cite{tian2017errdoc} & ESEC/FSE'17 & Source & C/C++ & \ding{52}
      &           &          & \ding{52}    & 
      &           &          & \ding{52}    &
      &           &
      & Error Handling       & P
      & Error Handling       & P
      & Func.     & MH
      & --        & --
      \\ \hline

      HyperGI\cite{mesecan2021hypergi} & ASE'21 & Source & C/C++ & \ding{56}
      &           &          &             & \ding{52}
      & \ding{52} &          &              & 
      &           &
      & Information Flow Leakage            & P \& T
      & Information Flow Leakage            & P \& T
      & Prog.     & --
      & --        & --
      \\ \hline

      Fix2Fit\cite{gao2019fix2fit} & ISSTA'19 & Source & C/C++ & \ding{52}
      &           &          &             & 
      & \ding{52} &          &              & 
      &           & \ding{52}
      & --        & --
      & GP        & P \& T
      & Prog.     & MH
      & PP \& T   & --
      \\ \hline

      Shaw et al.\cite{shaw2014automatically} & DSN'14 & Source & C/C++ & \ding{56}
      & \ding{52} &           &             & 
      &           & \ding{52} &             &
      &           &
      & BO        & P
      & BO        & P
      & Stat.     & SH
      & --        & --
      \\ \hline

      Exterminator\cite{novark2007exterminator} & PLDI'07 & Binary & C/C++ & \ding{56}
      &           &           &             & \ding{52}
      &           & \ding{52} &             &
      &           &
      & BO, Dangling Pointer  & P \& E
      & BO, Dangling Pointer  & P \& E
      & Prog.     & --
      & --        & --
      \\ \hline
      Angelix\cite{mechtaev2016angelix} & ICSE'16 & Source & C/C++ & \ding{52}
      &           &          & \ding{52}    & 
      &           &          & \ding{52}    &
      &           &
      & --        & P
      & --        & P
      & Prog.     & MH
      & --        & --
      \\ \hline

      CIntFix\cite{cheng2019cintfix} & TC'19  & Source & C & \ding{52}
      & \ding{52} &           &             & 
      &           & \ding{52} &             &
      &           & \ding{52}
      & IE        & P
      & IE        & P
      & Prog.     & MH
      & PP \& T   & --
      \\ \hline

      CDRep\cite{ma2016cdrep} & AsiaCCS'16  & Source & Java & \ding{56}
      &           &           &             & 
      &           & \ding{52} &             &
      &           &
      & --        & --
      & CM        & P
      & File.     & MH
      & --        & --
      \\ \hline

      NPEFix\cite{durieux2017npefix} & SANER'17  & Source & Java & \ding{52}
      &           &           &             & 
      &           & \ding{52} &             &
      &           & \ding{52}
      & --        & --
      & NPE       & P \& T
      & Prog.     & MH
      & PP \& T   & -
      \\ \hline

      Weimer et al.\cite{weimer2009automatically} & ICSE'09 & Source & C & \ding{56}
      &           &          &             & \ding{52}
      & \ding{52} &          &              & 
      &           & \ding{52}
      & GP        & P \& T
      & GP        & P \& T
      & Stat.     & MH
      & PP \& T   & --
      \\ \hline

      PAR\cite{kim2013par} & ICSE'13 & Source & Java & \ding{56}
      &           &           &             & \ding{52}
      &           & \ding{52} &             &
      &           & \ding{52}
      & GP        & P
      & GP        & P
      & Stat.     & SH
      & PP \& T   & --
      \\ \hline

      SymlogRepair\cite{liu2023symlogrepair} & ESEC/FSE'23 & Source & -- & \ding{52}
      &           &          & \ding{52}    & 
      &           &          & \ding{52}    &
      &           &
      & NPD, Data Leakage    & P
      & NPD, Data Leakage    & P
      & Prog.     & MH
      & --        & --
      \\ \hline

% ---------- Smart Contract -------------
      \multirow{3}*{SMARTSHIELD\cite{zhang2020smartshield}} & \multirow{3}*{SANER'20} & \multirow{3}*{Bytecode} & \multirow{3}*{Solidity} & \multirow{3}*{\ding{56}}
      & \multirow{3}*{\ding{52}}          &                         &             & 
      &           & \multirow{3}*{\ding{52}} &  &
      & \multirow{3}*{\ding{52}}          & 
      & OOB,         & \multirow{3}*{P}
      & OOB,         & \multirow{3}*{P}
      & \multirow{3}*{Func.}     & \multirow{3}*{--}
      & \multirow{3}*{PP \& E}   & \multirow{3}*{--}\\

         &   &   &   &
      &     &   &   & 
      &     &   &   &
      &     &  
      & Unchecked External Call,   &     
      & Unchecked External Call,   & 
      &     &
      &     &\\

         &   &   &   &
      &     &   &   &
      &     &   &   &
      &     &  
      & Post-Call State Change   &
      & Post-Call State Change   &
      &     &
      &     &
      \\ \hline
    
      \multirow{2}*{SGUARD\cite{nguyen2021sguard}} & \multirow{2}*{S\&P'21} & \multirow{2}*{Source} & \multirow{2}*{Solidity} & \multirow{2}*{\ding{52}}
      &           &             & \multirow{2}*{\ding{52}}            & 
      &           &  \multirow{2}*{\ding{52}}  &  &
      &           &
      & RE, AR,  & \multirow{2}*{P}
      & RE, AR,  & \multirow{2}*{P}
      & \multirow{2}*{Func}      & \multirow{2}*{MH}
      & \multirow{2}*{--}         & \multirow{2}*{--} \\

      &   &   &   &
      &     &   &   &
      &     &   &   &
      &     &  
      & Dangerous tx.origin   &     
      & Dangerous tx.origin   &
      &     &
      &     &
      \\ \hline

      EVMPatch\cite{rodler2021evmpatch} & USENIX SEC'21 & Bytecode & Solidity & \ding{52}
      &           &           &             & 
      &           & \ding{52} &             &
      &           & \ding{52} 
      & --          & --
      & AC, IO          & P
      & Func.     & --
      & PP \& T   & --
      \\ \hline

      \multirow{2}*{Elysium\cite{ferreira2022elysium}} & \multirow{2}*{RAID'22} & \multirow{2}*{Bytecode} & \multirow{2}*{Solidity} & \multirow{2}*{\ding{52}}
      & \multirow{2}*{\ding{52}}          &           &             & 
      &           & \multirow{2}*{\ding{52}} &  &
      & \multirow{2}*{\ding{52}}           & 
      & RE, AC, AR,        & \multirow{2}*{P}
      & RE, AC, AR,        & \multirow{2}*{P}
      & \multirow{2}*{Func.}               & \multirow{2}*{--}
      & \multirow{2}*{PP \& T}             & \multirow{2}*{--} \\

      &   &   &   &
      &     &   &   &
      &     &   &   &
      &     &  
      & Unchecked Calls, DoS  &
      & Unchecked Calls, DoS  &
      &     &
      &     &
    \\ \hline

      DeFinery\cite{tolmach2022definery} & ASE'22 & Source & Solidity & \ding{52}
      &           &         & \ding{52}            & 
      & \ding{52} &        &                & 
      & \ding{52}          &
      & GP        & P
      & GP        & P
      & Prog.     & MH
      & PP \& T   & --
      \\ \hline

      \multirow{2}*{SmartFix\cite{so2023smartfix}} & \multirow{2}*{ESEC/FSE'23} & \multirow{2}*{Source} & \multirow{2}*{Solidity} & \multirow{2}*{\ding{52}}
      &           & \multirow{2}*{\ding{52}}        &             & 
      &           & \multirow{2}*{\ding{52}} &  &
      & \ding{52} &
      & IO, Ether-Leak, RE,   & \multirow{2}*{P}
      & IO, Ether-Leak, RE,   & \multirow{2}*{P}
      & \multirow{2}*{Prog.}          & \multirow{2}*{MH}
      & \multirow{2}*{PP}             & \multirow{2}*{--}\\

      &   &   &   &
      &     &   &   &
      &     &   &   &
      &     &  
      & Suicidal, Dangerous tx.origin   &
      & Suicidal, Dangerous tx.origin   &
      &     &
      &     &
      \\ \hline
    
      ContractTinker\cite{wang2024contracttinker} & ASE'24 & Source & Solidity & \ding{52}
      & \ding{52} &         &             & 
      &           &          &              & \ding{52}
      & \ding{52} &
      & GP        & P
      & GP        & P
      & Func.     & --
      & PP        & \ding{52}
      \\ \hline

    \end{tabular}
}
\begin{tablenotes}
\scriptsize
\item[1] {\bf Accessibility}: \ding{56} = the code is not open-sourced or the open-source link is not accessible; \ding{52} = the code is open-sourced and accessible.
\item[2] {\bf Applicability}: IO = Integer Overflow; IE = Integer Errors; BO = Buffer Overflow; NPD = Null Pointer Dereference; NPE = Null Pointer Exception; ML = Memory Leak; \\ DF = Double Free; UAF = Use After Free; RL = Resource Leak;  BC = Bad Cast; OOB = Out-of-bound; AC = Access Control; MRV = Memory-related Vulnerability; \\ CM = Cryptographic Misuses; DoS = Denial of Service; RE = Reentrancy; AR = Arithmetic; GP = General Purpose; -- = not available or unknown.
\item[3] {\bf Input features}: P = Program; E = Exploit; HP = Human Patch; PP = Patched Program; T = Test Case; -- = not available or unknown.
\item[4] {\bf Granularity}: the granularity of AVR-generated patches, where Prog. = Program; File. = File; Stat. = Statement.
\item[5] {\bf Edit type}: SH = Single-Hunk; MH = Multi-Hunks; -- = not available or unknown.
\item[6] {\bf Rank algorithm}: \ding{52} = a method uses a rank algorithm to sort the generated patches; -- = a method does not use any rank algorithm to sort the generated patches.
\end{tablenotes}
\end{threeparttable}
\label{tab: AVR-Comparison}
\end{table*}

%-------------------------------------------------------------------------------
%----------------------------------- 3.1 ---------------------------------------

\subsection{Value-flow Analysis}
This approach employs code analysis tools (e.g., Joern \cite{yamaguchi2014joern}) to generate program representations, such as {\em Control Flow Graphs} (CFGs) and {\em Data Flow Graphs} (DFGs), and then analyzes these graphs to infer violations of security properties and localize vulnerabilities. This approach is type-sensitive as different vulnerability types require different inference rules. 

There are 13 methods in this approach.
(i) LeakFix\cite{gao2015leakfix} and MemFix\cite{lee2018memfix} leverage CFGs to analyze allocation/deallocation states of objects and identify memory leaks.
(ii) IntPTI\cite{cheng2017intpti} and IntPatch\cite{zhang2010intpatch} 
analyze types and value ranges of expressions to identify potential integer errors. 
(iii) ContractTinker \cite{wang2024contracttinker} uses program slices to generate contextual dependency graphs and extract vulnerable code snippets. 
(iv) SMARTSHIELD\cite{zhang2020smartshield} analyzes control flows and data manipulations to derive bytecode-level control flow and data flow dependencies for matching vulnerability patterns. 
(v) Rupair\cite{hua2021rupair} traces data flows (from allocation points 
to usage points) to recognize buffer overflow vulnerability paths.
(vi) FixMeUp\cite{son2013fix} analyzes data flow and control flow dependencies via interprocedural program slicing, and then uses them to recognize missing access control logic and extract {\em Access Control Templates} (ACTs). 
(vii) SkyPort\cite{shi2022skyport} employs static value-flow analysis to extract the semantic logic of vulnerability injection. 
(viii) \cite{shaw2014automatically} identifies unsafe library functions via  control-flow, data-flow, and pointer analysis.
(ix) CIntFix\cite{cheng2019cintfix} identifies (tolerable) C integer errors by leveraging the shortest-path from each node to a security-critical node in the use-def graph.
(x) Elysium\cite{ferreira2022elysium} uses taint analysis to conduct bytecode-level value-flow analysis and infer the context needed by  vulnerability patching. 
(xi) SAVER\cite{hong2020saver} uses the {\em Object Flow Graph} (OFG) to track the event flow of heap objects and identify memory error patterns and their associated code paths.
(xii) RLFixer\cite{Utture2023RLFixer} tracks propagation paths of resource objects 
to identify how they escape the context of their creation methods.
(xiii) AutoPAG\cite{lin2007autopag} uses static dataflow-based backward taint propagation to identify the statements that cause out-of-bounds vulnerabilities.

This approach often incurs prohibitive computational overheads because 
tracking value propagation in complex programs can lead to an exponential path growth, and often fails to accurately parse and/or analyze code.

\subsection{Formal Analysis}

This approach uses formal analysis \cite{woodcock2009formal} to identify the security properties that are violated and determine unsafe program states.
There are seven methods in this approach.
(i) Remedy\cite{chida2022remedy} constructs a semantic model of regex to identify potentially unsafe states, while using {\em Deterministic Finite Automaton} (DFA) to represent violations of security properties and assist patch synthesis. 
(ii) \cite{fang2016web} uses DFA to model web application inputs and identify potential code injection.
(iii) TAPFixer\cite{yu2024tapfixer} uses model checking to determine whether security properties hold.
(iv) VULMET\cite{xu2020vulmet} reasons the weakest precondition to convert patch-incurred changes into hot patch constraints.
(v) SmartFix\cite{so2023smartfix} uses a formal verification tool, VeriSmart \cite{VeriSmart}, to perform mathematical proofs on contracts, while using regression detection assertions to generate patch verification conditions.
(vi) FootPatch\cite{van2018footpatch} uses Separation Logic \cite{matthew2008separation} to model program heap states and uses the Frame Inference technique \cite{josh2005symbolic} to detect violations of heap safety properties.
(vii) CONCH\cite{xing2024conch} uses 
inference rules 
to localize code triggering null pointer dereference errors.

This approach requires comprehensive modeling and verification of a target program, and thus encounters two challenges:
(i) the inherent difficulty in creating accurate models for real-world programs, and (ii) the potential state space explosion that can render exhaustive verification infeasible.

\subsection{Symbolic Execution}

This approach uses symbolic variables and constraint solvers to explore program paths, derive path conditions, and identify the input conditions that can trigger vulnerability.

There are 11 methods in this approach. (i) Angelix\cite{mechtaev2016angelix} uses controlled symbolic execution to extract angelic forests (which is a representation of program behaviors) and capture repair constraints associated with suspicious expressions.
(ii) Errdoc\cite{tian2017errdoc} analyzes vulnerability information related to error handling code.
(iii) SymlogRepair\cite{liu2023symlogrepair} uses {\em Symbolic Execution of Datalog} (SEDL) to describe path constraints related to vulnerabilities.
(iv) KARMA\cite{chen2017karma} performs symbolic execution on both vulnerable functions and candidate functions in parallel to determine whether they are semantically equivalent.
(v) Senx\cite{huang2019senx} uses a hybrid execution engine
to run programs at the LLVM IR instruction level and gathers vulnerability information.
(vi) OSSPatcher\cite{duan2019osspatcher} uses a multi-path exploration 
to extract function-level binary characteristics.
(vii) Definery\cite{tolmach2022definery} uses symbolic execution to derive path constraints and state effects of smart contracts, and uses symbolic summaries to identify the execution paths that violate user-defined functional properties.
(viii) PatchWeave\cite{shariffdeen2020patchweave} uses symbolic execution to identify vulnerability constraints and maps them to semantically equivalent locations in a target program.
(ix) SGUARD\cite{nguyen2021sguard} uses bounded symbolic execution to generate path-complete symbolic traces and identify dependencies related to four types of vulnerabilities.
(x) ExtractFix \cite{gao2021extractfix} and CrashRepair \cite{shariffdeen2024crashrepair} use symbolic execution 
to identify fix locations via constraint propagation and enable root cause analysis 
\cite{zhang2019sosp, yagemann2021arcus, yagemann21bunkerbuster}.
(xi) BovInspector\cite{gao2016bovinspector} uses symbolic execution 
and bidirectional reachability analysis to narrow down the scope of vulnerability path exploration.

This approach requires continuous exploration of a program's execution paths and obtaining constraints via a constraint solver. Thus, it encounters two challenges: (i) %Path Explosion: 
the exponential growth of execution paths (incurred by loops, branches, or recursion) 
forces incomplete exploration, potentially overlooking vulnerabilities; and (ii)
complex operations (e.g., cryptographic functions, floating-point arithmetic) often generate constraints that are beyond the reach of current SMT solvers, causing timeouts or false negatives.

\subsection{Statistical Analysis}

This approach uses statistical features to identify correlations between vulnerability patterns and code characteristics.
It then uses these correlations to analyze code, recognize their potential vulnerabilities, and determine their locations. There are three methods in this approach.
(i) Exterminator \cite{novark2007exterminator} employs randomized heap memory layout and probabilistic canary checks 
to differentiate buffer overflows and dangling pointer errors to
identify the memory operations  responsible for these vulnerabilities.
(ii) HyperGI\cite{mesecan2021hypergi} uses information entropy to quantify information leakage and uses program slices to locate the information leakage point.
(iii) PAR\cite{kim2013par}, VFix\cite{xu2019vfix}, and  \cite{weimer2009automatically} employ statistical analysis  and congestion values \cite{Chuzhoy2012graph} for root cause analysis \cite{blazytko2020aurora, park2024benzene, xu2024racing} and vulnerability localization.

This approach demands numerous tests for identifying statistical correlations. It often incurs high false-positives because it uses statistical correlations rather than program semantics.

%-------------------------------------------------------------------------------
\section{Automated Patch Generation}
\label{section: patch-generation}
%-------------------------------------------------------------------------------

\subsection{Search-based Patch Generation}

This approach formulates patch generation as a search problem so that code snippets are iteratively modified and recombined to produce candidate patches. It often uses search algorithms (e.g., genetic programming \cite{koza1992genetic}) to identify 
repairs 
from typically a vast patch space. 
Existing methods in this approach can be divided into two categories:
\begin{itemize}
[leftmargin=.32cm,noitemsep,topsep=0pt]
\item {\bf Random search}.
These methods 
explore the patch space through
random perturbations/mutations, meaning they
lack explicit objective functions for optimization. For instance, GenProg \cite{le2012genprog} iteratively generates
program variants via mutation and crossover, evaluates them via test cases, and selects the best-performing variants that 
pass all tests.

\item {\bf Optimization-driven search}.
These methods
use objective functions (e.g., test pass rate \cite{gao2019fix2fit} and code quality metrics \cite{mesecan2021hypergi}) to steer the exploration while leveraging heuristic rules or optimization techniques to narrow down the search
space.
Five methods belong to this category. 
(i) Fix2Fit\cite{gao2019fix2fit} uses fuzzing and the test pass rate to refine the partitioning of the patch space. (ii) HyperGI\cite{mesecan2021hypergi} uses a fitness function to balance security
and functionality preservation to guide genetic algorithms in generating 
patches.
(iii) Definery \cite{tolmach2022definery} uses valid and invalid execution traces to direct the search. 
(iv) \cite{weimer2009automatically} prunes the search space by discarding the variants that fail all test cases and applying weighted sampling. (v) TapFixer \cite{yu2024tapfixer} uses negated-property reasoning to perform abstraction-refinement and identify patches.
\end{itemize}

This approach is largely agnostic to vulnerability types because it mutates code.
However, the {\em random search} methods suffer from a low efficiency incurred by the vast search space with a high likelihood of timeout \cite{bui24apr4vul}. 
The {\em optimization-driven search} methods encounter two challenges: (i) the search space explosion incurred by the combinatorial complexity of program mutations,
%results in high computational overhead; 
and (ii) the patch overfitting that the generated patches can pass the test cases but cannot fix the vulnerability in question.

% ========== Type 1: Template-based ==============
\subsection{Template-based Patch Generation}

This approach uses abstract patch
templates to synthesize candidate patches, where templates are defined manually or derived from (e.g.) historical vulnerability fixes \cite{ma2017vurle}.
Existing methods in this approach can be divided into three categories:

\begin{itemize}
[leftmargin=.32cm,noitemsep,topsep=0pt]
\item {\bf Leveraging patterns}.
These methods 
use pre-defined patterns to generate candidate patches.
Nine methods belong to this category. 
(i) BovInspector \cite{gao2016bovinspector} uses mapping between vulnerable APIs and their secure counterparts to address buffer overflow vulnerabilities. 
(ii) Rupair \cite{hua2021rupair} fixes buffer overflows via argument lifting and inserting guards composed of specific statement sequences.
(iii) RegexScalpel\cite{li2022regexscalpel} identifies vulnerable regex patterns to fix ReDoS vulnerabilities. 
(iv) \cite{shaw2014automatically} uses mapping between unsafe library functions and %alternative 
safe functions to prevent buffer overflow vulnerabilities.
(v) FixMeUp \cite{son2013fix} takes explicit mapping between sensitive operations and correct access-control checks as input.
(vi) CONCH \cite{xing2024conch} leverages three abstract patch templates to conduct NULL pointer checks.

(vii) Binary-level repair methods \cite{novark2007exterminator, chen2013safestack, duck2020e9patch} employ predefined byte-level patch templates.
(viii) Smart contract repair methods\cite{so2023smartfix, zhang2020smartshield, nguyen2021sguard, rodler2021evmpatch, ferreira2022elysium} use predefined patch templates specific to vulnerability types.
(ix) Other methods use their own abstract patch templates, which may be derived from secure coding practices \cite{Coker2013cint, zhang2022Seader, Utture2023RLFixer} or historical vulnerability patches \cite{lin2007autopag, ma2017vurle, muntean2019intrepair, lee2022NPEX,shahoor2023leakpair, yang2023tsbport, shariffdeen2020patchweave, shariffdeen2021fixmorph, shi2022skyport}.
\item {\bf Leveraging rules}.
These methods
use pre-defined rules to generate candidate patches, where pre-defined rules are transformation rules or logic rules that 
typically involve grammatical analysis or check the presence of a statement before applying an abstract patch template. Four methods belong to this category. 
(i) CIntFix\cite{Coker2013cint} 
fixes C integer errors by elevating the precision of arithmetic operations according to a set of code transformation rules.
(ii) NPEFix \cite{durieux2017npefix} uses
nine strategies to repair null pointer exceptions.
(iii) CDRep \cite{ma2016cdrep} defines seven rules to fix cryptographic misuse vulnerabilities.
(iv) VFix \cite{xu2019vfix} defines two sets of rules 
to determine how to add NULL pointer checks and initialization operations under different conditions. 
\item {\bf Leveraging template mutations}.
These methods
adapt patch templates to contexts via template mutations. Two methods belong to this category. 
(i) PAR \cite{kim2013par} extracts 10 common fix templates from developer patches, and then 
mutates these templates to identify high-fitness variants and iteratively generate plausible patches. 
(ii) SmartFix\cite{so2023smartfix} defines six atomic repair templates 
for five distinct vulnerability types
to produce various combinations of the six atomic repair templates (guided by the number of alarms that are reduced in the validation process).
\end{itemize}

This approach can generate quality candidate patches for specific types of vulnerabilities. 
Its effectiveness depends on the quality of the abstract patch templates, and the approach may fail due to its poor adaptability in dealing with 
abstract patch templates 
(Appendix \ref{Appendix: Explain for Template} presents two examples to illustrate this adaptability issue).
One mitigation to this problem is to
design post-processing rules  
\cite{xing2024conch}.

% ========== Type 2: Semantics-based ==============
\subsection{Semantics-based Patch Generation}

This approach leverages the security constraint(s) violated by a vulnerability to generate a candidate patch that satisfies the constraint(s).
Existing methods in this approach rely on either static analysis or dynamic execution of test cases or exploits to identify the security constraint(s) that are violated by a vulnerability. These methods fall under two categories:
\begin{itemize}
[leftmargin=.32cm,noitemsep,topsep=0pt]
\item {\bf Static analysis}.
These methods \cite{hong2020saver, van2018footpatch, cheng2017intpti, lee2018memfix, gao2015leakfix, zhang2010intpatch, fang2016web, chida2022remedy}
use static analysis and repair techniques to generate candidate patches.
Methods in this category include: 
(i) SAVER \cite{hong2020saver} uses object flow graphs and variable states to construct and solve constraints and generate candidate patches.
(ii) FootPatch \cite{van2018footpatch} uses Separation Logic \cite{reynolds2002sl} to infer specifications and address memory-related vulnerabilities.
(iii) \cite{fang2016web} solves constraints inferred from DFA to generate patches for string vulnerabilities in PHP.

\item {\bf Dynamic analysis}.
These methods \cite{huang2019senx, gao2021extractfix, zhang2022vulnfix, shariffdeen2024crashrepair, shariffdeen2021cpr, chen2017karma, xu2020vulmet, duan2019osspatcher, liu2023symlogrepair, mechtaev2016angelix} often execute test cases or exploits to detect crashes or unstable behaviors. 
Methods in this category include: 
ExtractFix \cite{gao2021extractfix} uses sanitizer-defined rules to derive crash-free constraints and transform them to candidate patches 
\cite{mechtaev2018symbolic}. 
(ii) Senx \cite{huang2019senx} uses expert-defined security properties and concolic execution \cite{sen2007concolic} to identify the security properties that are violated by vulnerabilities, and then generates predicates to synthesize candidate patches. 
(iii) VulnFix \cite{zhang2022vulnfix} mutates program states to infer patch invariants 
at potential fix locations, and then uses these invariants to guide the generation of candidate patches.
\end{itemize}

This approach encounters three challenges: 
(i) program analysis methods (e.g., symbolic execution) are time-consuming and thus not scalable;
(ii) the resulting accuracy depends on the precision of program analysis methods/tools;
and 
(iii) the patch overfitting problem, namely that the resulting patches are only applicable to specific vulnerability types. 
Note that (iii) may be alleviated via the constraint space partition technique \cite{shariffdeen2021cpr} or using more test cases \cite{zhang2022vulnfix}.

% ========== Type 3: Semantics-based ==============
\subsection{Learning-based Patch Generation}

This approach typically uses deep learning or {\em Large Language Models} (LLMs) to transform vulnerable code into non-vulnerable code.
Existing methods in this approach can be divided into three categories: 

\begin{itemize}
[leftmargin=.32cm,noitemsep,topsep=0pt]
\item {\bf Training}.
These methods \cite{harer2018learning, zhou2022spvf, fu2024vqm} often use the {\em Neural Machine Translation} (NMT) technique \cite{stahlberg2020neural} and demand a sufficient amount of training data, which may not be available in practice. This demand of data may be alleviated by leveraging Generative Adversarial Networks (GANs)\cite{goodfellow2020generative}, as shown in
\cite{harer2018learning}, 
or by 
encoding vulnerabilities security properties, as shown in
\cite{zhou2022spvf}.

% ===============================================
\item{\bf Fine-tuning}.
These methods
involve additional training on top of a pre-trained model, typically by leveraging a smaller, domain-specific vulnerability dataset. Four methods fall under this category.
(i) VRepair \cite{chen2023vrepair} and SeqTrans \cite{chi2023seqtrans}use a specialized vulnerability repair dataset to fine-tune a Transformer model \cite{ashish2017transformer} trained with a bug fix corpus.
(ii) VQM\cite{fu2024vqm} fine-tunes a Vision Transformer model \cite{carion2020end} to learn code changes. 
(iii) VulRepair \cite{fu2022vulrepair} incorporates pre-trained CodeT5 \cite{wang2021codet5} into AVR to achieve a high repair capability.
(iv) VulMaster\cite{zhou2024vulmaster} augments its fine-tuning dataset with CWE knowledge to improve repair accuracy.

\item{\bf Prompt engineering}.
These methods
use carefully crafted prompts to guide LLMs in patch generation.
Four methods belong to this category.
(i) \cite{pearce2023examining} uses reports from \cite{CodeQL} to construct prompts and evaluate LLMs' patch generation capability in zero-shot settings.
(ii) PPatHF \cite{pan2024ppathf} incorporates human-crafted patches into prompts to guide LLMs in patch generation, in a zero-shot setting.
(iii) \cite{liu2024exploring} incorporates historical patches into prompts to assess LLMs in few-shot settings, showing marginal improvement over the results obtained in the zero-shot setting.
(iv) ContractTinker \cite{wang2024contracttinker} uses {\em Chain-of-Thought} (CoT) to guide LLMs in patch generation.
\end{itemize}

This approach initially adopts APR techniques for AVR purposes without accounting for unique characteristics of vulnerabilities. It is known that accounting for vulnerability-specific characteristics can enhance the quality of the resulting candidate patches \cite{zhou2024vulmaster}. 
This approach typically addresses vulnerabilities at the function level, which may not be adequate due to the prevalence of cross-function vulnerabilities \cite{li24crossfunction}.
Moreover, most learning-based patch generation methods (e.g., \cite{chen2023vrepair,fu2022vulrepair, fu2024vqm}) use special tokens to highlight vulnerable statements in the data pre-processing step, allowing the resulting models to generate patches for these specific statements. However, the evaluation of these methods is conducted on the same pre-processed data (e.g., \cite{chen2023vrepair,fu2022vulrepair, fu2024vqm}), rendering their usefulness questionable.

%-------------------------------------------------------------------------------
\section{Automated Patch Validation}
\label{section: patch-validation}
%-------------------------------------------------------------------------------

\subsection{Static Analysis-based Validation}

This approach uses static analysis to determine whether a patched program is no longer vulnerable.
Existing methods in this approach can be divided into two categories: 
\begin{itemize}
[leftmargin=.32cm,noitemsep,topsep=0pt]
\item{\bf Reinspection}. 
These validation methods
use static analysis to rescan a program after applying a candidate patch to it. 
Methods in this category include:
(i) Elysium\cite{ferreira2022elysium} leverages Osiris, Mythril, and Oyente\cite{loi2016oyente} for patch validation.
(ii) BovInspector \cite{gao2016bovinspector} leverages the rules defined by Fortify \cite{Fortify} to validate a patched program.
(iii) SmartFix\cite{so2023smartfix} develops a patch verifier on top of VeriSmart\cite{VeriSmart} for patch validation.
(iv) RLFixer \cite{Utture2023RLFixer} re-runs the resource-leak detector on a patched program to ensure that the previously detected leaks have been resolved. 
(v) FootPatch \cite{van2018footpatch} leverages Infer \cite{FBInfer} to validate a patched program.  
(vi) ContractTinker\cite{wang2024contracttinker} uses GPT-4 for patch validation. (vii) SMARTSHIELD\cite{zhang2020smartshield} leverages three state-of-the-art smart contract analysis tools, namely
Securify\cite{petar2018securify}, Osiris\cite{ferreira2018osiris}, and Mythril\cite{mythril}, for patch validation.
\item{\bf Constraint validation}.
These validation methods
use SMT to verify that a patched program indeed satisfies the desired security constraints. Methods in this category include:
(i) ExtractFix \cite{gao2021extractfix} uses SMT to recalculate the satisfiability of security constraints with candidate patches.
(ii) NPEX \cite{lee2022NPEX}, IntRepair \cite{muntean2019intrepair}, and Definery\cite{tolmach2022definery} use static symbolic execution to validate patched programs.
(iii) FixMeUp\cite{son2013fix} uses access control templates and data dependences 
to identify the missing access control checks.
\end{itemize}

This approach often incurs high false-negatives 
because the rules are often limited and symbolic executions often encounter the problem of path explosion (i.e., forcing a limited depth of exploration).

\subsection{Dynamic Analysis-based Validation} 

This approach runs an exploit to confirm that a vulnerability can no longer be triggered. Existing methods under this approach can be divided into two categories:

\begin{itemize}
[leftmargin=.32cm,noitemsep,topsep=0pt]
\item \noindent{\bf Triggering test}.
These validation methods 
compile a patched program and then run the associated exploit against the resulting executable, such that a failure in compromising the executable means the patch is valid. Methods in this category include:  
(i) SafeStack\cite{chen2013safestack}, and OSSPatcher\cite{duan2019osspatcher}, Senx \cite{huang2019senx}, AutoPaG \cite{lin2007autopag}, and VULMET\cite{xu2020vulmet} apply a candidate patch to a vulnerable program and retest the patched version via a known exploit.
(ii) IntPTI\cite{cheng2017intpti}, 
CIntFix\cite{cheng2019cintfix}, NPEFix\cite{durieux2017npefix}, 
Fix2Fit\cite{gao2019fix2fit}, 
PAR\cite{kim2013par}, Genprog\cite{le2012genprog}, RegexScalpel\cite{li2022regexscalpel}, 
LeakPair \cite{shahoor2023leakpair}, 
\cite{weimer2009automatically},
and
VFix \cite{xu2019vfix} validate a candidate patch by executing positive and negative test cases to provide a nuanced view of the patch's effectiveness.
(iii) EVMPatch\cite{rodler2021evmpatch} replays attacks against patched contracts to show the attacks are thwarted.

\item \noindent{\bf Fuzzing test}.
These validation methods 
conduct more comprehensive security checks on candidate patches. Methods in this category include: 
(i) Rupair\cite{hua2021rupair} uses the idea of trace validation in fuzzing to check the equivalence between a patched program and its original version to ensure that the patch introduces no additional security or functionality impacts.
(ii) CrashRepair \cite{shariffdeen2024crashrepair} and PatchWeave\cite{shariffdeen2020patchweave} validate a candidate patch via differential fuzzing \cite{gulzar2019perception}. \end{itemize}

This approach is effective in verifying whether a vulnerability has been fixed. However, it cannot guarantee that the patch does not alter the functionality or semantics of the vulnerable program, which demands comprehensive testing.
Moreover, this approach is hindered by the availability of exploits, in contrast to APR where adequate test cases are often available.

%-------------------------------------------------------------------------------

%-------------------------------------------------------------------------------
\section{\textsc{Vul4C}: Benchmark Dataset for Repairing C/C++ Vulnerabilities}
\label{sec:benchmark dataset}

%-------------------------------------------------------------------------------
\begin{table*}[!htbp]
    \centering
    \caption{Comparison between \textsc{Vul4C}, our benchmark dataset, and existing C/C++ vulnerability datasets}
    \label{tab: Comparison of our evaluation dataset with existing vulnerability datasets}
    \footnotesize
    \begin{threeparttable}
    \resizebox{\textwidth}{!}{
    \begin{tabular}{|l|c c c : c : c|c|c|c c|c|c|c|}
    \hline
       {\bf{Dataset\tnote{1}}} & \#Software & \#Vulnerabilities & \#CWE types & \#Exploits & \#Patch & Fidelity & Granularity & VTIIC\tnote{2} & VTORC\tnote{3} & Ground truth\tnote{4} & Testability\tnote{5} \\ \hline
       
        SATE IV\cite{okun2013report} 
        & --                & 41,171            & 116               
        & 0                 & 0
        & Synthesized       & Function           
        & 0$\%$             & 0$\%$             & 0$\%$ (0/41,171)
        & 0$\%$ (0/41,171)         
        \\ \hline
  
        CVEfixes\cite{bhandari2021cvefixes} 
        & 563               & 3,543            & 133   
        & 0                 & 3,574  
        & Real-world        & File        
        & 0$\%$             & 0$\%$             & 0$\%$ (0/3,543)
        & 0$\%$ (0/3,543)          
        \\ \hline
        
        Big-Vul\cite{fan2020ac} 
        & 348               & 3,754             & 91
        & 0                 & 4,432 
        & Real-world        & File        
        & 0$\%$             & 0$\%$             & 0$\%$ (0/3,754) 
        & 0$\%$ (0/3,754)         
        \\ \hline

        CrossVul\cite{nikitopoulos21crossvul} 
        & 388               & 2,167              & 100              
        & 0                 & 2,379
        & Real-world        & File           
        & 0$\%$             & 0$\%$             & 0$\%$ (0/2,167)
        & 0$\%$ (0/2,167)         
        \\ \hline

        DiverseVul\cite{chen23diversevul} 
        & 1,179              & 2,957              & 128  
        & 0                 & 7,514  
        & Real-world        & Function        
        & 0$\%$             & 0$\%$             & 0$\%$ (0/2,957)
        & 0$\%$ (0/2,957)     
        \\ \hline

        MegaVul\cite{ni24megavul}   
        & 1,062              & 8,476             & 176               
        & 0                 & 9,288
        & Real-world        & Function           
        & 0$\%$             & 0$\%$             & 0$\%$ (0/8,476)
        & 0$\%$ (0/8,476)         
        \\ \hline
       
        ReposVul\cite{wang24reposvul} 
        & 601               & 4,196            & 151
        & 0                 & 4,699 
        & Real-world        & Repository       
        & 0$\%$           & 0$\%$             & 0$\%$ (0/4,196)     
        & 0$\%$ (0/4,196)            
        \\ \hline 
        
        CB-REPAIR\cite{pinconschi2021comparative} \tnote{6} 
        & 55                & 55                & 36
        & 55                & 0 
        & Synthesized       & Program       
        & 100$\%$           & 0$\%$             & 0$\%$ (0/55)     
        & 100$\%$ (55/55)            
        \\ \hline 
        
        ExtractFix\cite{gao2021extractfix}   
        & 6                 & 23                & 9  
        & 20                & 23  
        & Real-world        & Program        
        & 100$\%$           & 100$\%$           & 87$\%$ (20/23)
        & 0$\%$ (0/23)            
        \\ \hline
        
        VulnLoc \cite{shen2021vulnloc} 
        & 10                & 36               & 10   
        & 36                & 36 
        & Real-world        & Program          
        & 100$\%$           & 100$\%$          & 100$\%$ (36/36)
        & 0$\%$ (0/36)             
        \\ \hline

        LinuxFlaw \cite{mu2018linuxflaw} 
        & 124               & 291             & 17   
        & 332               & 105
        & Real-world        & Program          
        & 62.5$\%$          & 62.5$\%$        & 34$\%$ (100/291)
        & 0$\%$ (0/291)             
        \\ \hline

        {\bf{\textsc{Vul4C}}} 
        & \textbf{23}       & \textbf{144}     & \textbf{19}   
        & \bf{144}          & \bf{144}
        & Real-world        & Program         
        & 100$\%$           & 100$\%$          & 100$\%$ (144/144) 
        & 47$\%$ (68/144)         
        \\ \hline
    \end{tabular}
    }
    \begin{tablenotes}
        \scriptsize
        \item[1] For vulnerability datasets containing examples of multiples languages, we only report the statistics associated with C/C++ vulnerabilities.
        \item[2] VTIIC: Vulnerability-Triggering Input Information Coverage, namely the \% of vulnerabilities in the dataset that come with information on how to trigger a vulnerability.
        \item[3] VTORC: Vulnerability-Triggering Output Results Coverage, namely the \% of vulnerabilities in the dataset that come with post-exploitation output information.
        \item[4] Ground truth is the \% of vulnerabilities in the vulnerability dataset with exploit, patch, vulnerability-triggering input information, and vulnerability-triggering output results.
        \item[5] Testability means the \% of vulnerabilities in the vulnerability dataset that come with test cases. % or test modules.
        \item[6] CB-REPAIR only provides vulnerable programs but not the number of vulnerabilities in each program, forcing us to count \#Vulnerabilities and \#Exploits as \#Software.
     \end{tablenotes}
    \end{threeparttable}
\end{table*}

An AVR benchmark should provide (i) 
real-world vulnerabilities and their patches and (ii) at least one vulnerability trigger for each vulnerability and the corresponding test reports. Table \ref{tab: Comparison of our evaluation dataset with existing vulnerability datasets} summarizes existing C/C++ vulnerability datasets, including: 
(i) two synthetic datasets \cite{okun2013report, pinconschi2021comparative};
(ii) six datasets \cite{bhandari2021cvefixes, fan2020ac, nikitopoulos21crossvul, chen23diversevul, ni24megavul, wang24reposvul} that lack exploits, making them useful for model training but not for evaluating semantics-based patch generation methods;
(iii) two datasets \cite{gao2021extractfix, shen2021vulnloc} that cover a very small number, and few types, of vulnerabilities; 
(iv) one dataset, LinuxFlaw \cite{mu2018linuxflaw}, that contains 105 patches and five distinct exploit trigger methods, but four of them are not suitable for evaluating AVR
because AVR often leverages fuzzing for patch generation 
\cite{zhang2022vulnfix}.
All these datasets, except CB-REPAIR \cite{pinconschi2021comparative}, do not include test cases. 
This highlights the need of benchmark and prompts us to construct one, dubbed \textsc{Vul4C}, via the following four steps.

\noindent{\bf Step 1: Collecting vulnerabilities}. We collect existing datasets \cite{gao2021extractfix, shen2021vulnloc, shariffdeen2020patchweave, mu2018linuxflaw}, including their associated exploits. Since Senx \cite{huang2019senx} does not release any dataset, we re-collect the dataset as described in \cite{huang2019senx}. 
To accommodate vulnerabilities that do not belong to the datasets \cite{gao2021extractfix, shen2021vulnloc, shariffdeen2020patchweave, mu2018linuxflaw}, we crawl online blogs 
\cite{Asarubbo}.
Moreover, we accommodate small programs (i.e., smaller than 5K lines of code), which are just as important as large programs but are excluded from the existing datasets mentioned above, by searching vulnerabilities from the NVD spanning between 2010 and 2023. This leads to the identification of seven vulnerabilities with exploits.
We remove redundant vulnerabilities according to their CVE IDs, leading to 239 unique vulnerabilities in total.

\noindent{\bf Step 2: Collecting patches}.
For the 239 vulnerabilities mentioned above, we collect their patches via their reference links described in the NVD. If a reference link is missing or invalid, the vulnerability is eliminated.
In total, we eliminate 61 vulnerabilities, leading to 178 vulnerabilities.

\noindent{\bf Step 3: Testing vulnerability exploitability}. Given the 178 vulnerabilities, we (re-)build the vulnerable programs to which they belong. We use the associated exploits to assure that the vulnerabilities can be exploited by leveraging sanitizers
to generate error reports on whether an exploitation is successful or not \cite{zhang2022vulnfix,gao2021extractfix, song2019soksanitizing}.
Examples of sanitizers include AddressSanitizer (ASAN) \cite{serebryany2012addresssanitizer}, UBSAN \cite{clang-ubsan}, and Low-fat Pointer \cite{duck2016heap, duck2017stack}.
If a vulnerability is not successfully exploited, then it is eliminated. This leads to the elimination of 34 vulnerabilities, resulting in a benchmark of 144 vulnerabilities, which correspond to 19 CWE types and 23 software products (cf. Appendix \ref{appendix: Details of Benchmark} for details). 

\noindent{\bf Step 4: Collecting test cases}.
To collect test cases for the 23 software products, we use their built-in test modules, which contain test scripts and test inputs. We execute these test modules 
to ensure that the associated test cases are valid.
In total, we are able to collect test cases for 11 software product, as the other 12 software products do not have any test module in their repository or their test cases fail.

\section{Evaluation of AVR Tools}
\label{section: evaluation}

\subsection{Experiment Design}
\label{section: exp design}

\noindent{\bf Collecting AVR tools}. 
It is known to be difficult to evaluate AVR tools via real-world vulnerabilities \cite{pinconschi2021comparative, li2025sok}. This is reaffirmed with the fact that we are only able to evaluate nine out of the 37 AVR tools collected by this study (cf. Appendix \ref{Appendix: Applicability of AVR Tools}). 

For C/C++ programs, we evaluate seven AVR tools recently published in top-tier conferences or journals for repairing vulnerabilities: 
VRepair \cite{chen2023vrepair}, 
VulRepair \cite{fu2022vulrepair}, 
VQM \cite{fu2024vqm}, VulMaster \cite{zhou2024vulmaster}, 
Senx \cite{huang2019senx}, ExtractFix \cite{gao2021extractfix}, and 
VulnFix \cite{zhang2022vulnfix}. 
Among them, Senx \cite{huang2019senx} is not open-source but we obtain its binary from the authors; VulMaster \cite{zhou2024vulmaster} is open-source but missing the data preprocessing code in the public repository, which we obtain from the authors (but only for C/C++).

For AVR tools geared towards Java programs, we consider two tools, Seader \cite{zhang2022Seader} and SeqTrans \cite{chi2023seqtrans}, but not VulMaster \cite{zhou2024vulmaster} because we could not obtain its data preprocessing code from the authors. 

\begin{table}[!htbp]
    \centering
    \footnotesize
    \caption{The number of vulnerabilities for testing AVR tools}
    \vspace{0.2cm}
    \label{tab: Number of Evaluable Vulnerabilities for AVR Tools}
    \resizebox{0.48\textwidth}{!}{
    \begin{tabular}{|c|c|c|c|}
        \hline
        {\bf Benchmark}   &{\bf Tool}                                        & {\bf Applicable reasons}                                 & {\bf \# Tests}       \\ \hline
         \multirow{8}{*}{
         \textsc{Vul4C}}  &{VulRepair} \cite{fu2022vulrepair}                &  \multirow{4}{*}{\makecell[c]{Only apply to single-file
                                                                                    \\vulnerabilities they can process}}                & \multirow{4}{*}{81}  \\ \cline{2-2}
        ~                 & {VRepair} \cite{chen2023vrepair}                 & ~                                                        &  ~                   \\ \cline{2-2}
        ~                 & {VQM} \cite{fu2024vqm}                           & ~                                                        &  ~                   \\ \cline{2-2}
        ~                 & {VulMaster} \cite{zhou2024vulmaster}             & ~                                                        &  ~                   \\ \cline{2-4}
        ~                 & \multirow{2}{*}{VulnFix \cite{zhang2022vulnfix}} & \multirow{2}{*}{\makecell[c]{Only apply to vulnerabilities
                                                                                    \\crashed by UBSAN or ASAN}}                        & \multirow{2}{*}{135} \\ 
        ~                 & ~                                                & ~                                                        & ~                    \\ \cline{2-4}
        ~                 & {ExtractFix} \cite{gao2021extractfix}            & \multirow{2}{*}{\makecell[c]{Only apply to vulnerabilities with their applicable 
                                                                                    \\types and compiled on their special compiler}}    & 69                   \\ \cline{2-2} \cline{4-4}
        ~                 & {Senx} \cite{huang2019senx}                      & ~                                                        & 85                   \\ \hline
        \multirow{2}{*}{
        \textsc{Vul4J}}   & {Seader}\cite{zhang2022Seader}                   & Only apply to compilable vulnerabilities                 & 68                   \\ \cline{2-4}
        ~                 & {SeqTrans}\cite{chi2023seqtrans}                 & Only apply to single-line and compilable vulnerabilities & 15                   \\ \hline
    \end{tabular}
    }
\end{table}
\noindent{\bf Benchmarks}.
We use \textsc{Vul4C} to evaluate the seven AVR tools and the third-party  \textsc{Vul4J} \cite{bui2022vul4j} to evaluate the two AVR tools.
Table \ref{tab: Number of Evaluable Vulnerabilities for AVR Tools} summarizes the number of vulnerabilities that are suitable for evaluating the nine AVR tools. The four learning-based AVR tools  \cite{chen2023vrepair, fu2022vulrepair, fu2024vqm, zhou2024vulmaster} evaluated by \textsc{Vul4C} can only repair single-file vulnerabilities, leading to 81 vulnerabilities in each case. VulnFix\cite{zhang2022vulnfix} can only repair vulnerabilities with UBSAN or ASAN sanitizers, leading to 135 vulnerabilities. Both ExtractFix\cite{gao2021extractfix} and Senx\cite{huang2019senx} rely on specific compiler versions and can only repair certain vulnerability types, leading to 69 and 85 vulnerabilities, respectively.
For the two AVR tools geared towards Java programs, Seader\cite{zhang2022Seader} is evaluated by all the 68 compilable vulnerabilities in {\sc Vul4J}, while SeqTrans\cite{chi2023seqtrans} can only fix vulnerabilities requiring single-line modifications (leading to 15 vulnerabilities).

\noindent{\bf Training for AVR tools}. 
Among the seven AVR tools geared towards C/C++ programs, four\cite{chen2023vrepair, fu2022vulrepair, zhou2024vulmaster, fu2024vqm} belong to learning-based patch generation, 
meaning that we need to train their models. For this purpose, we use the union of the Big-Vul dataset \cite{fan2020ac} and the CVEfixes \cite{bhandari2021cvefixes} dataset, removing the duplicates between them and the vulnerabilities that are already contained our benchmark \textsc{Vul4C}. We use 80\% of the resulting dataset for training and 20\% for validation.
Among the two AVR tools\cite{zhang2022Seader, chi2023seqtrans} geared towards Java programs,  one\cite{chi2023seqtrans} belongs to learning-based patch generation, for which we use the model
provided by its authors.

\noindent{\bf Experiments design}.
We conduct experiments for three purposes.
The first purpose is to evaluate the automated vulnerability analysis step of AVR tools, geared towards C/C++ programs and Java programs alike. 
For this purpose, we observe that it is difficult to quantify some characteristics (e.g., security properties). Thus, we focus on evaluating the accuracy of fix locations because they point out where a patch should be applied and thus are critical to the subsequent patch generation. 
We also observe that only two (of the nine) AVR tools provide their own vulnerability analysis \cite{huang2019senx, gao2021extractfix}
and that the non-modular design of Senx \cite{huang2019senx} makes it impossible to carve out its vulnerability analysis. Thus, we use a localization tool, VulnLoc \cite{shen2021vulnloc}, %as baseline 
to 
identify fix locations. 
For each vulnerability, we set the maximum execution time as four hours to prevent them from running indefinitely, which is the default time used in \cite{shen2021vulnloc}. We use following metrics to evaluate the competency of the vulnerability analysis step.
\begin{itemize}
[leftmargin=.32cm,noitemsep,topsep=0pt]
\item {\bf\em File-level accuracy}. 
Let $x$ denote the number of test vulnerabilities for which an AVR tool correctly localizes the files that need to be fixed
and $y$ the total number of test vulnerabilities that are used to evaluate the AVR tool.
This metric is defined as $x/y \times 100\%$.

\item {\bf\em Statement-level accuracy}. 
Let $x$ denote the number of test vulnerabilities for which an AVR tool correctly localizes the fix locations 
in terms of the lines of code (or statements) that need to be fixed
and $y$ the total number of test vulnerabilities that are used to evaluate the AVR tool. This metric is defined as $x/y \times 100\%$.
\end{itemize}

The second purpose is to evaluate the automated patch generation step of AVR tools via a unified automated patch validation step.
For learning-based patch generation methods \cite{chen2023vrepair, fu2022vulrepair, zhou2024vulmaster, fu2024vqm, chi2023seqtrans}, we set the beam size as 50, meaning that each AVR tool generates 50 candidate patches for each applicable vulnerability. 
We set their maximum execution time as one hour for each vulnerability, as suggested in \cite{shariffdeen2021cpr, noller2022trust}. We use following metrics for evaluation.
\begin{itemize}
[leftmargin=.32cm,noitemsep,topsep=0pt]

\item {\bf\em Patch restoration rate}. 
Let $x$ denote the number of candidate patches that can be successfully applied to their respective test vulnerabilities (i.e., successfully modify the vulnerable code but does not guarantee that the modified code fixed the vulnerability)
and $y$ the total number of candidate patches. This metric is defined as $x/y \times 100\%$.

\item {\bf\em Patch compilation rate}.
Let $x$ denote the number of candidate patches that can be successfully compiled 
and $y$ the total number of candidate patches.
This metric is defined as $x/y \times 100\%$.

\item {\bf\em Test pass rate}.
Let $x$ be the number of candidate patches that successfully pass triggering tests (via exploits) and functions tests (via test cases)
and $y$ the total number of candidate patches.
It is defined as $x/y \times 100\%$.

\end{itemize}

The third purpose is to manually evaluate the validity of plausible patches that passed the automated patch validation step of AVR.
This is important because a plausible patch that passes automated patch validation still can fail to fix a vulnerability.
For this purpose, we compare an plausible patch with the human-crafted patch in question.
This allows us to
consider semantically equivalent patches. Specifically, we use following metrics.
\begin{itemize}
[leftmargin=.32cm,noitemsep,topsep=0pt]
\item {\bf\em Consistency rate}. Let $x$ denote the number of test vulnerabilities for which an AVR tool generates at least one candidate patch that is the same as
the human-crafted ground-truth patch 
and $y$ the total number of test vulnerabilities. It is defined as $x/y \times 100\%$.

\item {\bf\em Success rate}. 
Let $x$ denote the number of test vulnerabilities for which an AVR tool generates at least one candidate patch that is the same or semantically equivalent
to the human-crafted ground-truth patch
and $y$ the total number of test vulnerabilities that are used to evaluate the AVR tool. This metric is defined as $x/y \times 100\%$.
\end{itemize}
This involvement of domain experts (i.e., co-authors of the present paper) is a common practice in APR and AVR research, as shown in (e.g.) \cite{yin2024thinkrepair, yang2023tsbport, zhang2022vulnfix}. As elaborated in the ``Ethics Considerations'' below, there are no ethical concerns, nor IRB is required.

\noindent{\bf Experimental platform and parameters}. 
Our experiments are conducted on a computer with four 16-core 2.90GHz Intel(R) Xeon(R) Gold 6226R CPUs, 256 GB RAM, and two NVIDIA RTX A6000 GPUs.

\subsection{Experimental Results for AVR Tools Geared Towards C/C++ Programs}

\textbf{\begin{table*}[!htbp]
    \centering
    \footnotesize
    \caption{Evaluation results of the applicable AVR tools' C/C++ vulnerability localization capabilities 
    }
    \label{tab: results of localization competency}
    \begin{tabular}{|l|c|c|c|c|c|c|}
        \hline
        {\bf Tool}                     & ExtractFix \cite{gao2021extractfix} & VulnLoc \cite{shen2021vulnloc} (top-1) & VulnLoc \cite{shen2021vulnloc} (top-3) & VulnLoc \cite{shen2021vulnloc} (top-5)  \\ \hline
        {File-level accuracy}      & 68.4\% (13/19)                      & 41.8\% (38/91)                                                  & 51.6\% (47/91)                      & 59.3\% (54/91) \\ \hline   
        {Statement-level accuracy} & 31.6\% (6/19)                       & 7.7\% (7/91)                                                    & 13.2\% (12/91)                      & 14.3\% (13/91) \\ \hline
    \end{tabular}
\end{table*}}

\subsubsection{Evaluating Vulnerability Analysis (Localization)}
\label{sec: evaluating localiztion}

Table \ref{tab: results of localization competency} summarizes the vulnerability analysis (localization) results of  ExtractFix\cite{gao2021extractfix} and VulnLoc\cite{shen2021vulnloc}
where we use the top-$k$ ($k=1,3,5$) locations in the sorted fix locations returned by VulnLoc. We observe that the total number of fix locations returned by
ExtractFix
is much smaller than 144 (i.e., the number of vulnerabilities in our benchmark dataset). This discrepancy is incurred by the fact that it is only applicable to some vulnerability types or may not produce results at all. 
Experimental results show that ExtractFix and VulnLoc, both leveraging dynamic analysis techniques to trigger vulnerabilities, achieve a high file-level accuracy (68.4\% for ExtractFix and 59.3\% for VulnLoc) in localizing fix locations. However, their statement-level accuracy remains low, indicating that these tools can effectively identify the files that require modification for patching purposes, but they struggle to precisely pinpoint the exact statements that need to be fixed.

Nevertheless, ExtractFix exhibits a higher statement-level accuracy (31.6\%) than VulnLoc, despite being limited to memory-related vulnerabilities. This relatively high precision can be attributed to ExtractFix's use of symbolic execution, which incorporates program semantics into its analysis. In contrast, VulnLoc relies on statistical methods and fuzzing-generated test cases, making its results dependent on the quality of the test cases; this explains its low effectiveness. Neither of these two tools provide explanations for the failed analyses (i.e., why they mistakenly identify the claimed fix locations).
This lack of explainability hinders failure analysis.

\begin{observation}
Existing vulnerability analysis methods cannot accurately localize vulnerabilities to statement(s). 
\end{observation}

% ------------------------------------------------------------------------------------

\subsubsection{Evaluating Patch Generation Capabilities}

Table \ref{tab: Evaluation results of passing rate during patch validation} summarizes the number of patches, which are generated by 7 AVR tools and 2 APR tools,
that pass the automated patch validation. 
We observe that the semantics-based patch generator VulnFix \cite{zhang2022vulnfix} outperforms the others with a 96.0\% test pass rate, perhaps because its goal is to find a patch invariant to show that an exploit is thwarted. As a result, this approach, and thus VulnFix, generates candidate patches by finding solutions that satisfy the relevant constraints by, for instance, adding conditional checks to prevent a vulnerability from being triggered. Semantics-based patch generation methods \cite{gao2021extractfix, huang2019senx} outperform learning-based patch generation methods \cite{chen2023vrepair, fu2022vulrepair, fu2024vqm, zhou2024vulmaster} in all three metrics.

\begin{table*}[!htbp]
    \centering
    \caption{Evaluation results of the applicable AVR tools' patch validation outcomes}
    \label{tab: Evaluation results of passing rate during patch validation}
    \footnotesize
    \begin{tabular}{|l|c|c|c|}
        \hline
        {\bf Tool}   &  {\bf Patch restoration rate}  & {\bf Patch compilation rate}  & {\bf Test pass rate}  \\ \hline            
        VulnFix \cite{zhang2022vulnfix}                 & 100.0\% (25/25)               & 100.0\% (25/25)               & 96.0\% (24/25)        \\ \cline{1-4}
        ExtractFix \cite{gao2021extractfix}             & 100.0\% (16/16)               & 50.0\% (8/16)                 & 18.8\% (3/16))          \\ \cline{1-4}
        Senx \cite{huang2019senx}                       & 100.0\% (19/19)               & 52.6\% (10/19)                & 15.8\% (3/19)         \\ \cline{1-4}
        VRepair \cite{chen2023vrepair}                  & 31.3\% (1269/4050)            & 2.6\% (106/4050)              & 0.2\% (10/4050)       \\ \cline{1-4}
        VulRepair \cite{fu2022vulrepair}                & 66.5\% (2695/4050)            & 12.5\% (507/4050)             & 4.3\% (175/4050)      \\ \cline{1-4}
        VQM \cite{fu2024vqm}                            & 69.9\% (2829/4050)            & 11.8\% (479/4050)             & 0.8\% (31/4050)        \\ \cline{1-4}
        VulMaster \cite{zhou2024vulmaster}              & 78.9\% (3197/4050)            & 13.9\% (564/4050)             & 1.6\% (66/4050)       \\ \hline
    \end{tabular}
   % }
\end{table*}

\begin{observation}
AVR tools leveraging semantics-based patch generation methods demonstrate superior effectiveness in generating high-quality patches when
compared with the AVR tools leveraging learning-based patch generation methods.
\end{observation}

Table \ref{tab: Evaluation results of the generation competency} summarizes the results with the 7
applicable AVR tools while deferring (i) the detailed results with respect to different vulnerability types, edit types, and each vulnerability in \textsc{Vul4C},
and (ii) an example for explaining semantic equivalence, to Appendix \ref{sec:extra evaluation}. Analysis will be provided below.

\begin{table}[!htbp]
    \centering
    \footnotesize
    \vspace{-0.2cm}
    \caption{Evaluation results of the applicable AVR tools' candidate patch generation capabilities for C/C++ vulnerabilities
    }
    \label{tab: Evaluation results of the generation competency}
    \begin{tabular}{{|l|c|c|}}
        \hline
        {\bf Tool}                                      & {\bf Consistency rate}    & {\bf Success rate}    \\ \hline
         VulnFix \cite{zhang2022vulnfix}                 & 5.9\% (8/135)             & 10.4\% (14/135)       \\ \cline{1-3}
         ExtractFix \cite{gao2021extractfix}             & 1.4\% (1/69)              & 1.4\% (1/69)          \\ \cline{1-3}
         Senx \cite{huang2019senx}                       & 0.0\% (0/85)              & 0.0\% (0/85)          \\ \cline{1-3}
         VRepair \cite{chen2023vrepair}                & 0.0\% (0/81)              & 0.0\% (0/81) \\ \cline{1-3}         
         VulRepair \cite{fu2022vulrepair}                & 0.0\% (0/81)              & 0.0\% (0/81)          \\ \cline{1-3}
         VQM \cite{fu2024vqm}                            & 0.0\%(0/81)               & 0.0\%(0/81)           \\ \cline{1-3}
         VulMaster \cite{zhou2024vulmaster}              & 2.5\%(2/81)               & 2.5\%(2/81)           \\ \hline
    \end{tabular}
\end{table}

\begin{table*}[tb]
    \centering
    \footnotesize
    \caption{Statistics 
    of causes of failures for {\it semantics-based} patch generation methods
    in two steps: patch generation and patch validation, where $x$ is the number of test vulnerabilities that failed and $y$ is the total number of test vulnerabilities. Note that a test vulnerability may fail for reasons, as there can be multiple candidate patches for a vulnerability.
    }
    \label{tab: Failure for semantics-based patch generation methods}
    \begin{tabular}{|l|c|c|c|c|}
        \hline                       
        \multirow{2}{*}{\bf Tool}                     & {\bf Patch generation}             & \multicolumn{3}{c|}{\bf Patch validation}                  \\ \cline{2-5}
         ~                                            & Fail to generate candidate patches & Compilation error & Test failure & Semantic inequivalence  \\ \hline
        VulnFix \cite{zhang2022vulnfix}               & 81.48\% (110/135) & 0.00\% (0/135) & 0.74\% (1/135) & 7.41\% (10/135)                           \\ \hline
        ExtractFix \cite{gao2021extractfix}           & 82.61\% (57/69)   & 7.25\% (5/69)  & 7.25\% (5/69)  & 1.45\% (1/69)                             \\ \hline
        Senx \cite{huang2019senx}                     & 77.65\% (66/85)   & 10.59\% (9/85) & 8.23\% (7/85)  & 3.53\% (3/85)                             \\ \hline
    \end{tabular}
\end{table*}

To understand why the plausible patches  fail, we manually analyze them and summarize the reasons in Tables \ref{tab: Failure for semantics-based patch generation methods} and \ref{tab: Failure reasons for learning-based patch generation methods}. 
Specifically, for semantics-based patch generation methods\cite{zhang2022vulnfix, gao2021extractfix, huang2019senx}, the resulting plausible patches fail due to the following reasons.  

First, they fail because of the analysis technique they use. 
For instance, VulnFix \cite{zhang2022vulnfix}, which relies on inferring patch invariants via snapshot fuzzing, fails for the two main reasons: (i) fuzzing-related errors, including missing necessary resources (18 cases) or not being able to generate candidate patch invariants (39 cases) in the course of snapshot fuzzing; (ii) failure in reducing the candidate patch invariants to a single invariant (20 cases), as shown in the example of Listing \ref{Appendix: CVE-2017-15021}. 
Senx \cite{huang2019senx} fails because of incorrect parsing of pointer variables, as shown in the examples in Appendix \ref{Appendix: CVE-2016-9387}.
ExtractFix \cite{gao2021extractfix} fails because of incorrect use of array variables, as shown in the examples in Appendix \ref{Appendix: CVE-2017-7598}. 
When applying them to our benchmark, they fail to generate candidate patches or the patched program cannot be compiled successfully, which are the main reason of failure.

\begin{observation}
The limited effectiveness of semantics-based patch generation tools primarily stems from their limited scalability. 
\end{observation}

\begin{lstlisting}[float=ht, language = diff, 
linewidth=.46\textwidth, 
xleftmargin=.02\textwidth,
label=example_developer, caption={The manual patch for CVE-2017-6838 (audiofile)}, captionpos=b,
                    linebackgroundcolor={%
                        \ifnum\value{lstnumber}>4
                            \ifnum\value{lstnumber}<7
                                \color{red!20}
                            \fi
                        \fi
                        \ifnum\value{lstnumber}>6
                            \ifnum\value{lstnumber}<12
                                \color{green!20}
                            \fi
                        \fi},]
@@ -323,8 +350,11 @@ bool copyaudiodata (AFfilehandle infile, AFfilehandle outfile, int trackid)
 {
 	int frameSize = afGetVirtualFrameSize(infile, trackid, 1);
 
-	const int kBufferFrameCount = 65536;
-	void *buffer = malloc(kBufferFrameCount * frameSize);
+	int kBufferFrameCount = 65536;
+	int bufferSize;
+	while (multiplyCheckOverflow(kBufferFrameCount, frameSize, &bufferSize))
+		kBufferFrameCount /= 2;
+	void *buffer = malloc(bufferSize);
 
 	AFframecount totalFrames = afGetFrameCount(infile, AF_DEFAULT_TRACK);
 	AFframecount totalFramesWritten = 0;
\end{lstlisting}

\begin{lstlisting}[float=ht, language = diff, 
linewidth=.46\textwidth, 
xleftmargin=.02\textwidth,
label=plausible_patch_vulnfix_1, caption={The plausible patch generated by VulnFix for CVE-2017-6838 (audifile)}, captionpos=b,
                    linebackgroundcolor={%
                        \ifnum\value{lstnumber}>4
                            \ifnum\value{lstnumber}<6
                                \color{green!20}
                            \fi
                        \fi},]
@@ -324,6 +324,7 @@
 	int frameSize = afGetVirtualFrameSize(infile, trackid, 1);
 
 	const int kBufferFrameCount = 65536;
+    if (!(frameSize < 32768)) exit(1);
 	void *buffer = malloc(kBufferFrameCount * frameSize);
 
 	AFframecount totalFrames = afGetFrameCount(infile, AF_DEFAULT_TRACK);
\end{lstlisting}

Second, all these methods suffer from the patch overfitting problem. More specifically, they generate if-condition patches, which render vulnerability-triggering locations unreachable and thus lead to plausible but invalid patches. 
Still, the plausible but invalid patches may alleviate the repairing burdens on developers because they can serve as a temporary mitigation to prevent vulnerability exploitation while developers only need to make minor adjustments on this basis to obtain valid patches.
To see this, we consider Listings \ref{example_developer} and \ref{plausible_patch_vulnfix_1}. For CVE-2017-6838, the human-crafted patch checks the size of buffer allocation via function \texttt{multiplyCheckOverflow}, then limits the allocation of \texttt{buffer} in 65536/2 = 32678 to make the program run correctly while VulnFix\cite{zhang2022vulnfix} generates a plausible patch. Its conditions are equivalent to the human-crafted patch, but VulnFix chooses to end the program with exit code \texttt{1}. Thus, a developer can still get the correct conditions from the plausible patches.

\begin{observation}
The plausible patches generated by semantics-based patch generation methods can serve as a temporary mitigation to prevent vulnerability exploitation, and developers can obtain valid patches with only minor modifications.
\end{observation}

\begin{table*}[!htbp]
    \centering
    \caption{Statistics of the causes of failures for learning-based patch generation methods
    }
    \footnotesize
    \label{tab: Failure reasons for learning-based patch generation methods}
    \begin{tabular}{|l|c|c|c|c|}
        \hline
        {\bf Cause of failures}                & {\bf VulRepair \cite{fu2022vulrepair}} & {\bf VRepair \cite{chen2023vrepair}} & {\bf VQM \cite{fu2024vqm} } & {\bf VulMaster \cite{zhou2024vulmaster} } \\ \hline
        Error in generating context tokens	   &81.78\% (3312/4050)	    &91.53\% (3707/4050) &75.28\%(3049/4050) &62.91\%(2548/4050) \\ \hline
        Error in matching context	           &6.12\% (248/4050)	    &0.00\% (0/4050)     &3.90\%(158/4050)   &8.49\%(344/4050) \\ \hline
        Syntax error	                       &8.89\% (360/4050)	    &3.48\% (141/4050)   &8.05\%(326/4050)   &13.31\%(539/4050)	 \\ \hline
        Unknown tokens	                       &0.00\% (0/4050)	        &1.21\% (49/4050)    &0.00\% (0/4050)    &0.00\% (0/4050)   \\ \hline
        Other reasons	                       &3.21\% (130/4050)	    &3.98\% (161/4050)	 &12.77\%(517/4050)  &15.23\%(617/4050)   \\ \hline
    \end{tabular}
\end{table*}

AVR tools leveraging learning-based patch generation methods\cite{chen2023vrepair, fu2022vulrepair} exhibit failures at various validation stages (e.g., patch placement and patch compilation, as shown in Table \ref{tab: Evaluation results of passing rate during patch validation}). 
To understand the causes of these patch generation failures, we analyze the 4,050 candidate patches (i.e., 81 vulnerabilities $\times$ 50 candidate patches for each vulnerability) generated in our experiments. 
Since \cite{fu2022vulrepair, chen2023vrepair, fu2024vqm, zhou2024vulmaster} do not follow the AVR workflow we propose (missing validation process), their original evaluation is based on comparing code sequence equivalence, which is inherent to their data preprocessing pipelines. When placing their generated patches (i.e., code fragments) back into the vulnerable code, issues may arise. This is the main reason for their failure. Note that all the four tools employ the three tokens to mark contexts of the patch designed by VRepair\cite{chen2023vrepair}. If the first three tokens are predicted by models incorrectly, the patch can not be placed (i.e, error in generating context tokens). Even with correct context, mismatches may still cause patch placement failures (i.e., error in matching context) due to the error patch placement strategy designed by VRepair\cite{chen2023vrepair}. The model may also introduce other issues, such as generating syntactically incorrect sequences or producing unknown tokens. We show 4 examples accounting for those 4 reasons in Appendix \ref{Appendix: CVE-2020-26208} to \ref{Appendix: CVE-2016-9828}. So the code sequence they generate may not actually fix the vulnerabilities, even though it is the same as the ground truth they process. 
% While APR tools \cite{huang2025ntr, chen2021sequencer} are consistently tested on common benchmarks Defects4J \cite{Just2014Defects4J} that assess complete patch deployment (placement, compilation, and testing), they have a 100\% patch restoration rate. 
This incomplete assessment and evaluation framework may stem from the lack of an evaluation benchmark.

\begin{observation}
AVR tools leveraging learning-based patch generation methods lack rigorous evaluation methodologies. The processed token sequences fail to accurately reflect their actual repair effectiveness due to the absence of comprehensive benchmarking in the AVR field.
\end{observation}

\subsection{Experimental Results for AVR Tools Geared Towards Java Programs}

Table \ref{tab: Evaluation results of the generation competency for java} summarizes the experimental results. We observe that the two AVR tools for Java programs also perform poorly, which can be understood as follows. 
First, 
the poor performance of template-based patch generation methods can be attributed to the diversity of templates and their narrow scope of applicability. Even though Seader \cite{zhang2022Seader} performs well in addressing cryptography-related API misuses, it performs poorly when dealing with other types of vulnerabilities. Second, learning-based patch generation methods \cite{chi2023seqtrans} usually generate syntactically incorrect, and thus useless patches (e.g., presence of undefined macro and variables).

\begin{observation}
Current AVR tools for Java are incompetent.
\end{observation}

\begin{table}[!htbp]
    \centering
    \footnotesize
\caption{Evaluation of AVR tools for Java programs}
    \label{tab: Evaluation results of the generation competency for java}
    \begin{tabular}{{|l|c|c|c|c|}}
        \hline
        {\bf Tool}                            & {\bf Seader} \cite{zhang2022Seader} & {\bf SeqTrans} \cite{chi2023seqtrans} \\ \hline     
        {\bf Patch restoration rate} & 100\% (8/8)                   & 100\% (750/750)           \\ \hline 
        {\bf Patch compilation rate} & 0\% (0/8)                     & 0.8\% (6/750)      \\ \hline       
        {\bf Test pass rate}         & 0\% (0/8)                     & 0.1\% (1/750)        \\ \hline         
        {\bf Consistency rate}       & 0\% (0/68)                    & 6.7\%(1/15)          \\ \hline      
        {\bf Success rate}           & 0\% (0/68)                    & 6.7\%(1/15)          \\ \hline     
    \end{tabular}
\end{table}
%-------------------------------------------------------------------------------
\section{Future Research Directions}
\label{section: discussion}
%-------------------------------------------------------------------------------

\noindent{\bf Direction 1: Developing effective vulnerability analysis techniques for patch generation}.
Table \ref{tab: AVR-Comparison} shows that many AVR tools lack the support of competent vulnerability analysis.
Many learning-based patch generation methods \cite{chen2023vrepair, fu2022vulrepair, fu2024vqm, zhou2024vulmaster, chi2023seqtrans} attempt to avoid vulnerability analysis and will inevitably fail;
in contrast, semantics-based patch generation methods intend to leverage vulnerability analysis and often result in high-quality patches (e.g., a higher patch compilation rate and a higher test pass rate as shown in Table \ref{tab: Evaluation results of passing rate during patch validation}).
%representing the right research direction.
Thus, future research should strive to design better
automated vulnerability analysis techniques.
Notably, some vulnerability detection and analysis methods, such as root cause analysis \cite{park2024benzene, xu2024racing} and learning-based detection \cite{li2021vuldeelocator, mirsky2023vulchecker},
have made significant progress in identifying fix locations.

\noindent{\bf Direction 2: Integrating multiple AVR methods}.
Since different AVR methods have their own strengths and weaknesses, it is possible to take advantage of their strengths, such as: (i) search-based methods can generate patches with a high test pass rate when guided by objectives (e.g., test pass rate); (ii) template-based patch generation methods can use matching or similarity techniques to deal with simple vulnerabilities, such as integer overflow, buffer overflow, and NULL pointer dereference because they are usually patched via type conversion or if-else checks; (iii) semantics-based patch generation methods are more suitable for memory-related vulnerabilities because they can leverage the constraints that are violated by overflows to generate if-else checks, thereby enhancing memory security; and (iv)
learning-based patch generation methods can be applied to uncommon and complex scenarios, thus incorporating templates into learning-based methods \cite{xia2022less, lin2024one} might overcome the limited generalization capability of template-based methods while mitigating the weakness of random generation used by learning-based methods.

\noindent{\bf Direction 3: Leveraging LLMs to improve the effectiveness of AVR}.
LLMs with extensive parameters, such as GPT-4, have showed significant value in software security \cite{hou2024survey}, including vulnerability detection \cite{sun2024gptscan}, penetration testing \cite{deng2024pentestgpt}, and fuzzing \cite{meng2024large}. There are also empirical studies \cite{liu2024exploring, pearce2023examining} in AVR that prove their effectiveness in assisting software security.
LLMs excel in analyzing vulnerability code, such as identifying critical variables and key code snippets to help understand vulnerabilities, and parsing and interpreting documents to learn secure coding practices to support AVR. LLMs can help generate repair suggestions \cite{zhang2024vuladvisor}, which can be used as input to learning-based AVR methods.
LLMs can interact with external tools in automated analysis and thus vulnerability repair \cite{wu2023autogen}.
To enhance LLMs in AVR, it is necessary to build a comprehensive vulnerability knowledge base or knowledge graph, which would include essential information such as vulnerability types, historical repair cases, and developer guidelines. Moreover, LLMs can leverage the {\em Retrieval-Augmented Generation} (RAG) technology \cite{lewis2020retrieval} to access and integrate relevant information from the knowledge base and deliver targeted and context-aware solutions.

\noindent{\bf Direction 4: Balancing security and functionality in search-based, semantics-based, and template-based patch generation methods}.
Most patch generation methods prioritize rapid treatment of known vulnerabilities, often at the expense of preserving a program's functionality. Search-based methods try to balance security and functionality via testing, but may lead to functional impairments because test cases can hardly cover all possible scenarios.
Semantics-based methods %\cite{shariffdeen2024crashrepair, gao2021extractfix,xu2019vfix} 
might fix vulnerabilities by adding simple conditional statements, but can cause functionality issues; moreover, developers often prefer nuanced repair strategies, such as refining loop conditions to eliminate vulnerabilities while maintaining a program's functionality. 
Template-based methods may suffer from the same problem because predefined templates may lack the flexibility in addressing complex logic, diverse coding styles, or novel attack vectors, resulting in ineffective patches or even new security risks. Furthermore, applying templates often requires significant contextual adjustments to avoid the disruption of the other parts of a program, adding complexity to the repair process.
Future research should emphasize minimizing the impact of repairs on program logic.

\noindent{\bf Direction 5: Designing effective automated vulnerability validation methods}.
Designing effective automated vulnerability validation methods is a challenge. Although AVR tools have made progress in generating candidate patches, the lack of a robust and efficient validation mechanism remains a significant limitation. As a result, these tools often produce a large number of candidate patches that overwhelm developers in identifying the competent one.
Moreover, unverified patches may introduce new issues or fail to completely resolve a vulnerability, undermining users' trust in AVR tools \cite{noller2022trust}. Future research should develop advanced and efficient vulnerability validation methods.

%-------------------------------------------------------------------------------
\section{Conclusion}
\label{section: conclusion}
%-------------------------------------------------------------------------------

We have systematized AVR via its work flow of vulnerability localization, patch generation, and patch validation.  
To fairly evaluate AVR tools, we propose a benchmark dataset for C/C++  vulnerabilities. 
We apply the benchmark dataset to empirically evaluate nine AVR tools,
leading to important findings, such as: 
the accuracy of vulnerability localization largely affects repairing effectiveness but existing localization techniques are incompetent; fuzzing-based and symbolic execution-based AVR tools generate compilable patches but may suffer from the patch overfitting problem. 
We discuss five future research directions.

\section*{Acknowledgements}
We thank the anonymous reviewers for their comments,
which guided us in improving the paper. The authors affiliated with Huazhong University
of Science and Technology were supported by the National Natural Science Foundation
of China under Grant No. 62272187. Shouhuai Xu's research was not sponsored /
supported by any funding agency. Any opinions, findings, conclusions, or
recommendations expressed in this work are those of the authors and do not
reflect the views of the funding agencies in any sense.

\section*{Ethics Considerations}

We have read the ethics considerations discussions in the conference call for papers, the detailed submission instructions, and the guidelines for ethics documentation.

We thank the anonymous reviewers for their insightful comments. Our experience is that leveraging domain experts to evaluate technical matters does not require IRB approval because human subjects are not the target of studies. 
This is also shown in prior studies that leverage domain experts to verify the effectiveness of APR or AVR, such as \cite{yin2024thinkrepair, yang2023tsbport, zhang2022vulnfix}.
Nevertheless, we always stick to ethical practice when conducting this and other studies. Inspired by the comments, we now
clarify the scope of human involvement in two aspects.

First, in the course of constructing the benchmark, we (i.e., co-authors of this paper) strictly followed the public fuzzing reports to reproduce the experiments associated with the collected vulnerable software and their exploits. Note that for verifying their sanitizer outputs, it is necessary to leverage the reports because semantics-based AVR requires the information provided by the sanitizer for localization purposes.

Second, 
we (i.e., co-authors) manually evaluate the success rate because automated patch validation cannot guarantee correctness.
Moreover, simply comparing plausible patches against the ground-truth patches 
(crafted by developers) may overlook semantically equivalent patches. Therefore, we also introduce the success rate metric to quantify the degree at which plausible patches are aligned with the ground-truth ones. When semantic equivalence is involved (see Appendix \ref{Appendix: Explanation for Semantic Equivalence} for examples), it is a common practice to evaluate plausible patches of APR and AVR with human involvement, which is why we (i.e., co-authors) examine manually. 
Again, this practice does not need IRB approval based on our experience and the literature (see, e.g., \cite{yin2024thinkrepair, yang2023tsbport, zhang2022vulnfix}).

\section*{Open Science}

We have released our benchmark dataset, experimental code, and results at \url{https://doi.org/10.5281/zenodo.15609776}.

\bibliographystyle{plain}
% Conference Version
% \bibliography{reference}
% Full Version
\bibliography{reference_detailed}

\newpage

\begin{appendices}
\section*{Appendix}

\section{Vulnerability Location vs. Fix Location}
\label{Appendix: fix location v.s vulnerability location}

We use examples to show that fix location may or may not be the same as vulnerability location. 
For the former, Listing \ref{location_consistency} shows a buffer overflow vulnerability (CVE-2024-26889 \cite{CVE-2024-26889}) caused by the \texttt{strcpy} function. The patch is to replace \texttt{strcpy} with \texttt{strscpy}, showing fix location and vulnerability location are identical.

\begin{lstlisting}[float=ht, language = diff, label=location_consistency, linewidth=.46\textwidth, 
xleftmargin=.02\textwidth,
caption={An example of CVE-2024-26889 for showing vulnerability fix location is the same as vulnerability location.}, captionpos=b,
                    linebackgroundcolor={%
                        \ifnum\value{lstnumber}>4
                            \ifnum\value{lstnumber}<6
                                \color{red!20}
                            \fi
                        \fi
                        \ifnum\value{lstnumber}>5
                            \ifnum\value{lstnumber}<7
                                \color{green!20}
                            \fi
                        \fi},]
@@ -908,7 +908,7 @@ int hci_get_dev_info(void __user *arg)
 	else
 		flags = hdev->flags;
        
-	strcpy(di.name, hdev->name);
+	strscpy(di.name, hdev->name, sizeof(di.name));
 	di.bdaddr   = hdev->bdaddr;
 	di.type     = (hdev->bus & 0x0f) | ((hdev->dev_type & 0x03) << 4);
 	di.flags    = flags;
\end{lstlisting}

\begin{lstlisting}[float=ht, language = diff, linewidth=.46\textwidth, 
xleftmargin=.02\textwidth,
label=location_inconsistency, caption={An example of CVE-2013-1428 for showing fix location and vulnerability location are different.}, captionpos=b,
                    linebackgroundcolor={%
                        \ifnum\value{lstnumber}>4
                            \ifnum\value{lstnumber}<8
                                \color{green!20}
                            \fi
                        \fi},]
@@ -394,27 +394,30 @@ void receive_tcppacket(connection_t *c, const char *buffer, int len)
void receive_tcppacket(connection_t *c, const char *buffer, int len) {
    vpn_packet_t outpkt;

+   if(len > sizeof outpkt.data)
+   return;
+
    outpkt.len = len;
    if(c->options & OPTION_TCPONLY)
        outpkt.priority = 0;
    else
        outpkt.priority = -1;
    memcpy(outpkt.data, buffer, len);

    receive_packet(c->node, &outpkt);
}
\end{lstlisting}

For the latter, Listing \ref{location_inconsistency} shows another 
buffer overflow vulnerability (CVE-2013-1428 \cite{CVE-2013-1428}) caused by the \texttt{memcpy} function. The patch is to add the colored statements to check the copy length, showing fix location is different from  vulnerability location. 

\section{On Adaptability of Abstract Templates}
\label{Appendix: Explain for Template}

We justify the limited adaptability of abstract templates as follows.
First, abstract templates are not applicable to different vulnerability types. Listing \ref{template_1} presents an abstract template for fixing cryptographic misuse, where the fix is to replace an insecure hash function with a secure one (i.e., replacing variable \texttt{\$s1} in the {\tt insecureset} of hash functions with the one in the {\tt secureset} of hash functions (i.e., SHA-256)). However, this template would not work for other vulnerability types, such as Null Pointer Exception (NPE) because NPE stems from dereferencing null pointers rather than cryptographic misuse.

\begin{lstlisting}[float=ht, language=diff, label=template_1, frame=lrtb, linewidth=.46\textwidth, 
xleftmargin=.02\textwidth,
caption={An abstract template used by Seader\cite{zhang2022Seader}}, captionpos=b,]
\\ Abstract Template
MessageDigest.getInstance($sl); 
\\ Operation Set for $sl
insecureset={MD2, MD5, SHA-1}; 
secureset={SHA-256};
\end{lstlisting}

Second, abstract templates are, in general, not applicable even to the same type of vulnerabilities. 
Listing \ref{template_2} presents an abstract template for fixing the NULL Pointer Dereference (NPD) vulnerability by adding a NULL check, where \texttt{obj} should be replaced with the variable causing NPD. Listing \ref{CVE-2017-13710} presents a developer-crafted patch to CVE-2017-13710 (NPD vulnerability), where the fix includes a resource release operation (Line 16). When applying this abstract template to fix CVE-2017-13710, the fix only adds a \texttt{return} statement under the if-condition, causing the missing resource release operation to execute prior to the execution of the \texttt{return} statement and thus a failure in fixing the vulnerability.

\begin{lstlisting}[float=ht, language=diff, linewidth=.46\textwidth, 
xleftmargin=.02\textwidth,
label=template_2, frame=lrtb, caption={An abstract template used by VFix\cite{xu2019vfix}}, captionpos=b, linebackgroundcolor={%
                        \ifnum\value{lstnumber}>1
                            \ifnum\value{lstnumber}<5
                                \color{green!20}
                            \fi
                        \fi}
                        ,]
\\ Add NULL pointer check to avoid NPD 
+ if (obj == null){
+     return ;
+ }
\end{lstlisting}

\begin{lstlisting}[float=ht, language=diff, linewidth=.46\textwidth, 
xleftmargin=.02\textwidth,
label=CVE-2017-13710, frame=lrtb, caption={A patch for CVE-2017-13710}, captionpos=b, numbers=left,
linebackgroundcolor={%
                        \ifnum\value{lstnumber}>4
                            \ifnum\value{lstnumber}<6
                                \color{red!20}
                            \fi
                        \fi
                        \ifnum\value{lstnumber}>5
                            \ifnum\value{lstnumber}<7
                                \color{green!20}
                            \fi
                        \fi
                        \ifnum\value{lstnumber}>9
                            \ifnum\value{lstnumber}<20
                                \color{green!20}
                            \fi
                        \fi}
]
@@ -742,12 +742,22 @@ setup_group (bfd *abfd, Elf_Internal_Shdr *hdr, asection *newsect)
    {
      Elf_Internal_Shdr *shdr = elf_tdata (abfd)->group_sect_ptr[i];
      Elf_Internal_Group *idx;
-     unsigned int n_elt;
+     bfd_size_type n_elt;
      if (shdr == NULL)
        continue;
      idx = (Elf_Internal_Group *) shdr->contents;
+     if (idx == NULL || shdr->sh_size < 4)
+       {
+         /* See PR 21957 for a reproducer.  */
+         /* xgettext:c-format */
+         _bfd_error_handler (_("%B: group section '%A' has no contents"),
+                 abfd, shdr->bfd_section);
+         elf_tdata (abfd)->group_sect_ptr[i] = NULL;
+         bfd_set_error (bfd_error_bad_value);
+         return FALSE;
+       }
      n_elt = shdr->sh_size / 4;
      /* Look through this group's sections to see if current
\end{lstlisting}

\section{Vulnerability Types and Software Products Involved in \textsc{Vul4C}}
\label{appendix: Details of Benchmark}

Table \ref{tab: CWE Types Description for Vul4C} lists the 19 vulnerability types contained in \textsc{Vul4C}. As mentioned above, it contains 144 vulnerabilities. The top-5 vulnerability types are: 35 Buffer Overflow vulnerabilities (CWE-119), 29 Out-of-bounds Read vulnerabilities (CWE-125), 16 NULL Pointer Dereference vulnerabilities (CWE-476), 14 Out-of-bounds Write vulnerabilities (CWE-787), and 11 Divide By Zero vulnerabilities (CWE-369). 

\begin{table}[!htbp]
    \centering
%    \footnotesize
    \caption{The 19 vulnerability types contained in \textsc{Vul4C}.}
    \label{tab: CWE Types Description for Vul4C}
    \resizebox{.47\textwidth}{!}{
    \begin{threeparttable}
    \begin{tabular}{|c|c|c|}
        \hline
        {\bf CWE}         & {\bf Description}                                                            & {\bf \#CVE} \\ \hline
        % CWE-119           &Improper Restriction of Operations within the Bounds of a Memory Buffer       \\ \hline
        CWE-119           &Buffer Overflow                                                               & 35 \\ \hline
        CWE-120           &Buffer Copy without Checking Size of Input                                    & 1  \\ \hline 
        CWE-125           &Out-of-bounds Read                                                            & 29 \\ \hline 
        CWE-189           &Numeric Errors                                                                & 2  \\ \hline 
        CWE-190           &Integer Overflow or Wraparound                                                & 9  \\ \hline 
        CWE-191           &Integer Underflow                                                             & 1  \\ \hline 
        CWE-20            &Improper Input Validation                                                     & 6  \\ \hline 
        CWE-369           &Divide By Zero                                                                & 11 \\ \hline 
        CWE-415           &Double Free                                                                   & 1  \\ \hline 
        CWE-416           &Use After Free                                                                & 4  \\ \hline 
        CWE-476           &NULL Pointer Dereference                                                      & 16 \\ \hline 
        CWE-617           &Reachable Assertion                                                           & 2  \\ \hline 
        CWE-682           &Incorrect Calculation                                                         & 1  \\ \hline 
        CWE-704           &Incorrect Type Conversion or Cast                                             & 1  \\ \hline 
        CWE-770           &Allocation of Resources Without Limits or Throttling                          & 1  \\ \hline 
        CWE-787           &Out-of-bounds Write                                                           & 14 \\ \hline 
        CWE-835           &Loop with Unreachable Exit Condition                                          & 4  \\ \hline 
        CWE-843           &Access of Resource Using Incompatible Type                                    & 1  \\ \hline 
        NVD-CWE-Other 
        %Other
        \tnote{1}  & Vulnerability Types Not Covered in NVD                                                                 & 5  \\ \hline
    \end{tabular}
    \begin{tablenotes}
        %\footnotesize
        \item[1] 
        Note that NVD-CWE-Other is an entry in the NVD CWE Slice Categories \cite{NVDCategories}, meaning that this vulnerability type is not covered in the NVD database.
    \end{tablenotes}
    \end{threeparttable}
    }
\end{table}

Table \ref{tab: Software and corresponding software for Vul4C} lists all the 23 software products and their corresponding repository links, where ``\ding{52}'' denotes that a test case runs successfully and ``\ding{56}'' otherwise. Among the 23 products, 12 are client software, nine are libraries, and two are utilities. Moreover, the libtiff library contains the most vulnerabilities (i.e., 24), followed by the binutils utility software (18) and the jasper client software (17). 

\begin{table*}[!htbp]
    \centering
%    \footnotesize
    \caption{The 23 software products with vulnerabilities contained in \textsc{Vul4C}}
    \label{tab: Software and corresponding software for Vul4C}   
    \begin{tabular}{|c|c|c|c|c|}
        \hline
        {\bf Type} & {\bf Software}    & {\bf Repository Links}                                                & {\bf \#CVE} & {\bf Test Suites} \\ \hline
        \multirow{2}{*}{Utilities} &binutils          &\url{https://sourceware.org/git/binutils-gdb}             & 18     &  \ding{56} \\ \cline{2-5} 
                                   &elfutils          &\url{https://sourceware.org/pub/elfutils}                 & 5      &  \ding{52} \\ \hline 
        \multirow{12}{*}{Clients}  &audiofile         &\url{https://github.com/mpruett/audiofile}                & 12     &  \ding{52} \\ \cline{2-5} 
                                   &bento4            &\url{https://github.com/axiomatic-systems/Bento4}         & 6      &  \ding{56} \\ \cline{2-5} 
                                   &gilcc             &\url{https://github.com/trgil/gilcc}                      & 1      &  \ding{56} \\ \cline{2-5}
                                   &graphicsmagick    &\url{http://hg.code.sf.net/p/graphicsmagick/code/}        & 3      &  \ding{52} \\ \cline{2-5} 
                                   &imagemagick       &\url{https://github.com/ImageMagick/ImageMagick}          & 4      &  \ding{56} \\ \cline{2-5} 
                                   &imageworsener     &\url{https://github.com/jsummers/imageworsener}           & 8      &  \ding{52} \\ \cline{2-5} 
                                   &jasper            &\url{https://github.com/jasper-software/jasper}           & 17     &  \ding{56} \\ \cline{2-5} 
                                   &jhead             &\url{https://github.com/Matthias-Wandel/jhead}            & 2      &  \ding{56} \\ \cline{2-5}
                                   &ngiflib           &\url{https://github.com/miniupnp/ngiflib}                 & 4      &  \ding{52} \\ \cline{2-5}
                                   &openjpeg          &\url{https://github.com/uclouvain/openjpeg}               & 5      &  \ding{56} \\ \cline{2-5} 
                                   &qpdf              &\url{https://github.com/qpdf/qpdf}                        & 3      &  \ding{52} \\ \cline{2-5} 
                                   &zziplib           &\url{https://github.com/gdraheim/zziplib}                 & 3      &  \ding{56} \\ \hline 
        \multirow{9}{*}{Libraries} &libarchive        &\url{https://github.com/libarchive/libarchive}            & 3      &  \ding{52} \\ \cline{2-5} 
                                    &libcroco          &\url{https://gitlab.gnome.org/Archive/libcroco}          & 1      & \ding{56} \\ \cline{2-5} 
                                    &libdwarf          &\url{https://github.com/davea42/libdwarf-code}           & 3      & \ding{56} \\ \cline{2-5} 
                                    &libjpeg           &\url{https://github.com/libjpeg-turbo/libjpeg-turbo}     & 4      & \ding{56} \\ \cline{2-5} 
                                    &libming           &\url{https://github.com/libming/libming}                 & 9      & \ding{56} \\ \cline{2-5} 
                                    &libsndfile        &\url{https://github.com/libsndfile/libsndfile}           & 3      & \ding{52} \\ \cline{2-5} 
                                    &libtiff           &\url{https://github.com/vadz/libtiff}                    & 24     & \ding{52} \\ \cline{2-5} 
                                    &libxml2           &\url{https://gitlab.gnome.org/GNOME/libxml2}             & 4      & \ding{52} \\ \cline{2-5} 
                                    &libzip            &\url{https://github.com/nih-at/libzip}                   & 2      & \ding{52} \\ \hline 
    \end{tabular}
\end{table*}

\section{More Results on AVR Tools}
\label{sec:extra evaluation}

\subsection{Justification on the Nine AVR Tools}
\label{Appendix: Applicability of AVR Tools}

As discussed in Section \ref{section: exp design},
we evaluate nine AVR tools, including eight open-source tools and one closed-source tool (i.e., Senx \cite{huang2019senx}, for which we obtain its binary from the authors).
Among them, seven (including six open-source tools and the closed-source Senx \cite{huang2019senx}) are for C/C++ programs and two for Java programs. In what follows we discuss why the other $37-8=29$ open-source AVR tools are not evaluated.

\begin{itemize}
[leftmargin=.32cm,noitemsep,topsep=0pt]
\item {\bf Language incompatibility}. There are 10 AVR tools that are geared towards Solidity \cite{nguyen2021sguard, ferreira2022elysium, so2023smartfix, wang2024contracttinker, rodler2021evmpatch, tolmach2022definery, chida2022remedy}, JavaScript \cite{shahoor2023leakpair} or other programming languages \cite{yu2024tapfixer, liu2023symlogrepair} than C/C++ and Java, meaning that these 10 cannot be evaluated with \textsc{Vul4C} or \textsc{Vul4J}.

\item {\bf Binary code}. One tool \cite{duck2020e9patch}is geared towards binary code rather than source code, meaning it cannot be evaluated with \textsc{Vul4C} or \textsc{Vul4J}.

\item {\bf Vulnerability type}. Two AVR tools \cite{lee2022NPEX, durieux2017npefix} geared towards 
the NPE vulnerabilities in Java. However, \textsc{Vul4J} does not contain any NPE vulnerability,
meaning NPEFix\cite{durieux2017npefix} and NPEX\cite{lee2022NPEX} cannot be evaluated with \textsc{Vul4J}.

\item {\bf Insufficient documentation}. There are 11 AVR tools \cite{le2012genprog, mechtaev2016angelix, gao2016bovinspector, cheng2019cintfix, tian2017errdoc, shariffdeen2021cpr, lee2018memfix, van2018footpatch, hong2020saver, shariffdeen2024crashrepair, cheng2017intpti} that lack sufficient documentation or instruction, making it infeasible for us to repeat their experiments. 
Specifically, \cite{le2012genprog, mechtaev2016angelix, gao2016bovinspector, cheng2019cintfix, tian2017errdoc} only present evaluation results on simple code snippets (e.g., arithmetic operations) rather than programs;
MemFix \cite{lee2018memfix} does not offer any guidance on how to use it; CPR \cite{shariffdeen2021cpr} requires experts-crafted program constraints without providing instructions on how to do it; FootPatch\cite{van2018footpatch} only provides documentation on building, but not running; SAVER \cite{hong2020saver} requires detailed information about memory objects, which are not available to us; CrashRepair\cite{shariffdeen2024crashrepair}does not provide instructions on how to write the configuration file when testing new vulnerabilities; the results of IntPTI\cite{cheng2017intpti} cannot be replicated by us. 

\item{\bf Limited scenarios}. There are four AVR tools \cite{shariffdeen2020patchweave, shariffdeen2021fixmorph, yang2023tsbport, pan2024ppathf} that are geared towards vulnerability patch transplantation, meaning that they aim to fix the same vulnerability in different software or different versions by adapting a given patch to different scenarios.

\item{\bf Other reasons}. There is one AVR tool, Fix2Fit \cite{gao2019fix2fit}, which may generate thousands of candidate patches from a given exploit owing to the use of fuzzing. Since it does not provide any method to prioritize the candidate patches, it is infeasible to evaluate this large number of candidate patches.
\end{itemize}

\subsection{On Semantic Equivalence of Patches}
\label{Appendix: Explanation for Semantic Equivalence}

If a plausible patch is semantically equivalent to a human-crafted ground-truth patch, the plausible patch would not introduce any functionality errors or additional vulnerabilities in the patched program. We give an example in Listings \ref{example_human_patch} and \ref{example_semantic_equiv} to explain this. 

\begin{lstlisting}[float=ht, linewidth=.46\textwidth, 
xleftmargin=.02\textwidth,
language = diff, label=example_human_patch, caption={A human-crafted ground-truth patch to CVE-2017-9043 (binutils)}, captionpos=b,
                    linebackgroundcolor={%
                        \ifnum\value{lstnumber}>4
                            \ifnum\value{lstnumber}<9
                                \color{red!20}
                            \fi
                        \fi
                        \ifnum\value{lstnumber}>8
                            \ifnum\value{lstnumber}<21
                                \color{green!20}
                            \fi
                        \fi},]
@@ -16948,10 +16948,18 @@ print_gnu_build_attribute_name (Elf_Internal_Note * pnote)
     {
     case GNU_BUILD_ATTRIBUTE_TYPE_NUMERIC:
       {
-	unsigned int   bytes = pnote->namesz - (name - pnote->namedata);
-	unsigned long  val = 0;
-	unsigned int   shift = 0;
-	char *         decoded = NULL;
+	unsigned int        bytes = pnote->namesz - (name - pnote->namedata);
+	unsigned long long  val = 0;
+	unsigned int        shift = 0;
+	char *              decoded = NULL;
+
+	/* PR 21378 */
+	if (bytes > sizeof (val))
+	  {
+	    error (_("corrupt name field: namesz of %lu is too large for a numeric value\n"),
+		   pnote->namesz);
+	    return FALSE;
+	  }
 
 	while (bytes --)
 	  {
\end{lstlisting}

Listing \ref{example_human_patch} presents a 
human-crafted ground-truth patch to CVE-2017-9043 (binutils). The basic idea is to add an if-statement as follows: If the value of \verb|bytes| is greater than \verb|sizeof(val)|, then an error will occur when running the program. Note that the type of \verb|val| is \verb|unsigned long long| after patching, meaning that the value of \verb|sizeof(val)| is 8. 

\begin{lstlisting}[float=ht, linewidth=.46\textwidth, 
xleftmargin=.02\textwidth,
language = diff, label=example_semantic_equiv, caption={A  semantically equivalent patch generated by VulnFix \cite{zhang2022vulnfix} for CVE-2017-9043 (binutils)}, captionpos=b,
                    linebackgroundcolor={%
                        \ifnum\value{lstnumber}>4
                            \ifnum\value{lstnumber}<6
                                \color{green!20}
                            \fi
                        \fi},]
@@ -16933,6 +16933,7 @@
 	unsigned long  val = 0;
 	unsigned int   shift = 0;
 	char *         decoded = NULL;
+    if (!(bytes <= 8)) exit(1);
 
 	while (bytes --)
 	  {
\end{lstlisting}

Listing \ref{example_semantic_equiv} presents a plausible patch generated by VulnFix \cite{zhang2022vulnfix} for CVE-2017-9043. It is semantically equivalent to the human-crafted ground-truth patch presented in Listing \ref{example_human_patch} because it contains the if-statement \verb|if(!(bytes<=8))| such that an error will occur when the condition is met.

\subsection{Further Evaluation Results on AVR Tools}
\label{Appendix: Further Evaluation Results on AVR Tools}

\begin{table*}[!htbp]
    \centering
    \footnotesize
    \caption{Success rate of the seven AVR tools and two APR tools via the top-10 CWE types in \textsc{Vul4C}, where ``\texttt{-}'' means the tool is not applicable to the CWE.}
    \label{tab: Evaluation results over different CWE}
    \resizebox{\textwidth}{!}{
    \begin{tabular}{|c|c|c|c|c|c|c|c|c|c|}
        \hline
        \multirow{2}{*}{\bf CWE} & \multicolumn{7}{c|}{\bf AVR} &\multicolumn{2}{c|}{\bf APR} \\ \cline{2-10}
        ~ & {\bf VulRepair} \cite{fu2022vulrepair} & {\bf VRepair} \cite{chen2023vrepair} & {\bf \color{black} VQM} \cite{fu2024vqm} & {\bf \color{black} VulMaster} \cite{zhou2024vulmaster} &{\bf VulnFix} \cite{zhang2022vulnfix} & {\bf ExtractFix} \cite{gao2021extractfix} & {\bf Senx} \cite{huang2019senx} 
                   & {\bf CquenceR\cite{pinconschi2021comparative}} & {\bf NTR\cite{huang2025ntr}} \\ \hline
        CWE-119   &{\color{black}0.0\%(0/21)}  &{\color{black}0.0\%(0/21)}  &{\color{black}0.0\%(0/21)}  &{\color{black}4.8\%(1/21)}   &{\color{black}14.3\%(5/35)}  &{\color{black}5.9\%(1/17)}     &{\color{black}0.0\%(0/35)}        & {\color{black}0.0\%(0/17)}   & {\color{black}35.3\%(6/17)}       \\ \hline
        CWE-125   &0.0\%(0/13)                 &0.0\%(0/13)                &{\color{black}0.0\%(0/13)}  &{\color{black}0.0\%(0/13)}   &3.4\%(1/29)                &0.0\%(0/13)                   &0.0\%(0/24)                           &{\color{black}0.0\%(0/7)}    &{\color{black}0.0\%(0/7)} \\ \hline                                                                
                                                                                             
        CWE-476   &{\color{black}0.0\%(0/10)}  &{\color{black}0.0\%(0/10)}  &{\color{black}0.0\%(0/10)}  &{\color{black}0.0\%(0/10)}   &{\color{black}12.5\%(2/16)}   &{\color{black}0.0\%(0/7)} &-                                   &{\color{black}0.0\%(0/5)}   &{\color{black}0.0\%(0/5)}    \\ \hline
        CWE-787   &0.0\%(0/8)                 &0.0\%(0/8)                 &{\color{black}0.0\%(0/8)}   &{\color{black}0.0\%(0/8)}    &7.1\%(1/14)                 &0.0\%(0/10)                   &0.0\%(0/14)                       &{\color{black}0.0\%(0/5)}    &{\color{black}0.0\%(0/5)}    \\ \hline
        CWE-369   &0.0\%(0/7)                 &0.0\%(0/7)                 &{\color{black}0.0\%(0/7)}   &{\color{black}0.0\%(0/7)}    &36.4\%(4/11)                &0.0\%(0/7)                    &-                                 &{\color{black}0.0\%(0/9)}    &{\color{black}0.0\%(0/9)} \\ \hline
        CWE-190   &0.0\%(0/6)                 &0.0\%(0/6)                 &{\color{black}0.0\%(0/6)}   &{\color{black}0.0\%(0/6)}    &0.0\%(0/8)                 &0.0\%(0/5)                    &0.0\%(0/9)                         & -  & -   \\ \hline
       
        CWE-20    &0.0\%(0/2)                 &0.0\%(0/2)                 &{\color{black}0.0\%(0/2)}   &{\color{black}0.0\%(0/2)}    &25.0\%(1/4)                 &-                             &-                                 &{\color{black}0.0\%(0/1)}   &{\color{black}0.0\%(0/1)}\\ \hline
        CWE-416   &0.0\%(0/3)                 &0.0\%(0/3)                 &{\color{black}0.0\%(0/3)}   &{\color{black}0.0\%(0/3)}    & 0.0\%(0/4)                 &-                             &-                                 &-    
        &-     \\ \hline
        CWE-835   &0.0\%(0/1)                 &0.0\%(0/1)                 &{\color{black}0.0\%(0/1)}   &{\color{black}0.0\%(0/1)}    & 0.0\%(0/4)                 &-                             &-                                 &{\color{black}0.0\%(0/1)}   &{\color{black}0.0\%(0/1)}   \\ \hline
        CWE-189   &0.0\%(0/2)                 &0.0\%(0/2)                 &{\color{black}0.0\%(0/2)}   &{\color{black}0.0\%(0/2)}    & 0.0\%(0/2)                 &0.0\%(0/1)                    &-                                 &{\color{black}100.0\%(1/1)}  &{\color{black}100.0\%(1/1)}  \\ \hline
    \end{tabular}
    }
\end{table*}

Table \ref{tab: Evaluation results over different CWE} summarizes the {\em success rate} of the nine AVR tools over the top-10 vulnerability types in \textsc{Vul4C} (i.e., the types with the largest number of vulnerabilities). We observe that the AVR tools that target specific vulnerability types are not as effective as the general AVR tools (which are not designed towards specific vulnerability types). For instnace, ExtractFix \cite{gao2021extractfix} and Senx \cite{huang2019senx} are designed to address certain vulnerability types (e.g., memory-related vulnerabilities for ExtractFix; bad cast, buffer overflow and integer overflow vulnerabilities for Senx) and achieve a lower success rate than VulnFix\cite{zhang2022vulnfix} which is not geared towards specific vulnerability types. This phenomenon can be understood as follows: 
ExtractFix and Senx typically address memory-related vulnerabilities by simply adding if-return statements to prevent vulnerabilities from being triggered; whereas, VulnFix uses snapshot fuzzing and considers program states with respect to both positive and negative test cases to generate more precise condition ranges.

\subsection{Case Study of Patch Generation}
\label{sec:case study}

We provide case studies for the causes of failures. Cases \ref{Appendix: CVE-2017-15021} to \ref{Appendix: CVE-2017-7598} account for the 3 main causes of failures of semantics-based patch generation methods, and \ref{Appendix: CVE-2020-26208} to \ref{Appendix: CVE-2016-9828} discuss those of learning-based patch generation methods.

\subsubsection{Case Study: CVE-2017-15021 (binutils)} 

\label{Appendix: CVE-2017-15021}

VulnFix\cite{zhang2022vulnfix} generates candidate patch invariants via snapshot fuzzing, reduces them to a single invariant, and synthesizes it into a candidate patch. 
However, it may fail to reduce candidate patch invariants to a single one within limited fuzzing time, causing a failure. Take vulnerability CVE-2017-15021 in binutils as an example. It fails to produce a single invariant when  reaching the time limit (see Listing \ref{case_5_generate}).

\begin{lstlisting}[float=ht, linewidth=.46\textwidth, 
xleftmargin=.02\textwidth,
language = diff, label=case_5_generate, caption={ The output generated by Vulnfix \cite{zhang2022vulnfix} for CVE-2017-15021 (binutils) }, captionpos=b,,] 
FAIL (More than one or no patch invariants in the end)

Patch Invariants:
2
['_GSize_name - crc_offset >= 4', '_GSize_contents - crc_offset >= 4']
\end{lstlisting}

\subsubsection{Case Study: CVE-2016-9387 (jasper)} 

\label{Appendix: CVE-2016-9387}

Senx \cite{huang2019senx} fails when incorrectly parses pointer variables. Take CVE-2016-9387 in jasper as an example, Senx fails to generate \verb|dec->numhtiles| and \verb|dec->numvtiles| but instead generating \verb|dec@numhtiles| and \verb|dec@numvtiles| (see Listing \ref{case_6_generate}). 

\begin{lstlisting}[float=ht, linewidth=.46\textwidth, 
xleftmargin=.02\textwidth,
language = diff, label=case_6_generate, caption={The candidate patch generated by Senx\cite{huang2019senx} for CVE-2016-9387 (jasper) }, captionpos=b,
                    linebackgroundcolor={%
                        \ifnum\value{lstnumber}>4
                            \ifnum\value{lstnumber}<6
                                \color{green!20}
                            \fi
                        \fi},]
@@ -1231,7 +1232,10 @@ static int jpc_dec_process_siz(jpc_dec_t *dec, jpc_ms_t *ms)
 
 	dec->numhtiles = JPC_CEILDIV(dec->xend - dec->tilexoff, dec->tilewidth);
 	dec->numvtiles = JPC_CEILDIV(dec->yend - dec->tileyoff, dec->tileheight);
    +	if ((dec@numhtiles)*(dec@numvtiles) > 2147483647)
        dec->numtiles = dec->numhtiles * dec->numvtiles;
 	JAS_DBGLOG(10, ("numtiles = %d; numhtiles = %d; numvtiles = %d;\n",
 	  dec->numtiles, dec->numhtiles, dec->numvtiles));
 	if (!(dec->tiles = jas_alloc2(dec->numtiles, sizeof(jpc_dec_tile_t)))) {

\end{lstlisting}

\subsubsection{Case Study: CVE-2017-7598 (libtiff)} 
\label{Appendix: CVE-2017-7598}

ExtractFix\cite{gao2021extractfix} may fail due to incorrect using of array variables. Take vulnerability CVE-2017-7598 in libtiff as an example, ExtractFix fails to generate \verb|m.i[1]| as it generates \verb|m[1]| (see Listing \ref{case_7_generate}). 

\begin{lstlisting}[float=ht, linewidth=.46\textwidth, 
xleftmargin=.02\textwidth,
language = diff, label=case_7_generate, caption={ The candidate patch generated by ExtractFix\cite{gao2021extractfix} for CVE-2017-7598 (libtiff) }, captionpos=b,
                    linebackgroundcolor={%
                        \ifnum\value{lstnumber}>3
                            \ifnum\value{lstnumber}<5
                                \color{green!20}
                            \fi
                        \fi},]
    if (m.i[0]==0)
    	*value=0.0;
    else
+       if(m[1] != 0) return  (TIFFReadDirEntryErrOk);
    	*value=(double)m.i[0]/(double)m.i[1];

\end{lstlisting}

\subsubsection{Case Study: CVE-2020-26208 (jhead)}

\label{Appendix: CVE-2020-26208}

This case illustrates failure reason ``{\it Error in generating context tokens}'' for learning-based methods  \cite{fu2022vulrepair, chen2023vrepair, fu2024vqm, zhou2024vulmaster}, which rely on the first three context tokens to pinpoint where the candidate patches should make modifications (i.e., generating incorrect context tokens will render the candidate patches incorrect). Take vulnerability CVE-2020-26208 in jhead as an example. 
The first three tokens ``\verb|malloc|'', ``\verb|(|'', and ``\verb|itemlen|''  
in the processed target output (see Listing \ref{case_1_tgt}), aim to identify the fix location. However, the output  generated by VRepair (via these three tokens) \cite{fu2022vulrepair} is different from the target output (see Listing \ref{case_1_generate}), leading to incorrect candidate patch. 

\begin{lstlisting}[float=ht, language = diff, floatplacement=htbp, linewidth=.46\textwidth, 
xleftmargin=.02\textwidth,
label=case_1_tgt, caption={The target output for CVE-2020-26208 (jhead)}, captionpos=b, ]
    <S2SV_ModStart> malloc ( itemlen + 20
\end{lstlisting}

\begin{lstlisting}[float=ht, language = diff, 
linewidth=.46\textwidth, 
xleftmargin=.02\textwidth,
label=case_1_generate, caption={The candidate patch generated by VRepair \cite{fu2022vulrepair} for  CVE-2020-26208 (jhead)}, captionpos=b, ]
    <S2SV_ModStart> ) ; } if ( <unk> == NULL ) return FALSE ;
\end{lstlisting}

\subsubsection{Case Study: CVE-2017-15023 (binutils)}

\label{Appendix: CVE-2017-15023}

This case is used to illustrate failure reason ``{\it Error in matching
context}'' for learning-based methods.
Even when the context tokens generated by the learning-based patch generation methods \cite{fu2022vulrepair, chen2023vrepair, fu2024vqm, zhou2024vulmaster} are consistent with the target, there is no guarantee that the fix location pinpointed by the candidate patch is correct. This is because these methods utilize the first matching context as the fix location to apply the candidate patch. Take the vulnerability CVE-2017-15023 in binutils for example. 
In the fixed function, there are three lines that are partially consistent with the context 
``\verb|+= bytes_read ;|'' 
(see Listing \ref{case_2_diff}). 
The manual patches are applied to the third of them, while the output generated by the learning-based methods \cite{fu2022vulrepair, chen2023vrepair} is applied to the first one.

\begin{lstlisting}[float=h, linewidth=.46\textwidth, 
xleftmargin=.02\textwidth,
language = diff, label=case_2_diff, caption={The manual patch for CVE-2017-15023 (binutils)}, captionpos=b,
                    linebackgroundcolor={%
                        \ifnum\value{lstnumber}>3
                            \ifnum\value{lstnumber}<5
                                \color{red!60}
                            \fi
                        \fi
                        \ifnum\value{lstnumber}>5
                            \ifnum\value{lstnumber}<7
                                \color{red!60}
                            \fi
                        \fi
                        \ifnum\value{lstnumber}>10
                            \ifnum\value{lstnumber}<12
                                \color{green!60}
                            \fi
                        \fi
                        \ifnum\value{lstnumber}>11
                            \ifnum\value{lstnumber}<19
                                \color{green!20}
                            \fi
                        \fi},]
   for (formati = 0; formati < format_count; formati++)
     {
       _bfd_safe_read_leb128 (abfd, buf, &bytes_read, FALSE, buf_end);
       buf += bytes_read ; // Incorrect Location
       _bfd_safe_read_leb128 (abfd, buf, &bytes_read, FALSE, buf_end);
       buf += bytes_read ; // Incorrect Location
     }

   data_count = _bfd_safe_read_leb128 (abfd, buf, &bytes_read,
       FALSE, buf_end);
   buf += bytes_read; // Correct Location
+  if (format_count == 0 && data_count != 0)
+    {
+      _bfd_error_handler (_("Dwarf Error: Zero format count."));
+      bfd_set_error (bfd_error_bad_value);
+      return FALSE;
+    }
+
   for (datai = 0; datai < data_count; datai++)
     {
       bfd_byte *format = format_header_data;
\end{lstlisting}

\subsubsection{Case Study: CVE-2016-9827 (libming)}

\label{Appendix: CVE-2016-9827}

This case explains the ``{\it Syntax error}'' failure for learning-based methods \cite{fu2022vulrepair, chen2023vrepair}, which generate unmatched symbols (e.g., `\texttt{\}}' and `\texttt{)}') and make compilation fail. Considering vulnerability CVE-2016-9827 in libming (cf. Listing \ref{case_3_generate}), the absence of `\texttt{)}' in statement ``\verb|TEMP_FAILURE_RETRY(readBytes(f, length);|'' makes the patch generated by VulRepair\cite{fu2022vulrepair} fail to compile.

% \begin{lstlisting}[float=tb, language = diff, label=case_3_human, caption={ The manual patch for CVE-2016-9827 (libming) }, captionpos=b,
%                     linebackgroundcolor={%
%                         \ifnum\value{lstnumber}>4
%                             \ifnum\value{lstnumber}<6
%                                 \color{green!20}
%                             \fi
%                         \fi
%                         \ifnum\value{lstnumber}>5
%                             \ifnum\value{lstnumber}<7
%                                 \color{red!20}
%                             \fi
%                         \fi},]
% @@ -2752,7 +2752,7 @@ parseSWF_PROTECT (FILE * f, int length)
%    PAR_BEGIN (SWF_PROTECT);
 
%    if( length != 0 ) {
% -  	parserrec->Password = readBytes (f, length);
% +  	parserrec->Password = readString (f);
%    } else {
%    	parserrec->Password = NULL;
%    }
% \end{lstlisting}

\begin{lstlisting}[float=ht, linewidth=.46\textwidth, 
xleftmargin=.02\textwidth,
language = diff, label=case_3_generate, caption={ The candidate patch generated by VulRepair \cite{fu2022vulrepair} for CVE-2016-9827 (libming) }, captionpos=b,
                    linebackgroundcolor={%
                        \ifnum\value{lstnumber}>4
                            \ifnum\value{lstnumber}<6
                                \color{red!20}
                            \fi
                        \fi
                        \ifnum\value{lstnumber}>5
                            \ifnum\value{lstnumber}<8
                                \color{green!20}
                            \fi
                        \fi},]
@@ -2752,7 +2752,7 @@ parseSWF_PROTECT (FILE * f, int length)
   PAR_BEGIN(SWF_PROTECT);

   if (length != 0) {
-   parserrec->Password = readBytes (f, length);    
+   parserrec -> Password = 
        TEMP_FAILURE_RETRY( readBytes ( f , length ) ;
   } else {
    parserrec->Password = NULL;
   }

\end{lstlisting}

\subsubsection{Case Study: CVE-2016-9828 (libming)}

\label{Appendix: CVE-2016-9828}

This case is used to illustrate failure reason ``{\it Unknown tokens}'' for learning-based methods.
The learning-based patch generation methods \cite{fu2022vulrepair, chen2023vrepair, fu2024vqm, zhou2024vulmaster} may generate unknown tokens, which can lead to the failure of compilation. Take the vulnerability CVE-2016-9828 in libming for example. Due to the undefined token
\verb|<unk>|
(see Listing \ref{case_4_generate}), the candidate patch generated by VulRepair \cite{fu2022vulrepair} cannot be compiled.

\begin{lstlisting}[float=ht, linewidth=.46\textwidth, 
xleftmargin=.02\textwidth,
language = diff, label=case_4_generate, caption={The candidate patch generated by VRepair \cite{fu2022vulrepair} for CVE-2016-9828 (libming) }, captionpos=b,,]
<S2SV_ModStart> ; } } 
if ( stream -> blockName [ blockName ] == <unk> ) { return ; }
\end{lstlisting}

\section{Evaluation of APR Tools}
\label{Appendix: APR tools}

This section reports a side-product on evaluating four open-source APR tools, including two tools for C/C++ programs, namely NTR \cite{huang2025ntr} and CquenceR \cite{pinconschi2021comparative}, via our benchmark {\sc Vul4C} dataset (Appendix \ref{appendix: APR results for C/C++ programs}); and two APR tools for Java programs, namely  ThinkRepair \cite{yin2024thinkrepair} and SRepair \cite{xiang2024srepair}, via the third-party {\sc Vul4J} dataset (Appendix \ref{appendix: APR results for Java programs}). 
The experimental platform is the same as the one used for evaluating AVR tools. We use the following metrics (defined in Section \ref{section: exp design}) to evaluate these four APR tools, namely: {\em consistency rate}, {\em success rate}, {\em patch restoration rate}, {\em patch compilation rate}, and {\em test pass rate}. 

\subsection{Evaluating APR Tools for C/C++}
\label{appendix: APR results for C/C++ programs}

\noindent{\bf APR tools and parameters}.
The configurations of the two APR tools for C/C++ programs are as follows.

\begin{itemize}
[leftmargin=.32cm,noitemsep,topsep=0pt]
\item {\bf NTR} \cite{huang2025ntr}. NTR predicts fix templates by fine-tuning CodeT5 \cite{wang2021codet5}. 
We use the VulGen \cite{nong2023vulgen} dataset provided by the authors of \cite{huang2025ntr} to fine-tune NTR and conduct experiments with the single-hunk vulnerabilities in \textsc{Vul4C}. 
By using the same parameters as in \cite{huang2025ntr}, NTR generates 10 candidate patches for each predicted template. We use the Top 5 templates ranked by NTR to generate 50 candidate patches in total.
\item {\bf CquenceR} \cite{pinconschi2021comparative}. CquenceR is a C/C++ implementation of SequenceR \cite{chen2021sequencer}, which is based on the NMT architecture and uses a sequence-to-sequence model to generate patches. In our experiments, we set the beam size to 50 to generates 50 candidates, as in \cite{pinconschi2021comparative}. Note we are only able to evaluate it with the single-hunk vulnerabilities in \textsc{Vul4C} because CquenceR is only applicable to fixes to single-hunk vulnerabilities.
\end{itemize}

\begin{table*}[!htbp]
    \centering
    \footnotesize
    \caption{Evaluation results of APR tools for generating vulnerability patches}
    \label{tab: Evaluation results of APR tools for generating vulnerability patches}
    \begin{tabular}{{|l|c|c|c|c|}}
        \hline
        \multirow{2}{*}{\bf Tool}    & \multicolumn{2}{c|}{\bf C/C++}             &\multicolumn{2}{c|}{\bf Java}     \\ \cline{2-5}
        ~                            & CquenceR \cite{zhang2022Seader} & NTR \cite{chi2023seqtrans}  & ThinkRepair\cite{yin2024thinkrepair}       &SRepair\cite{xiang2024srepair}             \\ \hline
        {\bf Patch restoration rate} & 100.0\% (2600/2600)                   & 100.0\% (2550/2550)                  & 100\% (1619/1619)        & 100\% (2963/2963)  \\ \hline
         {\bf Patch compilation rate} & 28.1\% (731/2600)                     & 71.5\% (1822/2550)                   & 53.1\% (859/1619)
        & 62.6\% (1855/2963)  \\ \hline
        {\bf Test pass rate}         & 4.2\% (110/2600)                     & 8.2\% (208/2550)                       & 4.9\%(79/1619)
        & 9.2\% (273/2963)   \\ \hline
        {\bf Consistency rate}       & 1.9\% (1/52)                    & 11.8\% (6/51)                      & 6.1\%(2/33)
        & 6.3\% (4/63)        \\ \hline
        {\bf Success rate}           & 1.9\% (1/52)                    & 15.7\% (8/51)                      & 18.2\% (6/33)
        &14.3\% (9/63)        \\ \hline
    \end{tabular}
\end{table*}

\noindent{\bf Experimental Results}. Table \ref{tab: Evaluation results of APR tools for generating vulnerability patches} summarizes the results with the two APR tools for C/C++ programs, namely CquenceR\cite{pinconschi2021comparative} and NTR\cite{huang2025ntr}. We make two observations. First,  
NTR achieves a patch compilation rate and a test pass rate of 71.5\% and 8.2\%, respectively, which are respectively 43.4\% and 4\% higher than those of CquenceR (28.1\% and 4.2\%).
This discrepancy can be attributed to the following: CquenceR uses CodeT5 to generate candidate patches at the statement level, resulting in the lack of sufficient contextual information;
whereas, NTR employs StarCoder to generate candidate patches at the function level, resulting in the accommodation of more context information. 
Second, NTR also outperforms CquenceR in terms of consistency rate and success rate. Specifically, NTR achieves a consistency rate of 11.8\% and a success rate of 15.7\%, while CquenceR only scores 1.9\% in both metrics. This can be attributed to the fact that
StarCoder has a higher comprehension capability than CodeT5 while leveraging the template-based constraints  predicted by CodeT5 to mitigate uncertainty in its code generation;
whereas, CquenceR uses the candidate patches generated by CodeT5, but these patches are often different from human-crafted patches.

\begin{observation}
APR models with a stronger code comprehension capability perform better; leveraging multiple models can perform even better. 
\end{observation}

\subsection{Evaluating APR Tools for Java Programs}
\label{appendix: APR results for Java programs}

\noindent{\bf APR tools and parameters}. 
The configurations of the two APR tools for Java programs are as follows.
\begin{itemize}
[leftmargin=.32cm,noitemsep,topsep=0pt]
\item {\bf ThinkRepair} \cite{yin2024thinkrepair}. ThinkRepair is an LLM-based APR tool. It first collects pre-fixed knowledge by {\em Chain-of-Thought} (CoT) and then fixes bugs via CoT and few-shot learning. We use the same parameters as in \cite{yin2024thinkrepair}, and consider the first 50 resulting candidate patches. Note that patch generation may terminate prematurely when the prompt exceeds 4,096 tokens in the feedback iteration round, leading to less than 50 candidates. 
Since ThinkRepair is only applicable to single-function vulnerabilities (i.e., a vulnerability resides in a single program function), we select 33 vulnerabilities from the \textsc{Vul4J} dataset after removing the non-compilable projects.
\item {\bf SRepair} \cite{xiang2024srepair}. SRepair is a CoT-based function-level APR tool. %Different from ThinkRepair \cite{yin2024thinkrepair}, 
It uses repair suggestion to guide patch generation and demonstrates a higher performance than the closed-source tool ChatRepair \cite{xia2024chatrepair}, while applicable to multi-function vulnerabilities (i.e., vulnerabilities cutting across multiple program functions). Thus, we test it for 63 vulnerabilities from the \textsc{Vul4J} dataset after removing the non-compilable projects. Since GPT-3.5-turbo does not guarantee a sufficient number of repair suggestions, some vulnerabilities may yield fewer than 50 candidate patches.
\end{itemize}

\noindent{\bf Experimental Results}.
Table \ref{tab: Evaluation results of APR tools for generating vulnerability patches} 
also summarizes the experimental results for these two tools. We make two observations. 
First, SRepair achieves a higher patch compilation rate and a higher test pass rate than ThinkRepair. Specifically, SRepair achieves a patch compilation rate of 62.6\% and a test pass rate of 9.2\%; whereas, ThinkRepair achieves 53.1\% and 4.9\%, respectively. This discrepancy can be attributed to their CoT strategy: ThinkRepair directly incorporates test messages into prompt templates for patch generation; whereas, SRepair constructs a more sophisticated chain-of-thought process by first guiding LLM to comprehend the test message and generate a repair suggestion, and then using the suggestion to construct an informed prompt for patch generation.
Second, SRepair achieves a success rate of 14.3\%, which is 3.9\% lower than ThinkRepair's (18.2\%). This discrepancy can be attributed to the following: ThinkRepair only handles single-function vulnerabilities, while SRepair deals with more complex cross-function vulnerabilities.
More specifically, their CoT only accommodate names of failed test cases and exceptions in {\sc Vul4J},
namely lacking the vulnerability information (e.g., vulnerability types and locations) that cannot always be inferred by LLMs. This makes APRs often select incorrect fix locations and misidentify vulnerability types.

\begin{observation}
Employing detailed thought processes in the CoT framework can enhance LLMs' patch generation capabilities, while lacking vulnerability information leads to repair failures.
\end{observation}

\subsection{Further Evaluation Results on APR Tools}

Table \ref{tab: Evaluation results over different CWE} summarizes the {\em success rate} of the two APR tools over the top-10 vulnerability types in \textsc{Vul4C} (i.e., the types with the largest number of vulnerabilities). We observe that NTR's success rate on CWE-119 is significantly higher than that of CquenceR on CWE-119. This suggests that directly generating patches through training-based methods has a high error rate, and that using a deep learning model for simple template prediction yields a higher success rate. We also observe that NTR's success rate remains very low for the other vulnerability types (than CWE-119), suggesting that LLMs have varying capabilities in repairing different types of vulnerabilities. Nevertheless, LLMs can identify and fix memory buffer vulnerabilities.

\section{Detailed Results for AVR and APR}
\label{Appendix: Detailed Experiment}

\begin{table*}[!htbp]
    \centering
\caption{comparing patch generation capabilities of the seven AVR tools and two APR tools with respect to  \textsc{Vul4C}}
    \label{tab: Detailed Experiment}
    \resizebox{!}{0.475\textheight}{
    \begin{tabular}{|c|c|c|c|c|c|c|c|c|c|c|c|c|}
        \hline
        \multicolumn{4}{|c}{\textbf{Vulnerable program}} & \multicolumn{7}{|c|}{\textbf{AVR}} & \multicolumn{2}{c|}{\textbf{APR}} \\ \hline
        Software & CVE ID & CWE & Sanitizer & VRepair\cite{chen2023vrepair} & VulRepair\cite{fu2022vulrepair} & VQM\cite{fu2024vqm} & VulMaster\cite{zhou2024vulmaster} & VulnFix\cite{zhang2022vulnfix} & ExtractFix\cite{gao2021extractfix} & Senx\cite{huang2019senx} & CquenceR\cite{pinconschi2021comparative} & NTR\cite{huang2025ntr} \\ \hline
        \multirow{5}{*}{audiofile} & CVE-2017-6828 & CWE-119 & ASAN & -- & -- & -- & -- & \ding{56} & CE & \ding{56} & 50/15/4/0 & 50/46/10/1 \\ \cline{2-13}
        ~ & CVE-2017-6827 & CWE-119 & ASAN & -- & -- & -- & -- & \ding{56} & CE & \ding{56} & 50/16/4/0 & 50/46/5/1 \\ \cline{2-13}
        ~ & CVE-2017-6832 & CWE-119 & ASAN & -- & -- & -- & -- & \ding{56} & CE & \ding{56} & 50/15/0/0 & 50/46/2/0 \\ \cline{2-13}
        ~ & CVE-2017-6833 & CWE-369 & ASAN & -- & -- & -- & -- & \ding{56} & CE & -- & 50/18/0/0 & 50/46/0/0 \\ \cline{2-13}
        ~ & CVE-2017-6835 & CWE-369 & ASAN & -- & -- & -- & -- & \ding{56} & CE & -- & 50/18/0/0 & 50/46/0/0 \\ \cline{2-13}
        ~ & CVE-2017-6838 & CWE-190 & UBSAN & \ding{56} & 47/4/4/0 & 44/0/0/0 & 50/0/0/0 & 1/1/1/0 & CE & \ding{56} & -- & -- \\ \hline
        bento4 & CVE-2017-14640 & CWE-476 & ASAN & -- & -- & -- & -- & \ding{56} & CE & -- & 50/15/8/0 & 50/19/0/0 \\ \hline
        \multirow{18}{*}{binutils} & CVE-2017-14729 & CWE-119 & ASAN & 7/0/0/0 & 4/0/0/0 & 43/0/0/0 & 50/2/0/0 & \ding{56} & CE & 1/1/0/0 & 50/10/0/0 & 50/47/0/0 \\ \cline{2-13}
        ~ & CVE-2017-14745 & CWE-190 & ASAN & -- & -- & -- & -- & \ding{56} & CE & \ding{56} & -- & -- \\ \cline{2-13}
        ~ & CVE-2017-14939 & CWE-125 & ASAN & 34/0/0/0 & 6/0/0/0 & 50/0/0/0 & \ding{56} & \ding{56} & CE & \ding{56} & -- & -- \\ \cline{2-13}
        ~ & CVE-2017-14940 & CWE-476 & ASAN & 47/0/0/0 & 50/0/0/0 & \ding{56} & 50/2/0/0 & \ding{56} & CE & -- & 50/11/0/0 & 50/37/0/0 \\ \cline{2-13}
        ~ & CVE-2017-15020 & CWE-125 & ASAN & 11/7/0/0 & 40/0/0/0 & 49/10/0/0 & 50/20/0/0 & \ding{56} & CE & \ding{56} & -- & -- \\ \cline{2-13}
        ~ & CVE-2017-15021 & CWE-125 & ASAN & \ding{56} & 49/3/0/0 & 48/3/0/0 & 50/5/0/0 & \ding{56} & CE & \ding{56} & 50/13/0/0 & 50/49/0/0 \\ \cline{2-13}
        ~ & CVE-2017-15022 & CWE-476 & ASAN & 40/0/0/0 & 50/0/0/0 & \ding{56} & 50/0/0/0 & TO & CE & -- & -- & -- \\ \cline{2-13}
        ~ & CVE-2017-15023 & CWE-476 & ASAN & 1/1/0/0 & 50/30/30/0 & 46/0/0/0 & 50/0/0/0 & \ding{56} & CE & -- & 50/7/1/0 & 50/39/18/0 \\ \cline{2-13}
        ~ & CVE-2017-15024 & CWE-835 & ASAN & \ding{56} & \ding{56} & \ding{56} & 50/0/0/0 & \ding{56} & CE & -- & -- & -- \\ \cline{2-13}
        ~ & CVE-2017-15025 & CWE-369 & ASAN & 7/0/0/0 & 19/14/0/0 & 50/0/0/0 & 50/13/0/0 & 1/1/1/1 & CE & -- & 50/3/1/0 & 50/50/6/0 \\ \cline{2-13}
        ~ & CVE-2017-15938 & CWE-119 & ASAN & 15/0/0/0 & 50/1/0/0 & 50/0/0/0 & 48/1/0/0 & \ding{56} & CE & \ding{56} & -- & -- \\ \cline{2-13}
        ~ & CVE-2017-15939 & CWE-119 & ASAN & 4/1/0/0 & 49/24/0/0 & 50/0/0/0 & 50/0/0/0 & 1/1/1/1 & CE & \ding{56} & -- & -- \\ \cline{2-13}
        ~ & CVE-2017-6965 & CWE-119 & ASAN & \ding{56} & \ding{56} & \ding{56} & \ding{56} & \ding{56} & CE & \ding{56} & -- & -- \\ \cline{2-13}
        ~ & CVE-2017-9038 & CWE-125 & ASAN & 33/0/0/0 & 1/1/0/0 & \ding{56} & 38/0/0/0 & \ding{56} & CE & \ding{56} & 50/0/0/0 & 50/0/0/0 \\ \cline{2-13}
        ~ & CVE-2017-9040 & CWE-476 & ASAN & 9/0/0/0 & 45/0/0/0 & 1/0/0/0 & 50/0/0/0 & \ding{56} & CE & -- & -- & -- \\ \cline{2-13}
        ~ & CVE-2017-9042 & CWE-704 & UBSAN & 10/0/0/0 & 43/0/0/0 & 1/0/0/0 & 50/0/0/0 & \ding{56} & CE & \ding{56} & -- & -- \\ \cline{2-13}
        ~ & CVE-2017-9043 & CWE-20 & UBSAN & \ding{56} & 8/0/0/0 & 50/0/0/0 & 50/1/0/0 & 1/1/1/1 & CE & -- & -- & -- \\ \cline{2-13}
        ~ & CVE-2018-10372 & CWE-125 & ASAN & 17/0/0/0 & \ding{56} & 50/0/0/0 & 23/0/0/0 & \ding{56} & CE & \ding{56} & 50/15/0/0 & 50/0/0/0 \\ \hline
        \multirow{4}{*}{elfutils} & CVE-2017-7607 & CWE-125 & ASAN & 50/0/0/0 & 50/18/0/0 & 50/0/0/0 & 43/22/20/0 & \ding{56} & CE & CE & 50/40/0/0 & 50/50/0/0 \\ \cline{2-13}
        ~ & CVE-2017-7610 & CWE-125 & ASAN & \ding{56} & 4/0/0/0 & 43/0/0/0 & \ding{56} & \ding{56} & CE & CE & -- & -- \\ \cline{2-13}
        ~ & CVE-2017-7611 & CWE-125 & ASAN & \ding{56} & \ding{56} & 21/0/0/0 & \ding{56} & \ding{56} & CE & CE & 50/13/0/0 & 50/46/0/0 \\ \cline{2-13}
        ~ & CVE-2017-7612 & CWE-125 & ASAN & \ding{56} & 50/0/0/0 & 33/0/0/0 & 3/0/0/0 & \ding{56} & CE & CE & -- & -- \\ \hline
        graphicsmagick & CVE-2017-12937 & CWE-125 & ASAN & 27/0/0/0 & \ding{56} & \ding{56} & \ding{56} & 1/1/0/0 & \ding{56} & 1/1/0/0 & -- & -- \\ \hline
        \multirow{3}{*}{imagemagick} & CVE-2016-9556 & CWE-119 & ASAN & 50/10/8/0 & 50/42/42/0 & 15/6/6/0 & 6/0/0/0 & 1/1/1/0 & CE & 1/1/1/0 & 50/11/0/0 & 50/45/0/0 \\ \cline{2-13}
        ~ & CVE-2017-12876 & CWE-787 & ASAN & -- & -- & -- & -- & 1/1/1/0 & CE & 1/1/1/0 & -- & -- \\ \cline{2-13}
        ~ & CVE-2017-12877 & CWE-416 & ASAN & 24/0/0/0 & 50/0/0/0 & 9/0/0/0 & 50/0/0/0 & \ding{56} & CE & -- & -- & -- \\ \hline
        \multirow{3}{*}{imageworsener} & CVE-2017-7962 & CWE-369 & ASAN & 4/0/0/0 & 50/3/0/0 & 40/0/0/0 & 50/19/17/0 & 1/1/1/1 & \ding{56} & -- & 50/9/4/0 & 50/46/0/0 \\ \cline{2-13}
        ~ & CVE-2017-8325 & CWE-119 & ASAN & -- & -- & -- & -- & TO & \ding{56} & 1/1/0/0 & 50/0/0/0 & -- \\ \cline{2-13}
        ~ & CVE-2017-9206 & CWE-125 & ASAN & -- & -- & -- & -- & 1/1/1/0 & \ding{56} & \ding{56} & -- & -- \\ \hline
        \multirow{15}{*}{jasper} & CVE-2016-10248 & CWE-476 & ASAN & \ding{56} & 50/22/0/0 & 33/0/0/0 & 35/4/0/0 & \ding{56} & \ding{56} & -- & 50/8/0/0 & 50/32/0/0 \\ \cline{2-13}
        ~ & CVE-2016-10251 & CWE-190 & ASAN & 27/0/0/0 & 46/1/0/0 & 40/0/0/0 & 48/0/0/0 & \ding{56} & 1/0/0/0 & \ding{56} & -- & -- \\ \cline{2-13}
        ~ & CVE-2016-8691 & CWE-369 & ASAN & \ding{56} & 49/40/0/0 & 50/21/0/0 & 50/28/0/0 & 1/1/1/0 & \ding{56} & -- & 50/13/0/0 & 50/42/0/0 \\ \cline{2-13}
        ~ & CVE-2016-8692 & CWE-369 & ASAN & \ding{56} & 49/40/0/0 & 50/22/0/0 & 50/27/0/0 & \ding{56} & \ding{56} & -- & 50/11/0/0 & 50/42/0/0 \\ \cline{2-13}
        ~ & CVE-2016-8884 & CWE-476 & ASAN & 37/0/0/0 & 50/0/0/0 & 50/49/0/0 & 50/28/0/0 & \ding{56} & \ding{56} & -- & -- & -- \\ \cline{2-13}
        ~ & CVE-2016-8887 & CWE-476 & ASAN & 50/3/1/0 & 49/24/24/0 & 4/2/1/0 & 50/17/5/0 & \ding{56} & \ding{56} & -- & -- & -- \\ \cline{2-13}
        ~ & CVE-2016-9387 & CWE-190 & AE & \ding{56} & 50/9/9/0 & 50/32/4/0 & 42/5/1/0 & -- & 1/1/0/0 & 1/0/0/0 & -- & -- \\ \cline{2-13}
        ~ & CVE-2016-9388 & NVD-CWE-Other & AE & 29/21/0/0 & 32/0/0/0 & 50/0/0/0 & 46/8/0/0 & -- & \ding{56} & -- & -- & -- \\ \cline{2-13}
        ~ & CVE-2016-9391 & NVD-CWE-Other & AE & 10/0/0/0 & \ding{56} & \ding{56} & 25/20/0/0 & -- & \ding{56} & -- & -- & -- \\ \cline{2-13}
        ~ & CVE-2016-9394 & CWE-20 & AE & 10/0/0/0 & 49/2/2/0 & 44/0/0/0 & 7/0/0/0 & -- & -- & -- & -- & -- \\ \cline{2-13}
        ~ & CVE-2016-9396 & NVD-CWE-Other & AE & 31/1/1/0 & 50/0/0/0 & 47/3/2/0 & 48/9/3/0 & -- & \ding{56} & -- & 50/11/0/0 & 50/45/0/0 \\ \cline{2-13}
        ~ & CVE-2016-9398 & CWE-617 & AE & 45/0/0/0 & \ding{56} & 9/0/0/0 & 50/0/0/0 & -- & \ding{56} & -- & 50/21/0/0 & 50/33/0/0 \\ \cline{2-13}
        ~ & CVE-2016-9557 & CWE-190 & UBSAN & \ding{56} & 48/37/0/0 & 46/0/0/0 & 50/20/0/0 & \ding{56} & \ding{56} & \ding{56} & -- & -- \\ \cline{2-13}
        ~ & CVE-2016-9560 & CWE-787 & ASAN & 3/0/0/0 & 2/0/0/0 & 49/0/0/0 & 50/1/0/0 & \ding{56} & 2/1/0/0 & 1/1/0/0 & 50/13/0/0 & 50/32/0/0 \\ \cline{2-13}
        ~ & CVE-2017-6850 & CWE-476 & ASAN & 39/1/0/0 & 43/0/0/0 & \ding{56} & 37/23/0/0 & \ding{56} & \ding{56} & -- & -- & -- \\ \hline
        \multirow{3}{*}{libarchive} & CVE-2016-10349 & CWE-119 & ASAN & 46/13/0/0 & 44/5/4/0 & 50/48/0/0 & \ding{56} & \ding{56} & \ding{56} & \ding{56} & 50/19/0/0 & 50/49/0/0 \\ \cline{2-13}
        ~ & CVE-2016-10350 & CWE-119 & ASAN & 46/13/0/0 & 44/5/4/0 & 50/48/0/0 & \ding{56} & \ding{56} & \ding{56} & \ding{56} & 50/15/0/0 & 50/49/0/0 \\ \cline{2-13}
        ~ & CVE-2016-5844 & CWE-190 & UBSAN & 4/0/0/0 & 44/20/13/0 & 44/0/0/0 & 50/48/11/0 & \ding{56} & 1/1/1/0 & 1/0/0/0 & -- & -- \\ \hline
        libcroco & CVE-2017-7961 & CWE-119 & ASAN & 2/0/0/0 & 47/0/0/0 & 43/1/0/0 & 50/3/0/0 & \ding{56} & \ding{56} & \ding{56} & 50/9/0/0 & 50/0/0/0 \\ \hline
        \multirow{3}{*}{libjpeg} & CVE-2012-2806 & CWE-119 & ASAN & -- & -- & -- & -- & 1/1/1/1 & 1/0/0/0 & \ding{56} & -- & -- \\ \cline{2-13}
        ~ & CVE-2017-15232 & CWE-476 & ASAN & -- & -- & -- & -- & 1/1/1/1 & \ding{56} & -- & -- & -- \\ \cline{2-13}
        ~ & CVE-2018-19664 & CWE-125 & ASAN & -- & -- & -- & -- & \ding{56} & \ding{56} & \ding{56} & 50/21/0/0 & 50/46/0/0 \\ \hline
        \multirow{8}{*}{libming} & CVE-2018-8806 & CWE-416 & ASAN & 1/0/0/0 & \ding{56} & 50/0/0/0 & 19/0/0/0 & \ding{56} & CE & -- & -- & -- \\ \cline{2-13}
        ~ & CVE-2018-8964 & CWE-416 & ASAN & 1/0/0/0 & \ding{56} & 50/0/0/0 & 19/0/0/0 & \ding{56} & CE & -- & -- & -- \\ \cline{2-13}
        ~ & CVE-2016-9264 & CWE-119 & ASAN & 9/0/0/0 & 50/0/0/0 & 45/0/0/0 & 50/0/0/0 & 1/1/1/1 & CE & 1/0/0/0 & -- & -- \\ \cline{2-13}
        ~ & CVE-2016-9265 & CWE-369 & ASAN & 29/0/0/0 & \ding{56} & 41/0/0/0 & 50/4/0/0 & 1/1/1/1 & CE & -- & 50/6/0/0 & 50/44/16/0 \\ \cline{2-13}
        ~ & CVE-2016-9266 & CWE-189 & UBSAN & 48/0/0/0 & 50/0/0/0 & 50/1/1/0 & 50/0/0/0 & \ding{56} & CE & -- & -- & -- \\ \cline{2-13}
        ~ & CVE-2016-9827 & CWE-119 & ASAN & 30/3/0/0 & 39/0/0/0 & 50/0/0/0 & 38/8/0/0 & \ding{56} & CE & \ding{56} & 50/22/0/0 & 50/0/0/0 \\ \cline{2-13}
        ~ & CVE-2016-9828 & CWE-476 & ASAN & 12/0/0/0 & 21/0/0/0 & 42/0/0/0 & 50/0/0/0 & \ding{56} & CE & -- & 50/21/0/0 & 50/4/0/0 \\ \cline{2-13}
        ~ & CVE-2016-9829 & CWE-119 & ASAN & 15/4/0/0 & 50/0/0/0 & 50/1/0/0 & 50/5/0/0 & 1/1/1/0 & CE & 1/0/0/0 & -- & -- \\ \hline
        \multirow{23}{*}{libtiff} & CVE-2006-2025 & NVD-CWE-Other & ASAN/UBSAN & 43/0/0/0 & 50/0/0/0 & 48/0/0/0 & 50/0/0/0 & 1/1/1/0 & \ding{56} & -- & -- & -- \\ \cline{2-13}
        ~ & CVE-2010-2481 & CWE-119 & ASAN & 34/0/0/0 & 50/0/0/0 & 50/0/0/0 & 49/0/0/0 & \ding{56} & \ding{56} & \ding{56} & -- & -- \\ \cline{2-13}
        ~ & CVE-2013-4243 & CWE-119 & ASAN & 39/3/0/0 & 3/0/0/0 & \ding{56} & ~ & \ding{56} & \ding{56} & \ding{56} & 50/32/0/0 & 50/30/0/0 \\ \cline{2-13}
        ~ & CVE-2016-10092 & CWE-119 & ASAN & 1/0/0/0 & 50/34/14/0 & 50/48/0/0 & 50/26/0/0 & \ding{56} & \ding{56} & 1/0/0/0 & 50/26/20/0 & 50/44/10/6 \\ \cline{2-13}
        ~ & CVE-2016-10093 & CWE-119 & ASAN & \ding{56} & 42/0/0/0 & 46/0/0/0 & 50/1/0/0 & TO & \ding{56} & 1/1/0/0 & -- & -- \\ \cline{2-13}
        ~ & CVE-2016-10094 & CWE-189 & ASAN & 1/0/0/0 & 18/0/0/0 & 45/0/0/0 & 50/0/0/0 & \ding{56} & 2/0/0/0 & -- & 50/48/25/5 & 50/44/15/8 \\ \cline{2-13}
        ~ & CVE-2016-10266 & CWE-369 & ASAN & \ding{56} & 34/33/0/0 & 47/28/16/0 & 50/2/0/0 & \ding{56} & \ding{56} & -- & 50/16/0/0 & 50/47/0/0 \\ \cline{2-13}
        ~ & CVE-2016-10267 & CWE-369 & ASAN & -- & -- & -- & -- & \ding{56} & 1/1/0/0 & -- & -- & -- \\ \cline{2-13}
        ~ & CVE-2016-10268 & CWE-191 & ASAN & \ding{56} & 49/0/0/0 & 43/0/0/0 & 50/2/1/1 & \ding{56} & \ding{56} & \ding{56} & 50/19/6/0 & 50/44/12/4 \\ \cline{2-13}
        ~ & CVE-2016-10269 & CWE-125 & ASAN & -- & -- & -- & -- & \ding{56} & \ding{56} & 1/0/0/0 & -- & -- \\ \cline{2-13}
        ~ & CVE-2016-10270 & CWE-125 & ASAN & -- & -- & -- & -- & \ding{56} & \ding{56} & 1/0/0/0 & -- & -- \\ \cline{2-13}
        ~ & CVE-2016-10271 & CWE-119 & ASAN & 1/0/0/0 & 50/34/15/0 & 50/48/0/0 & 50/26/0/0 & \ding{56} & \ding{56} & 1/0/0/0 & 50/30/24/0 & 50/44/22/6 \\ \cline{2-13}
        ~ & CVE-2016-10272 & CWE-119 & ASAN & 1/0/0/0 & 50/34/14/0 & 50/48/0/0 & 50/26/0/0 & \ding{56} & \ding{56} & 1/1/0/0 & 50/20/0/0 & 50/44/0/0 \\ \cline{2-13}
        ~ & CVE-2016-5321 & CWE-119 & ASAN & \ding{56} & 47/0/0/0 & 1/1/1/0 & 50/24/3/1 & 1/1/1/1 & 2/2/2/1 & 1/1/0/0 & 50/25/13/0 & 50/43/16/11 \\ \cline{2-13}
        ~ & CVE-2016-9273 & CWE-125 & ASAN & \ding{56} & 16/7/0/0 & 18/6/0/0 & 50/9/5/0 & \ding{56} & 1/0/0/0 & \ding{56} & 50/3/0/0 & 50/44/14/0 \\ \cline{2-13}
        ~ & CVE-2016-9532 & CWE-125 & ASAN & 3/0/0/0 & 50/16/0/0 & 50/44/0/0 & 50/3/0/0 & \ding{56} & \ding{56} & 1/0/0/0 & -- & -- \\ \cline{2-13}
        ~ & CVE-2017-5225 & CWE-119 & ASAN & \ding{56} & 1/0/0/0 & 1/0/0/0 & 48/0/0/0 & \ding{56} & \ding{56} & 1/1/1/0 & -- & -- \\ \cline{2-13}
        ~ & CVE-2017-7595 & CWE-369 & ASAN & -- & -- & -- & -- & 1/1/1/1 & 1/1/0/0 & -- & 50/4/0/0 & 50/44/0/0 \\ \cline{2-13}
        ~ & CVE-2017-7598 & CWE-369 & UBSAN & 5/0/0/0 & 45/4/0/0 & 38/3/0/0 & 44/5/0/0 & \ding{56} & 1/0/0/0 & -- & -- & -- \\ \cline{2-13}
        ~ & CVE-2017-7599 & CWE-20 & UBSAN & -- & -- & -- & -- & \ding{56} & -- & -- & -- & -- \\ \cline{2-13}
        ~ & CVE-2017-7600 & CWE-20 & UBSAN & -- & -- & -- & -- & \ding{56} & -- & -- & -- & -- \\ \cline{2-13}
        ~ & CVE-2017-7601 & CWE-20 & UBSAN & -- & -- & -- & -- & 1/1/1/0 & -- & -- & 50/16/0/0 & 50/40/0/0 \\ \cline{2-13}
        ~ & CVE-2017-7602 & CWE-190 & UBSAN & 27/0/0/0 & 50/0/0/0 & \ding{56} & 41/0/0/0 & TO & 2/1/0/0 & \ding{56} & -- & -- \\ \hline
        \multirow{4}{*}{libxml2} & CVE-2012-5134 & CWE-119 & ASAN & 37/0/0/0 & 46/0/0/0 & 50/0/0/0 & 50/0/0/0 & 1/1/1/1 & \ding{56} & \ding{56} & 50/0/0/0 & 50/49/36/29 \\ \cline{2-13}
        ~ & CVE-2016-1838 & CWE-125 & ASAN & 47/0/0/0 & 50/0/0/0 & 50/0/0/0 & 38/24/0/0 & \ding{56} & \ding{56} & \ding{56} & -- & -- \\ \cline{2-13}
        ~ & CVE-2016-1839 & CWE-125 & ASAN & -- & -- & -- & -- & 1/1/1/1 & \ding{56} & \ding{56} & -- & -- \\ \cline{2-13}
        ~ & CVE-2017-5969 & CWE-476 & ASAN & -- & -- & -- & -- & 1/1/1/1 & \ding{56} & -- & -- & -- \\ \hline
        \multirow{2}{*}{libzip} & CVE-2017-12858 & CWE-415 & ASAN & 11/3/0/0 & 48/0/0/0 & 42/0/0/0 & 50/24/0/0 & \ding{56} & -- & -- & 50/4/0/0 & 50/47/0/0 \\ \cline{2-13}
        ~ & CVE-2017-14107 & CWE-770 & ASAN & -- & -- & -- & -- & \ding{56} & -- & -- & 50/0/0/0 & 50/0/0/0 \\ \hline
        \multirow{4}{*}{openjpeg} & CVE-2016-7445 & CWE-476 & ASAN & 49/22/0/0 & 50/0/0/0 & 50/4/0/0 & 7/7/0/0 & \ding{56} & \ding{56} & -- & -- & -- \\ \cline{2-13}
        ~ & CVE-2017-12982 & CWE-119 & ASAN & \ding{56} & 50/0/0/0 & 50/2/0/0 & 50/42/0/0 & \ding{56} & \ding{56} & \ding{56} & 50/21/0/0 & 50/44/0/0 \\ \cline{2-13}
        ~ & CVE-2017-14041 & CWE-787 & ASAN & \ding{56} & 13/0/0/0 & \ding{56} & \ding{56} & \ding{56} & \ding{56} & \ding{56} & 50/9/0/0 & 50/7/0/0 \\ \cline{2-13}
        ~ & CVE-2017-14151 & CWE-119 & ASAN & -- & -- & -- & -- & \ding{56} & \ding{56} & \ding{56} & 50/11/0/0 & 50/26/26/0 \\ \cline{2-13}
        ~ & CVE-2017-9210 & CWE-835 & ASAN & -- & -- & -- & -- & \ding{56} & CE & -- & 50/0/0/0 & 50/25/0/0 \\ \hline
        \multirow{3}{*}{zziplib} & CVE-2017-5974 & CWE-119 & ASAN & -- & -- & -- & -- & 1/1/1/0 & CE & \ding{56} & -- & -- \\ \cline{2-13}
        ~ & CVE-2017-5975 & CWE-787 & ASAN & -- & -- & -- & -- & 1/1/1/1 & CE & \ding{56} & -- & -- \\ \cline{2-13}
        ~ & CVE-2017-5976 & CWE-787 & ASAN & -- & -- & -- & -- & \ding{56} & CE & \ding{56} & -- & -- \\ \hline
        \multirow{2}{*}{jhead} & CVE-2020-26208 & CWE-787 & ASAN & 44/0/0/0 & 50/0/0/0 & 25/0/0/0 & 50/0/0/0 & 1/1/1/0 & \ding{56} & \ding{56} & 50/5/0/0 & 50/37/0/0 \\ \cline{2-13}
        ~ & CVE-2020-28840 & CWE-787 & ASAN & 2/0/0/0 & \ding{56} & 48/0/0/0 & 50/0/0/0 & \ding{56} & \ding{56} & \ding{56} & 50/12/0/0 & 50/0/0/0 \\ \hline
        \multirow{4}{*}{ngiflib} & CVE-2018-10677 & CWE-787 & ASAN & 4/0/0/0 & 48/0/0/0 & \ding{56} & 50/0/0/0 & \ding{56} & \ding{56} & \ding{56} & 50/11/0/0 & 50/43/0/0 \\ \cline{2-13}
        ~ & CVE-2018-10717 & CWE-787 & ASAN & 6/0/0/0 & 14/0/0/0 & 47/0/0/0 & 45/0/0/0 & \ding{56} & \ding{56} & \ding{56} & -- & -- \\ \cline{2-13}
        ~ & CVE-2019-16346 & CWE-787 & ASAN & \ding{56} & \ding{56} & 50/0/0/0 & 50/0/0/0 & \ding{56} & \ding{56} & \ding{56} & -- & -- \\ \cline{2-13}
        ~ & CVE-2019-16347 & CWE-787 & ASAN & \ding{56} & \ding{56} & 50/0/0/0 & 50/0/0/0 & \ding{56} & \ding{56} & \ding{56} & -- & -- \\ \hline
    \end{tabular}
    }
\end{table*}

\begin{table*}[!htbp]
    \centering
\caption{Comparing patch generation capabilities of the 2 AVR tools and 2 APR tools with respect to \textsc{Vul4J}}
    \label{tab: Detailed Experiment Java}
    \resizebox{!}{0.475\textheight}{
    \begin{tabular}{|c|c|c|c|c|c|c|c|}
    \hline
        \multicolumn{4}{|c|}{\bf Vulnerable program}& \multicolumn{2}{c|}{\bf AVR} &\multicolumn{2}{c|}{\bf APR} \\ \hline
        ID & CVE ID & CWE ID & Compilable & Seader\cite{zhang2022Seader} & SeqTrans\cite{chi2023seqtrans} & ThinkRepair\cite{yin2024thinkrepair} & SRepair\cite{xiang2024srepair} \\ \hline
        VUL4J-1 & CVE-2017-18349 & CWE-20 & \ding{52} & -- & -- & 50/19/0/0 & 50/33/0/0 \\ \hline
        VUL4J-2 & CVE-2017-5662 & CWE-611 & \ding{52} & -- & -- & -- & 50/22/11/0 \\ \hline
        VUL4J-3 & CVE-2015-0263 & Not Mapping & \ding{56} & -- & -- & -- & -- \\ \hline
        VUL4J-4 & CVE-2015-0264 & Not Mapping & \ding{56} & -- & -- & -- & -- \\ \hline
        VUL4J-5 & APACHE-COMMONS-001 & Not Mapping & \ding{52} & 1/0/0/0 & -- & 50/45/6/5 & 50/23/2/0 \\ \hline
        VUL4J-6 & CVE-2018-1324 & CWE-835 & \ding{52} & 1/0/0/0 & 50/0/0/0 & 50/22/6/1 & 40/37/6/6 \\ \hline
        VUL4J-7 & CVE-2018-11771 & CWE-835 & \ding{52} & 1/0/0/0 & -- & 50/49/0/0 & 50/50/0/0 \\ \hline
        VUL4J-8 & CVE-2019-12402 & CWE-835 & \ding{52} & 1/0/0/0 & -- & 50/42/1/0 & 50/32/1/0 \\ \hline
        VUL4J-9 & CVE-2020-1953 & Not Mapping & \ding{52} & -- & -- & -- & 43/9/7/0 \\ \hline
        VUL4J-10 & CVE-2013-2186 & CWE-20 & \ding{52} & -- & -- & 50/32/0/0 & 47/27/0/0 \\ \hline
        VUL4J-11 & CVE-2014-0050 & CWE-264 & \ding{52} & -- & -- & -- & 50/17/8/0 \\ \hline
        VUL4J-12 & CVE-2018-17202 & CWE-835 & \ding{52} & -- & -- & 47/47/22/0 & 49/38/26/3 \\ \hline
        VUL4J-13 & CVE-2018-17201 & CWE-835 & \ding{52} & -- & -- & 50/33/7/0 & 43/30/2/0 \\ \hline
        VUL4J-14 & CVE-2021-29425 & CWE-20 & \ding{52} & -- & -- & -- & 50/41/0/0 \\ \hline
        VUL4J-15 & CVE-2016-8739 & CWE-611 & \ding{56} & -- & -- & -- & -- \\ \hline
        VUL4J-16 & CVE-2015-5253 & CWE-264 & \ding{56} & -- & -- & -- & -- \\ \hline
        VUL4J-17 & HTTPCLIENT-1803 & Not Mapping & \ding{52} & -- & -- & 50/49/0/0 & 40/36/0/0 \\ \hline
        VUL4J-18 & CVE-2019-0225 & CWE-22 & \ding{52} & -- & -- & 50/0/0/0 & 39/39/18/0 \\ \hline
        VUL4J-19 & PDFBOX-3341 & Not Mapping & \ding{52} & -- & -- & 50/3/0/0 & 50/43/0/0 \\ \hline
        VUL4J-20 & CVE-2018-11797 & Not Mapping & \ding{52} & -- & -- & 50/0/0/0 & 45/39/6/2 \\ \hline
        VUL4J-21 & CVE-2014-8152 & CWE-254 & \ding{52} & 1/0/0/0 & -- & -- & 49/6/0/0 \\ \hline
        VUL4J-22 & CVE-2016-6802 & CWE-284 & \ding{52} & -- & -- & 50/3/0/0 & 50/47/0/0 \\ \hline
        VUL4J-23 & CVE-2016-5394 & CWE-79 & \ding{52} & -- & -- & -- & 50/45/0/0 \\ \hline
        VUL4J-24 & CVE-2016-6798 & CWE-611 & \ding{52} & -- & -- & -- & 50/26/0/0 \\ \hline
        VUL4J-25 & CVE-2017-15717 & CWE-79 & \ding{52} & -- & -- & -- & 50/26/0/0 \\ \hline
        VUL4J-26 & CVE-2016-4465 & CWE-20 & \ding{52} & -- & -- & 50/50/1/1 & 49/35/0/0 \\ \hline
        VUL4J-27 & CVE-2014-0112 & CWE-264 & \ding{52} & -- & -- & -- & 49/43/33/0 \\ \hline
        VUL4J-28 & CVE-2014-0113 & CWE-264 & \ding{52} & -- & -- & -- & 47/0/0/0 \\ \hline
        VUL4J-29 & CVE-2014-0116 & CWE-264 & \ding{52} & -- & -- & -- & 49/42/0/0 \\ \hline
        VUL4J-30 & CVE-2016-8738 & CWE-20 & \ding{52} & -- & 50/0/0/0 & 50/44/1/0 & 50/45/0/0 \\ \hline
        VUL4J-31 & CVE-2016-4436 & Not Mapping & \ding{52} & -- & -- & -- & 50/35/0/0 \\ \hline
        VUL4J-32 & CVE-2015-1831 & Not Mapping & \ding{52} & -- & -- & -- & -- \\ \hline
        VUL4J-33 & CVE-2016-3081 & CWE-77 & \ding{52} & -- & -- & -- & 49/28/0/0 \\ \hline
        VUL4J-34 & CVE-2016-2162 & CWE-79 & \ding{52} & -- & -- & -- & 35/30/0/0 \\ \hline
        VUL4J-35 & CVE-2014-7809 & CWE-352 & \ding{52} & -- & 50/0/0/0 & -- & -- \\ \hline
        VUL4J-36 & CVE-2018-8017 & CWE-835 & \ding{52} & -- & -- & -- & -- \\ \hline
        VUL4J-37 & CVE-2015-8581 & CWE-502 & \ding{56} & -- & -- & -- & -- \\ \hline
        VUL4J-38 & CVE-2014-4172 & CWE-74 & \ding{52} & -- & -- & -- & 47/37/27/0 \\ \hline
        VUL4J-39 & CVE-2018-1192 & CWE-200 & \ding{56} & -- & -- & -- & -- \\ \hline
        VUL4J-40 & CVE-2019-3775 & CWE-287 & \ding{52} & -- & -- & 50/18/0/0 & 38/29/0/0 \\ \hline
        VUL4J-41 & CVE-2018-1002200 & CWE-22 & \ding{52} & -- & -- & 50/37/0/0 & 50/46/0/0 \\ \hline
        VUL4J-42 & CVE-2017-1000487 & CWE-78 & \ding{52} & -- & -- & -- & 49/0/0/0 \\ \hline
        VUL4J-43 & CVE-2018-20227 & CWE-22 & \ding{52} & -- & -- & -- & 36/21/2/0 \\ \hline
        VUL4J-44 & CVE-2013-5960 & CWE-310 & \ding{52} & 1/0/0/0 & -- & -- & 50/32/6/0 \\ \hline
        VUL4J-45 & CVE-2018-1000854 & CWE-74 & \ding{52} & -- & -- & -- & 50/46/0/0 \\ \hline
        VUL4J-46 & CVE-2016-3720 & Not Mapping & \ding{52} & -- & -- & 50/49/0/0 & 50/30/17/14 \\ \hline
        VUL4J-47 & CVE-2016-7051 & CWE-611 & \ding{52} & -- & -- & 50/40/12/1 & 48/41/22/16 \\ \hline
        VUL4J-48 & CVE-2018-1000531 & CWE-20 & \ding{52} & 1/0/0/0 & -- & 47/40/0/0 & 37/34/2/0 \\ \hline
        VUL4J-49 & CVE-2018-1000125 & CWE-20 & \ding{52} & 1/0/0/0 & -- & -- & -- \\ \hline
        VUL4J-50 & CVE-2013-4378 & CWE-79 & \ding{52} & -- & 50/1/1/1 & 25/24/19/19 & 50/33/26/3 \\ \hline
        VUL4J-51 & CVE-2018-1000054 & CWE-918 & \ding{56} & -- & -- & -- & -- \\ \hline
        VUL4J-52 & CVE-2018-1000865 & CWE-269 & \ding{52} & -- & -- & 50/36/0/0 & 50/11/0/0 \\ \hline
        VUL4J-53 & CVE-2018-1999044 & CWE-835 & \ding{52} & -- & -- & 50/0/0/0 & 34/11/9/8 \\ \hline
        VUL4J-54 & CVE-2017-1000355 & CWE-502 & \ding{56} & -- & -- & -- & -- \\ \hline
        VUL4J-55 & CVE-2018-1000864 & CWE-835 & \ding{52} & -- & -- & 50/4/0/0 & 50/34/0/0 \\ \hline
        VUL4J-56 & CVE-2018-1000056 & CWE-918 & \ding{52} & -- & -- & -- & 50/18/1/0 \\ \hline
        VUL4J-57 & CVE-2018-1000089 & CWE-532 & \ding{52} & -- & -- & -- & 50/12/0/0 \\ \hline
        VUL4J-58 & CVE-2018-1000111 & CWE-863 & \ding{52} & -- & -- & -- & 50/24/0/0 \\ \hline
        VUL4J-59 & CVE-2015-6748 & CWE-79 & \ding{52} & -- & -- & 50/37/4/3 & 47/38/2/2 \\ \hline
        VUL4J-60 & CVE-2016-10006 & CWE-79 & \ding{52} & -- & -- & 50/44/0/0 & 50/40/0/0 \\ \hline
        VUL4J-61 & CVE-2018-1000820 & CWE-611 & \ding{52} & -- & -- & 50/37/0/0 & 50/0/0/0 \\ \hline
        VUL4J-62 & CVE-2018-18389 & CWE-287 & \ding{52} & -- & -- & 50/1/0/0 & 47/0/0/0 \\ \hline
        VUL4J-63 & CVE-2018-1000615 & Not Mapping & \ding{52} & -- & -- & 50/0/0/0 & 50/33/1/0 \\ \hline
        VUL4J-64 & CVE-2018-20157 & CWE-611 & \ding{52} & -- & -- & 50/8/0/0 & 49/31/20/0 \\ \hline
        VUL4J-65 & CVE-2018-19859 & CWE-22 & \ding{52} & -- & -- & 50/2/0/0 & 39/32/16/0 \\ \hline
        VUL4J-66 & CVE-2020-1695 & CWE-20 & \ding{52} & -- & -- & 50/0/0/0 & 50/50/0/0 \\ \hline
        VUL4J-67 & CVE-2018-1274 & CWE-770 & \ding{52} & -- & -- & -- & 48/25/0/0 \\ \hline
        VUL4J-68 & CVE-2017-8046 & CWE-20 & \ding{52} & -- & -- & -- & 50/24/0/0 \\ \hline
        VUL4J-69 & CVE-2016-9878 & CWE-22 & \ding{52} & -- & -- & 50/44/0/0 & 50/40/1/0 \\ \hline
        VUL4J-70 & CVE-2018-15756 & Not Mapping & \ding{52} & -- & -- & -- & 50/35/0/0 \\ \hline
        VUL4J-71 & CVE-2018-1272 & Not Mapping & \ding{52} & -- & 50/0/0/0 & -- & -- \\ \hline
        VUL4J-72 & CVE-2018-15801 & CWE-345 & \ding{56} & -- & -- & -- & -- \\ \hline
        VUL4J-73 & CVE-2019-11272 & CWE-522 & \ding{56} & -- & -- & -- & -- \\ \hline
        VUL4J-74 & CVE-2019-3795 & CWE-332 & \ding{56} & -- & -- & -- & -- \\ \hline
        VUL4J-75 & CVE-2016-4977 & CWE-19 & \ding{52} & -- & -- & -- & 50/6/0/0 \\ \hline
        VUL4J-76 & CVE-2018-1000850 & CWE-22 & \ding{52} & -- & -- & -- & 48/29/0/0 \\ \hline
        VUL4J-77 & CVE-2017-1000207 & CWE-502 & \ding{52} & -- & -- & -- & 29/6/1/1 \\ \hline
        VUL4J-78 & CVE-2019-10173 & CWE-502 & \ding{52} & -- & -- & -- & 48/42/0/0 \\ \hline
        VUL4J-79 & CVE-2018-1002201 & CWE-22 & \ding{52} & -- & -- & -- & 46/41/0/0 \\ \hline
    \end{tabular}
    }
\end{table*}

Table \ref{tab: Detailed Experiment} summarizes the detailed evaluation results on the patch generation capabilities of the seven AVR tools and the two APR tools via \textsc{Vul4C}. We show evaluation results which at least one AVR tool or APR tool generates valid candidate patches (i.e., can be successfully applied to vulnerabilities). 
Table \ref{tab: Detailed Experiment Java} summarizes the detailed evaluation results on the patch generation capabilities of the two AVR tools and two APR tools via \textsc{Vul4J},
where the results can be understood as follows: for the notation $a$/$b$/$c$/$d$, $a$ denotes the total number of valid candidate patches, $b$ denotes the total number of compilable patches, $c$ denotes the total number of patches that pass the test, $d$ denotes the number of correct patches; CE denotes that an AVR tool encounters a compilation error when testing a vulnerability; TO denotes that an AVR tool times out when testing a vulnerability,; - denotes that an AVR tool is not applicable to a vulnerability; \ding{55} denotes that an AVR tool fails to generate a patch for a vulnerability; AE denotes an assertion error defined by developer.

\end{appendices}

\end{document}